\RequirePackage[2020-02-02]{latexrelease}
\documentclass[preprint,aps,floatfix]{revtex4}
\DeclareUnicodeCharacter{00A0}{ }
\usepackage{graphicx}
\usepackage{textgreek}
\usepackage{verbatim} 
\usepackage{amssymb}
\usepackage{url}
\usepackage{amsmath,amssymb}
\usepackage{mathtools} 
\usepackage[symbol]{footmisc}
\usepackage{subcaption}
\usepackage{capt-of}
\usepackage{bigints}
\usepackage{romannum}
\usepackage{xcolor} % For \hl
\usepackage{soul} % For \hl
\AtBeginDocument{\pagenumbering{arabic}}

\def\be{\begin{eqnarray}}
\def\ee{\end{eqnarray}}
\def\nn{\nonumber}

\begin{document}

\title{Unveiling Skewness Dependence of Quark Wigner Distributions}

  \author{Sujit Jana${}^1$\footnote{d21ph010@phy.svnit.ac.in}, Vikash Kumar Ojha${}^1$\footnote{vko@phy.svnit.ac.in (corresponding author)}}
  \affiliation{${}^1$ Department of Physics, Sardar Vallabhbhai National Institute of Technology, Surat, 395 007, India.%\\ ${}^2$ Department of Physics, National Institute of Technology, Kurukshetra, Haryana, 136 119, India.
  }
\begin{abstract}
%\begin{abstract}
We present a detailed investigation of the skewness dependence of quark Wigner distributions within the light-front dressed quark model. % Wigner distributions provide a multidimensional phase-space representation of parton dynamics, unifying information from generalized parton distributions (GPDs) and transverse momentum-dependent distributions (TMDs). 
While previous studies have largely focused on the forward limit, we explore the impact of nonzero longitudinal momentum transfer (\(\xi \neq 0\)) on the full set of leading-twist quark Wigner distributions across various polarization configurations. % Analytic expressions for the corresponding generalized transverse momentum-dependent distributions (GTMDs) are derived, and the resulting 
%Wigner distributions are visualized in transverse impact parameter space, transverse momentum space, and mixed phase space. 
We observe characteristic distortions in the spatial and momentum correlations with increasing skewness, including the emergence of dipole and quadrupole patterns, asymmetries, and localization effects. These features reflect spin-orbit correlations and quantum interference between light-front wave function components with differing orbital angular momentum.% Our findings highlight the critical role of skewness in revealing the full structure of hadronic phase space %and provide theoretical guidance for future experimental and lattice QCD explorations of GTMD-sensitive observables.
%\end{abstract}

%We investigate the Wigner distribution of quark when momentum transferred to the target has non-zero transverse and longitudinal components. The density of quarks in both momentum and position spaces is examined as a function of skewness ($\xi$), which encodes information about the longitudinal component of the momentum transferred to the target. Additionally, we analyze the impact of quark and target polarization on the quark distributions.
\end{abstract}
\maketitle

%----------------Section----------------------%

\section{Introduction}
%\section{Introduction}

Understanding the internal structure of the nucleon in terms of its quark and gluon constituents remains a central pursuit in modern hadronic physics \cite{Accardi:2012qut,Anderle:2021wcy,gluck1998dynamical,gluck1995dynamical,martin1998parton,Lorce:2025aqp,Karr:2020wgh,Gao:2017yyd, Diehl:2023nmm, Constantinou:2020pek}. Over the past two decades, substantial progress has been made through the development of generalized parton distributions (GPDs) \cite{belitsky2005unraveling,diehl2003generalized,ji1997deeply,Mukherjee:2013yf} and transverse momentum-dependent distributions (TMDs) \cite{brodsky2002final,bacchetta2007semi,barone2002transverse,mulders1996complete,Boussarie:2023izj,Echevarria:2016scs}, each offering complementary perspectives on partonic correlations. While GPDs provide spatial imaging through deeply virtual Compton scattering, TMDs offer momentum-resolved information in semi-inclusive deep inelastic scattering. However, a fully unified, multidimensional picture requires going beyond these frameworks.

Wigner distributions, defined as quantum phase-space quasi-probability distributions, encapsulate the most general correlations between a parton's transverse position and momentum \cite{wigner1932quantum,ji2003viewing,Belitsky:2003nz, radhakrishnan2022wigner} . At leading twist, they encode the maximal amount of spatial and momentum information allowed by the uncertainty principle. These distributions are the Fourier transform of the generalized transverse momentum-dependent distributions (GTMDs) \cite{meissner2009generalized,lorce2013structure,Lorce:2011dv}, which reduce to GPDs and TMDs in specific limits. As such, Wigner distributions serve as a powerful tool for exploring parton dynamics at the interface of position, momentum, and spin.

Despite their theoretical richness, Wigner distributions remain challenging to access both computationally and experimentally. Existing studies often focus on the forward limit (\(\xi = 0\)), where \(\xi\) denotes the longitudinal momentum transfer (skewness) \cite{lorce2012quark,lorce2011quark,mukherjee2015wigner,mukherjee2014quark,kaur2018wigner,liu2015quark,chakrabarti2016wigner,chakrabarti2017quark,more2017quark,more2018wigner,Jana:2023btd}. However, the skewness parameter plays a crucial role in off-forward processes and captures essential information about longitudinal momentum correlations \cite{ojha2023quark, maji2022leading,Guo:2023ahv,Mamo:2022jhp,Rinaldi:2017roc,Chakrabarti:2024hwx}. A systematic investigation of nonzero skewness effects can unveil novel features such as spin-orbit couplings, interference between wave function components with differing orbital angular momentum, and modifications to the spatial-momentum correlation structure of quarks.

In this work, we aim to unveil the skewness dependence of quark Wigner distributions within the light-front dressed quark model. We consider all leading-twist Wigner distributions for quarks in various polarization configurations of the quark and the target. Using the analytic expressions of the leading twist GTMDs and numerically visualizing the resulting Wigner distributions in transverse impact parameter space, transverse momentum space, and mixed space, we explore how nonzero skewness deforms the underlying phase-space structure. In particular, we analyze the emergence of multipole patterns, asymmetries, and localization effects, and provide a possible physical interpretation of their origin.

The novelty of this study lies in its explicit inclusion of nonzero skewness in all Wigner components and its systematic treatment across multiple polarization channels. To our knowledge, this represents one of the most comprehensive analyses of skewness dependence in the Wigner framework using an analytically tractable model. Our findings bridge the gap between forward-limit studies and the broader, off-forward landscape relevant for exclusive processes. The results offer valuable guidance for ongoing efforts in lattice QCD \cite{Chen:2019lcm,Riberdy:2023awf,Mamo:2024vjh} and experimental programs aiming to access phase-space distributions through GTMD-sensitive observables \cite{Accardi:2012qut,Anderle:2021wcy,Bhattacharya:2018lgm,Bhattacharya:2023yvo}.

\section{Convention, Kinematics and Model \label{sec:Conven}}
We choose to use the light front coordinate system  ($x^+,x^-,x_\perp$), defining light-front time and longitudinal spatial variables as $x^\pm=x^0 \pm x^3$, while all additional conventions are detailed in \cite{harindranath1996introduction,zhang1994light}. We consider a quark dressed by a gluon, probed by a virtual photon that transfers a four-momentum \( \Delta \), with invariant momentum transfer squared given by \( t = \Delta^2 \). The longitudinal momentum transfer is characterized by the skewness parameter $
\xi = \frac{\Delta^+}{2P^+} $,
where \( P = \frac{1}{2}(p + p') \) is the average momentum of the initial (\( p \)) and final (\( p' \)) dressed quark states defined in symmetric frame as \cite{brodsky2001light}
 %We examine a system consisting of a quark dressed by a gluons interacting with a virtual photon that probes the system with energy transfer $t=\Delta^2$, where we define the transfer of longitudinal momentum as $\xi=\frac{\Delta^+}{2P^+}$. For kinematics, we depict the initial momentum of the dressed quark as $p$ and the final momentum as $p'$, as follows:
\begin{align}
    p=&\Big((1+\xi)P^+,\frac{\Delta_\perp}{2},\frac{m^2+\frac{{\Delta_\perp}^2}{4}}{(1+\xi)P^+}\Big),\\
    p'=&\Big((1-\xi)P^+,-\frac{\Delta_\perp}{2},\frac{m^2+\frac{{\Delta_\perp}^2}{4}}{(1-\xi)P^+}\Big).
\end{align}
%Within this context, the target's average momentum is expressed as $P=\frac{p+p'}{2}$, while the momentum transfer to the target is denoted as $\Delta=p-p'$. 
%The symbol $m$ stands for the target mass. The quark's average four-momentum is denoted by the symbol $k$, where $k^+=xP^+$.
The target mass is denoted by \( m \). The average four-momentum of the quark is represented by \( k \), with its longitudinal component given by \( k^+ = x P^+ \), where \( x \) is the longitudinal momentum fraction carried by the quark.

%At the partonic scale, a proton is a very complicated particle because it is a bound state of three valence quarks and gluons.
The analysis of strongly bound multiparticle systems in QCD remains a complex endeavor. To gain insight into the internal dynamics of partonic bound states, various simplified models have been employed. These include the chiral quark soliton model~\cite{lorce2011quark,lorce2012quark}, quark–diquark models~\cite{kaur2018wigner,brodsky2001light,kumar2015single,bacchetta2008transverse,bacchetta2016electron}, the dressed quark model~\cite{harindranath1999nonperturbative,harindranath1999orbital,zhang1993light}, and the AdS/QCD-inspired quark–diquark model~\cite{chakrabarti2016wigner,chakrabarti2017quark,maji2016light}.
%The analysis of a strongly-bound multiparticle state is naturally cumbersome. However, the bound state of partons has been studied using some simplified models of the quark's bound state, like the chiral quark soliton model \cite{lorce2011quark,lorce2012quark}, quark-diquark model \cite{kaur2018wigner,brodsky2001light,kumar2015single,bacchetta2008transverse,bacchetta2016electron}, dressed quark model \cite{harindranath1999nonperturbative,harindranath1999orbital,zhang1993light}, and ADS/QCD quark-diquark model \cite{chakrabarti2016wigner,chakrabarti2017quark,maji2016light}. 
%In our previous paper\cite{ojha2023quark}, we examined GTMDs and Wigner distributions for a quark in boost-invariant space using the dressed quark model.
In this article, we focus using the dressed quark model on the behavior of Wigner distribution with the change in the skewness. The state of the dressed quark with momentum $p$ and helicity $\sigma$ can be described as \cite{zhang1993light,mukherjee2014quark,ojha2023quark}

 \begin{align}
 \Big{| }p^+,p_\perp,\sigma  \Big{\rangle} = \Phi^\sigma(p) b^\dagger_\sigma(p)
 | 0 \rangle +
 \sum_{\sigma_1 \sigma_2} \int [dp_1]
 \int [dp_2] \sqrt{16 \pi^3 p^{+}}
 \delta^3(p-p_1-p_2) \nn \\ \Phi^\sigma_{\sigma_1 \sigma_2}(p;p_1,p_2) 
b^\dagger_{\sigma_1}(p_1) 
 a^\dagger_{\sigma_2}(p_2)  | 0 \rangle,
 \end{align}
 The Fock-state expansion is truncated to include only the two-particle quark--gluon sector. The functions \( \Phi^\sigma \) and \( \Phi^\sigma_{\sigma_1 \sigma_2} \) represent the light-front wave functions (LFWFs) for the single-particle and two-particle states, respectively. Throughout this work, we adopt the notation $
[dp] = \frac{dp^+\, d^2p_\perp}{\sqrt{16\pi^3}p^+}
$ for the integration measure.
%
%where we truncated the Fock state expansion upto two particle quark-gluon state. The functions $\Phi^\sigma$ and $\Phi^\sigma_{\sigma_1\sigma_2}$ are the light-front wave functions (LFWF) for a single particle and two particles state, respectively. Additionally, we use the notation $[dp]=\frac{dp^+d^2p_\perp}{\sqrt{16\pi^3}p^+}$. 
%
The LFWF can be conveniently expressed in terms of the Jacobi momenta \( (x_i, q_{i\perp}) \), defined as
\begin{align}
    p^+_i = x_i p^+, \quad q_{i\perp} = k_{i\perp} + x_i p_\perp,
\end{align}
with the constraints \( \sum_i x_i = 1 \) and \( \sum_i q_{i\perp} = 0 \). 
The relation between the two-particle light-front wave function \( \Phi^\sigma_{\sigma_1 \sigma_2} \) and the corresponding boost-invariant wave function \( \Psi^\sigma_{\sigma_1 \sigma_2}(x, q_\perp) \) is given by
\begin{align}
    \Psi^\sigma_{\sigma_1 \sigma_2}(x, q_\perp) = \Phi^\sigma_{\sigma_1 \sigma_2} \sqrt{P^+}.
\end{align}
The explicit form of the boost-invariant two-particle LFWF read \cite{harindranath1999orbital,more2017quark}
%
%
%The LFWF can be expressed using Jacobi momenta $(x_i,q_{i\perp})$, defined as 
%\begin{align}
    %p^+_i=x_ip^+,\;\;\;\;q_{i\perp}=k_{i\perp}+x_ip_\perp,
%\end{align}
%where $\sum_ix_i=1,\ \sum_iq_{i\perp}=0$.
%
%The relation between the two-particle light-front wave function and the boost-invariant LFWF is given by $\Psi^\sigma_{\sigma_1\sigma_2}(x,q_\perp)=\Phi^\sigma_{\sigma_1,\sigma_2}\sqrt{P^+}$. The boost-invariant two-particle LFWF is  \cite{harindranath1999orbital,more2017quark}
%
\begin{align}
    \Psi^{\sigma a}_{\sigma_1 \sigma_2}(x_,q_{\perp }) = 
\frac{1}{\Big[    m^2 - \frac{m^2 + (q_{\perp })^2 }{x} - \frac{(q_{\perp})^2}{1-x} \Big]}
\frac{g}{\sqrt{2(2\pi)^3}} T^a \chi^{\dagger}_{\sigma_1} \frac{1}{\sqrt{1-x}}
\nn \\ \Big[ 
-2\frac{q_{\perp}}{1-x}   -  \frac{(\sigma_{\perp}.q_{\perp})\sigma_{\perp}}{x}
+\frac{im\sigma_{\perp}(1-x)}{x}\Big]
\chi_\sigma (\epsilon_{\perp \sigma_2})^{*}.
\end{align}
The variables \( \sigma_1 \), \( \sigma_2 \), \( x \), \( m \), and \( \epsilon_{\perp}^{\sigma_2} \) denote the helicities of the quark and gluon, the longitudinal momentum fraction carried by the quark, the quark mass, and the transverse polarization vector of the gluon, respectively. The transverse polarization vectors of the gluon are defined as~\cite{diehl2003generalized,mukherjee2013generalized}
\begin{align}
\epsilon_\perp^{\pm} = \frac{1}{\sqrt{2}}(\mp 1,\,-i).
\end{align}
%
%The symbols $\sigma_1, \sigma_2, x, m, \epsilon_{\perp\sigma_2}$ represent the quark's helicity, gluon's helicity, a fraction of the target state's longitudinal momentum, quark's mass, and gluon's polarization vector, respectively. The definition of the photon's polarization vector is as follows \cite{diehl2003generalized,mukherjee2013generalized}:
%\begin{align}
%\epsilon_\perp^{\pm}=\frac{1}{\sqrt{2}}(\mp1,-i).
%\end{align}
 
%-------------------Section--------------------%

\section{Wigner distribution for non-zero skewness\label{sec:WD}}
%\section{Wigner Distribution for Non-Zero Skewness}
%
The Wigner distribution encodes the simultaneous information of partonic degrees of freedom in both momentum space \((x, k_\perp)\) and position space \((\sigma, b_\perp)\), thereby offering a six-dimensional phase-space representation of quark distributions inside hadrons. While the Wigner distributions have been extensively studied in various phenomenological models in the forward limit \((\xi = 0)\), their structure at non-zero skewness \((\xi \ne 0)\) remains relatively less explored and is essential for understanding the interplay between longitudinal momentum transfer and spatial localization in QCD.

The skewness parameter \(\xi\) represents the longitudinal momentum transfer between the initial and final hadronic states and is Fourier conjugate to the boost-invariant longitudinal impact parameter, defined as
\begin{align}
    \sigma = \frac{1}{2} b^- P^+,
\end{align}
where \(P^+ = (p^+ + p'^+)/2\) is the average light-front longitudinal momentum of the initial and final states. On the other hand, the transverse impact parameter \(\bm{b}_\perp\) is conjugate to the transverse momentum transfer variable \(\bm{D}_\perp\), defined as
\begin{align}
    \bm{D}_\perp = \frac{\bm{\Delta}_\perp}{1 - \xi^2},
\end{align}
which accounts for the non-trivial skewness dependence in the transverse structure.

To access the Wigner distributions, one starts from the fully unintegrated quark-quark correlator \(W^{[\Gamma]}(x,\xi,\bm{\Delta}_\perp,\bm{k}_\perp;S)\), which depends on the average quark momentum fraction \(x\), the skewness \(\xi\), the quark’s average transverse momentum \(\bm{k}_\perp\), and the transverse momentum transfer \(\bm{\Delta}_\perp\). The Wigner distribution is then obtained by performing a Fourier transform with respect to the transverse momentum transfer variable \(\bm{D}_\perp\), yielding
\begin{align}
    \rho^{[\Gamma]}(x,\xi,\bm{k}_\perp,\bm{b}_\perp;S)
    = \int \frac{d^2 \bm{D}_\perp}{(2\pi)^2} e^{-i \bm{D}_\perp \cdot \bm{b}_\perp} 
    W^{[\Gamma]}(x, \xi, \bm{\Delta}_\perp, \bm{k}_\perp; S), \label{wig1}
\end{align}
where \(S\) denotes the polarization state of the dressed quark, and \(\Gamma\) specifies the Dirac structure used to project different spin configurations. 

The quark-quark correlator in Eq.(\ref{wig1}) is defined as~\cite{meissner2009generalized}
\begin{align}
    W^{[\Gamma]}_{\lambda,\lambda'}(x,\xi,\bm{\Delta}_\perp,\bm{k}_\perp)
    = \frac{1}{2} \int \frac{dz^-}{2\pi} \frac{d^2 \bm{z}_\perp}{(2\pi)^2} 
    e^{i p \cdot z} \left\langle p', \lambda' \left| 
    \bar{\psi}(-\tfrac{z}{2}) \mathcal{W}_{[-\frac{z}{2},\frac{z}{2}]} \Gamma 
    \psi(\tfrac{z}{2}) \right| p, \lambda \right\rangle \bigg|_{z^+ = 0},
    \label{eq.qqc}
\end{align}
where \(|p, \lambda\rangle\) and \(|p', \lambda'\rangle\) are the initial and final dressed quark states with momenta \(p\) and \(p'\), respectively. The Wilson line \(\mathcal{W}_{[-\frac{z}{2},\frac{z}{2}]}\) ensures gauge invariance by connecting the two quark fields \(\psi(\tfrac{z}{2})\) and \(\bar{\psi}(-\tfrac{z}{2})\). In the light-front gauge \(A^+ = 0\), this Wilson line becomes unity, simplifying the analysis within perturbative models. The choice of \(\Gamma \in \{\gamma^+, \gamma^+ \gamma^5, i\sigma^{+j} \gamma^5\}\) corresponds to the unpolarized, longitudinally polarized, and transversely polarized quark distributions, respectively.

\subsection{Unpolarized Target\label{subsec:WDU}}
Following the convention adopted in \cite{lorce2012quark}, the Wigner distribution for an unpolarized target can be expressed as
\begin{align}
\rho_{UY}=\frac{1}{2}\Big[\rho^{[\Gamma]}(x,\xi,b_\perp,k_\perp,+\hat{e}_z)+\rho^{[\Gamma]}(x,\xi,b_\perp,k_\perp,-\hat{e}_z)\Big]\label{un_1},
\end{align}
where the subscript $Y=\{U,L,T\}$ indicates the distinct quark polarization. %corresponding gamma strutures $\Gamma=\{\gamma^+,\gamma^+\gamma^5,i\sigma^{+j}\gamma^5\}$. 
The quark-quark correlator  in Eq.(\ref{eq.qqc}), %that defines the Wigner distributions 
can be parameterized through GTMDs \cite{meissner2009generalized} %as shown below:
\begin{align}
    W^{[\gamma^+]}_{\lambda,\lambda'}=&\frac{1}{2m}\Bar{u}(p',\lambda')\Bigg[F_{1,1}-\frac{i\sigma^{i+}k_{i\perp}}{P^+}F_{1,2}-\frac{i\sigma^{i+}\Delta_{i\perp}}{P^+}F_{1,3}+\frac{i\sigma^{ij}k_{i\perp}\Delta_{j\perp}}{m^2}F_{1,4}\Bigg]u(p,\lambda)\label{eqn.up},\\
W^{[\gamma^+\gamma_5]}_{\lambda,\lambda'}=&\frac{1}{2m}\Bar{u}(p',\lambda')\Bigg[\frac{-i\epsilon^{ij}_{\perp}k_{i\perp}\Delta_{j\perp}}{m^2}G_{1,1}-\frac{i\sigma^{i+}\gamma_5 k_{i\perp}}{P^+}G_{1,2}-\frac{i\sigma^{i+}\gamma_5 \Delta_{i\perp}}{P^+}G_{1,3}+i\sigma^{+-}\gamma_5 G_{1,4}\Bigg]u(p,\lambda)\label{eqn.lp},\\
%\end{align}
%\begin{align}
W^{[i\sigma^{+j}\gamma_5]}_{\lambda\lambda'}=&\frac{1}{2m}\Bar{u}(p',\lambda')\Bigg[-\frac{i\epsilon^{ij}_\perp p^i_\perp}{m}H_{1,1}-\frac{i\epsilon^{ij}_\perp \Delta^i_\perp}{m}H_{1,2}+\frac{mi\sigma^{j+}\gamma^5}{P^+}H_{1,3}+\frac{p^j_\perp i \sigma^{k+}\gamma^5p^k_\perp}{mP^+}H_{1,4}\nn\\
&+\frac{\Delta^j_\perp i \sigma^{k+}\gamma^5p^k_\perp}{mP^+}H_{1,5}+\frac{\Delta^j_\perp i \sigma^{k+}\gamma^5\Delta^k_\perp}{mP^+}H_{1,6}
    +\frac{p^j_\perp i \sigma^{+-}\gamma^5}{m}H_{1,7}+\frac{\Delta^j_\perp i \sigma^{+-}\gamma^5}{m}H_{1,8}\Bigg]u(p,\lambda).\label{eqn.tp}
\end{align}
The functions $F_{1i}, G_{1i}, H_{1j}$, where $i=1,2,...,4$ and $j=1,2,...,8$, are GTMDs for the quark. Using the definition of Wigner distribution (Eqs.(\ref{wig1},\ref{un_1})) and the quark-quark correlator parametrization (Eqs.(\ref{eqn.up},\ref{eqn.lp},\ref{eqn.tp})), the Wigner distributions for unpolarized target with different polarization of quark can be obtained as \cite{maji2022leading, ojha2023quark}
\begin{align}
    \rho_{UU}(x,\xi,b_\perp,k_\perp)&=\frac{1}{2(2\pi)^5}\int\frac{d^2\Delta_\perp}{(1-\xi^2)^{\frac{3}{2}}}e^{-i\frac{\Delta_\perp\cdot b_\perp}{1-\xi^2}}F_{1,1}(x,\xi,\Delta_\perp,k_\perp),\\
    \rho_{UL}(x,\xi,b_\perp,k_\perp)&=\frac{1}{2(2\pi)^5}\int\frac{d^2\Delta_\perp}{(1-\xi^2)^{\frac{3}{2}}}e^{-i\frac{\Delta_\perp\cdot b_\perp}{1-\xi^2}}\frac{-i}{m^2}\epsilon^{ij}_\perp k^i_\perp\Delta^j_\perp G_{1,1}(x,\xi,\Delta_\perp,k_\perp),\\
    \rho^j_{UT}(x,\xi,b_\perp,k_\perp)&=\frac{1}{2(2\pi)^5}\int\frac{d^2\Delta_\perp}{(1-\xi^2)^{\frac{3}{2}}}e^{-i\frac{\Delta_\perp\cdot b_\perp}{1-\xi^2}}\frac{-i}{m^2}\epsilon^{ij}_\perp\Big[k^i_\perp H_{1,1}(x,\xi,\Delta_\perp,k_\perp)+\Delta^i_\perp H_{1,2}(x,\xi,\Delta_\perp,k_\perp)\Big].
\end{align}
Here j=1,2 represents the two components of the transverse polarization case.

%-----------------subsection---------------%

\subsection{Longitudinally polarized Target\label{subsec:WDL}}
For the longitudinally polarized target, the definition of the Wigner distribution is
\begin{align}
\rho_{LY}=\frac{1}{2}\Big[\rho^{[\Gamma]}(x,\xi,b_\perp,k_\perp,+\hat{e}_z)-\rho^{[\Gamma]}(x,\xi,b_\perp,k_\perp,-\hat{e}_z)\Big]  \label{long_1}. 
\end{align}
The subscript $Y$ represents the quark's polarization corresponding to $\Gamma$ as discussed in sec. \ref{subsec:WDU}. 
Again using the definition of Wigner distribution (Eqs.(\ref{wig1},\ref{long_1})) and the quark-quark correlator parametrization (Eqs.(\ref{eqn.up},\ref{eqn.lp},\ref{eqn.tp})), we obtain the Wigner distributions for longitudinally polarized target with different polarization of quark as
\begin{align}
     \rho_{LU}(x,\xi,b_\perp,k_\perp)&=\frac{1}{2(2\pi)^5}\int\frac{d^2\Delta_\perp}{(1-\xi^2)^{\frac{3}{2}}}e^{-i\frac{\Delta_\perp\cdot b_\perp}{1-\xi^2}}\frac{i}{m^2}\epsilon^{ij}_\perp k^i_\perp\Delta^j_\perp F_{1,4}(x,\xi,\Delta_\perp,k_\perp),\\
     \rho_{LL}(x,\xi,b_\perp,k_\perp)&=\frac{1}{2(2\pi)^5}\int\frac{d^2\Delta_\perp}{(1-\xi^2)^{\frac{3}{2}}}e^{-i\frac{\Delta_\perp\cdot b_\perp}{1-\xi^2}}\;2\;G_{1,4}(x,\xi,\Delta_\perp,k_\perp),\\
     \rho^j_{LT}(x,\xi,b_\perp,k_\perp)&=\frac{1}{2(2\pi)^5}\int\frac{d^2\Delta_\perp}{(1-\xi^2)^{\frac{3}{2}}}e^{-i\frac{\Delta_\perp\cdot b_\perp}{1-\xi^2}}\frac{2}{m}\Big[k^j_\perp H_{1,7}(x,\xi,\Delta_\perp,k_\perp)+\Delta^j_\perp H_{1,8}(x,\xi,\Delta_\perp,k_\perp)\Big].
\end{align}
%
%-----------------subsection------------------%
%
\subsection{Transversely polarized Target\label{subsec:WDT}}

For the transversely polarized target, the Wigner distribution is defined as
\begin{align}
\rho^i_{TY}=\frac{1}{2}\Big[\rho^{[\Gamma]}(x,\xi,b_\perp,k_\perp,+\hat{e}_i)-\rho^{[\Gamma]}(x,\xi,b_\perp,k_\perp,-\hat{e}_i)\Big].\label{trans_1}
\end{align}
%The subscript Y refers to the quark polarization corresponding to $\Gamma$ mentioned in sec. \ref{subsec:WDU}. 
Following the same approach as in previous two subsections and using Eqs.(\ref{wig1}, \ref{trans_1}, \ref{eqn.up}, \ref{eqn.lp}, \ref{eqn.tp}), we get the Wigner distributions for transversely polarized target with different polarization of quark as \cite{maji2022leading, ojha2023quark}
%
%Using the definition of Wigner distribution (Eqs.(\ref{wig_1}, \ref{trans_1})) and the quark-quark correlator parametrization (Eqs.(\ref{eqn.up}, \ref{eqn.lp}, \ref{eqn.tp})), the Wigner distributions for transversely polarized target with different polarization of quark can be obtained as

\begin{align}
    \rho^i_{TU}(x,\xi,b_\perp,k_\perp)=&\frac{1}{2(2\pi)^5}\int\frac{d^2\Delta_\perp}{(1-\xi^2)^{\frac{3}{2}}}e^{-i\frac{\Delta_\perp\cdot b_\perp}{1-\xi^2}}\frac{-i}{2m}\epsilon^{ij}_\perp\Big[\Delta^j_\perp(F_{1,1}(x,\xi,\Delta_\perp,k_\perp)-2(1-\xi^2)\nn\\
    &F_{1,3}(x,k_\perp,\Delta_\perp,\xi))-2(1-\xi^2)k^j_\perp F_{1,2}(x,\xi,\Delta_\perp,k_\perp)+\frac{\xi}{m^2}\epsilon^{k,l}_\perp k^k_\perp\Delta^l_\perp\Delta^j_\perp \nn\\
    &F_{1,4}(x,\xi,\Delta_\perp,k_\perp)\Big],\\
    \rho^i_{TL}(x,\xi,b_\perp,k_\perp)=&\frac{1}{2(2\pi)^5}\int\frac{d^2\Delta_\perp}{1-\xi^2}e^{-i\frac{\Delta_\perp\cdot b_\perp}{1-\xi^2}}\Big[\frac{-1}{2m^3(1-\xi^2)^{\frac{3}{2}}}\epsilon^{ij}_\perp\epsilon^{kl}_\perp k^k_\perp\Delta^l_\perp\Delta^j_\perp G_{1,1}(x,\xi,\Delta_\perp,k_\perp)\nn\\
    &+\frac{\sqrt{1-\xi^2}}{m}k^i_\perp G_{1,2}(x,\xi,\Delta_\perp,k_\perp)+\frac{1}{m\sqrt{1-\xi^2}}\Delta^i_\perp((1-\xi^2)\nn\\
    &G_{1,3}(x,\xi,\Delta_\perp,k_\perp)-\xi G_{1,4}(x,\xi,\Delta_\perp,k_\perp))\Big],
    \end{align}
    \begin{align}
    \rho^j_{TT}(x,\xi,b_\perp,k_\perp)=&\frac{1}{2(2\pi)^5}\int\frac{d^2\Delta_\perp}{(1-\xi^2)}e^{-i\frac{\Delta_\perp\cdot b_\perp}{1-\xi^2}}\epsilon^{ij}_\perp(-1)^j\Big[\frac{1}{2m^2\sqrt{1-\xi^2}}( k^i_\perp\Delta^i_\perp H_{1,1}(x,\xi,\Delta_\perp,k_\perp)\nn\\
    &+(\Delta^i_\perp)^2H_{1,2}(x,\xi,\Delta_\perp,k_\perp))+\sqrt{1-\xi^2}H_{1,3}(x,\xi,\Delta_\perp,k_\perp)+\frac{\sqrt{1-\xi^2}}{m^2}(k^j_\perp)^2\nn\\
    &H_{1,4}(x,\xi,\Delta_\perp,k_\perp)+\frac{1}{m^2\sqrt{1-\xi^2}}k^j_\perp\Delta^j_\perp((1-\xi^2)H_{1,5}(x,\xi,\Delta_\perp,k_\perp)-\xi\nn\\
    &H_{1,7}(x,\xi,\Delta_\perp,k_\perp))+\frac{1}{m^2\sqrt{1-\xi^2}}(\Delta^j_\perp)^2((1-\xi^2)H_{1,6}(x,\xi,\Delta_\perp,k_\perp)-\xi\nn\\
    &H_{1,8}(x,\xi,\Delta_\perp,k_\perp))\Big].
\end{align}

%%%%%%%%%%--------Section-------------%%%%%%%%%
\section{Numerical Analysis}
%\section{Discussion}

In this section, we present the plots for the quark Wigner distributions for various polarization configurations in the context of a light-front dressed quark model, incorporating nonzero longitudinal momentum transfer. The Wigner distributions have been plotted in three different representations: transverse impact parameter space ($\vec{b}_\perp$), transverse momentum space ($\vec{k}_\perp$), and the mixed space ($b_x, k_y$), allowing us to visualize the spatial and momentum correlations in a multidimensional phase space.

For the distributions in $\vec{b}_\perp$-space, we fixed the transverse momentum of the quark at $\vec{k}_\perp = 0.4 \, \hat{y}$ GeV, which allowed us to examine the distortion in the impact parameter plane arising from nonzero transverse momentum. Conversely, in $\vec{k}_\perp$-space, we set the impact parameter to $\vec{b}_\perp = 0.4 \, \hat{y}$ GeV$^{-1}$, enabling the study of momentum space structures influenced by transverse spatial localization. For the mixed-space Wigner distributions, where position and momentum coordinates are not Fourier conjugate, we integrated over $k_x$ and $b_y$ in the range $[0,\,0.4]$ to isolate correlations in the $(b_x, k_y)$ plane.

Throughout the analysis, we set the model quark mass to $m = 0.0033$ GeV and used a cutoff $\Delta_{\text{max}} = 1$ GeV for the transverse momentum transfer. The parton longitudinal momentum fraction $x$ was integrated over the DGLAP region ($\xi<x<1$), which corresponds to the physical regime where the active quark carries a non-negligible fraction of the parent hadron's momentum.

The resulting Wigner distributions exhibit characteristic patterns that reflect the interplay between transverse motion and spatial localization of the quark. In particular, dipole and quadrupole structures emerge in certain polarization combinations, indicating nontrivial correlations between position and momentum. These structures are sensitive to the choice of polarization and the inclusion of skewness ($\xi \neq 0$), which encodes the off-forward nature of the partonic configuration and enables access to deeper structural information, such as orbital angular momentum and spatial asymmetries.

 In Fig.~\ref{rho_UU} and Fig.~\ref{rho_UL} - Fig.~\ref{rho_TL} the distributions are presented in three different representations: the transverse impact parameter plane $(b_x, b_y)$, the transverse momentum plane $(k_x, k_y)$, and the mixed phase-space $(b_x, k_y)$. 
%
%---------- Subsection ------------------
%
\subsection{Unpolarized quark in unpolarized target}
\begin{figure}[!htp]
\begin{minipage}[c]{1\textwidth}
\small{(a)}\includegraphics[width=5cm,height=4cm,clip]{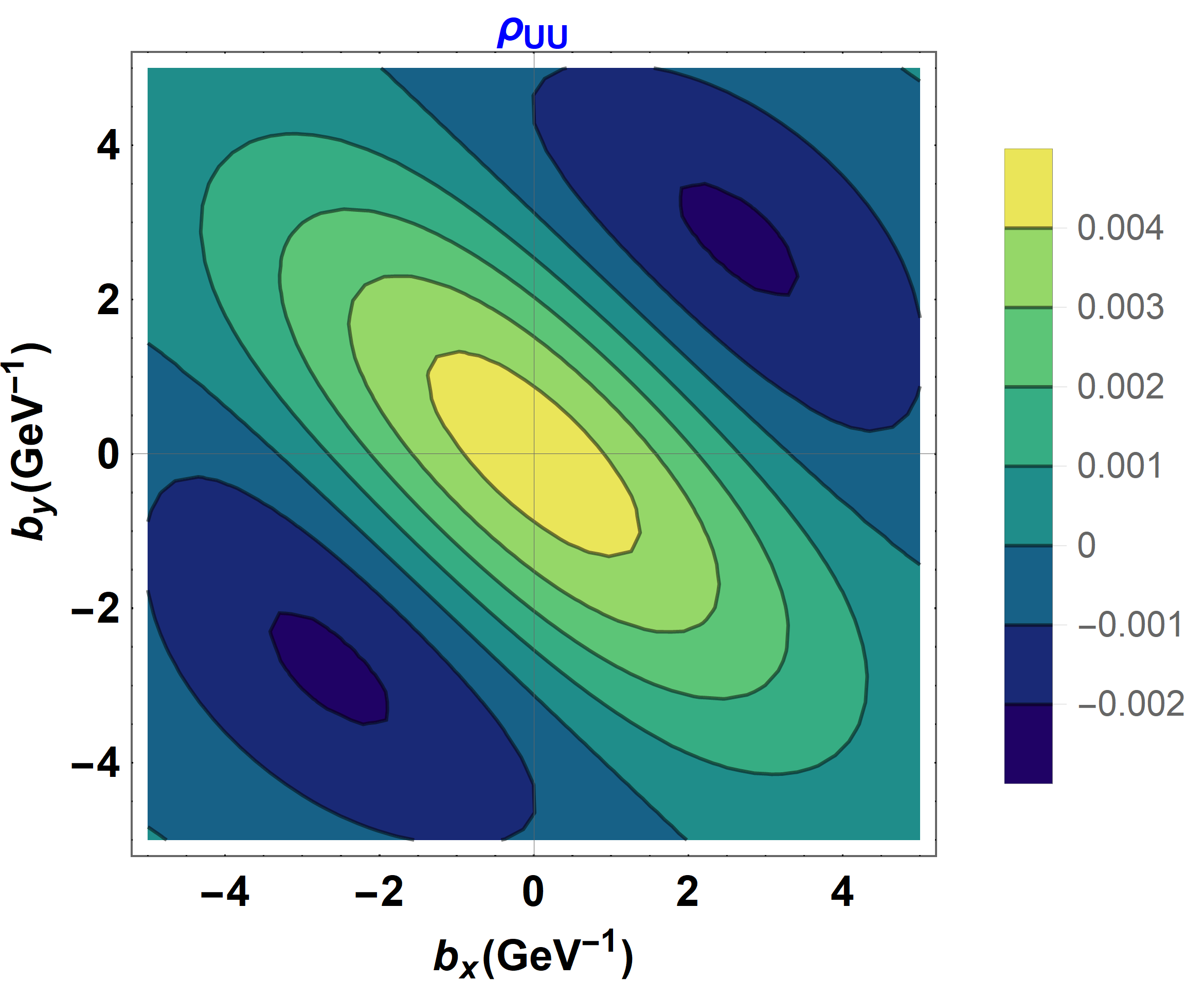}
\small{(b)}\includegraphics[width=5cm,height=4cm,clip]{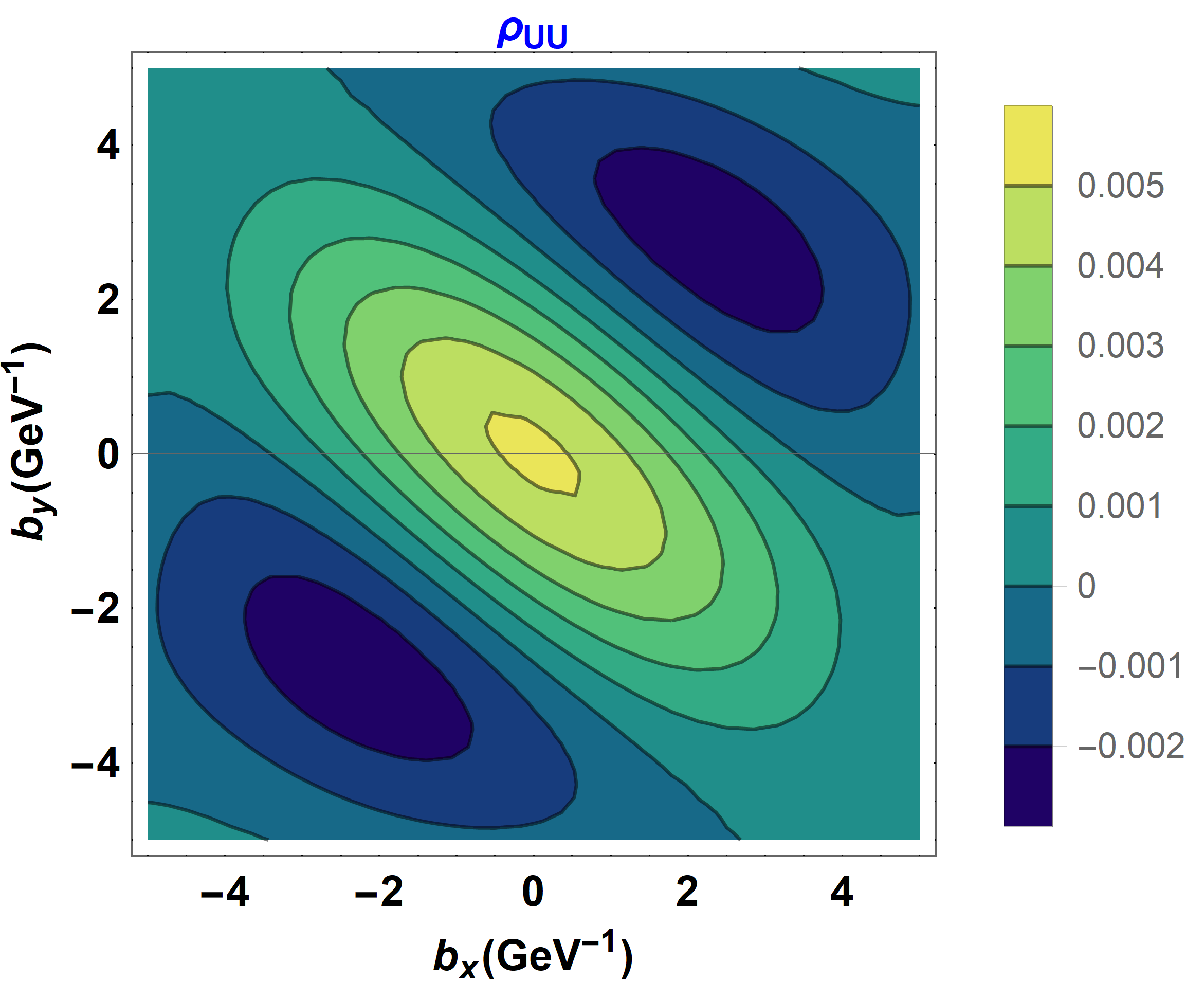} 
\small{(c)}\includegraphics[width=5cm,height=4cm,clip]{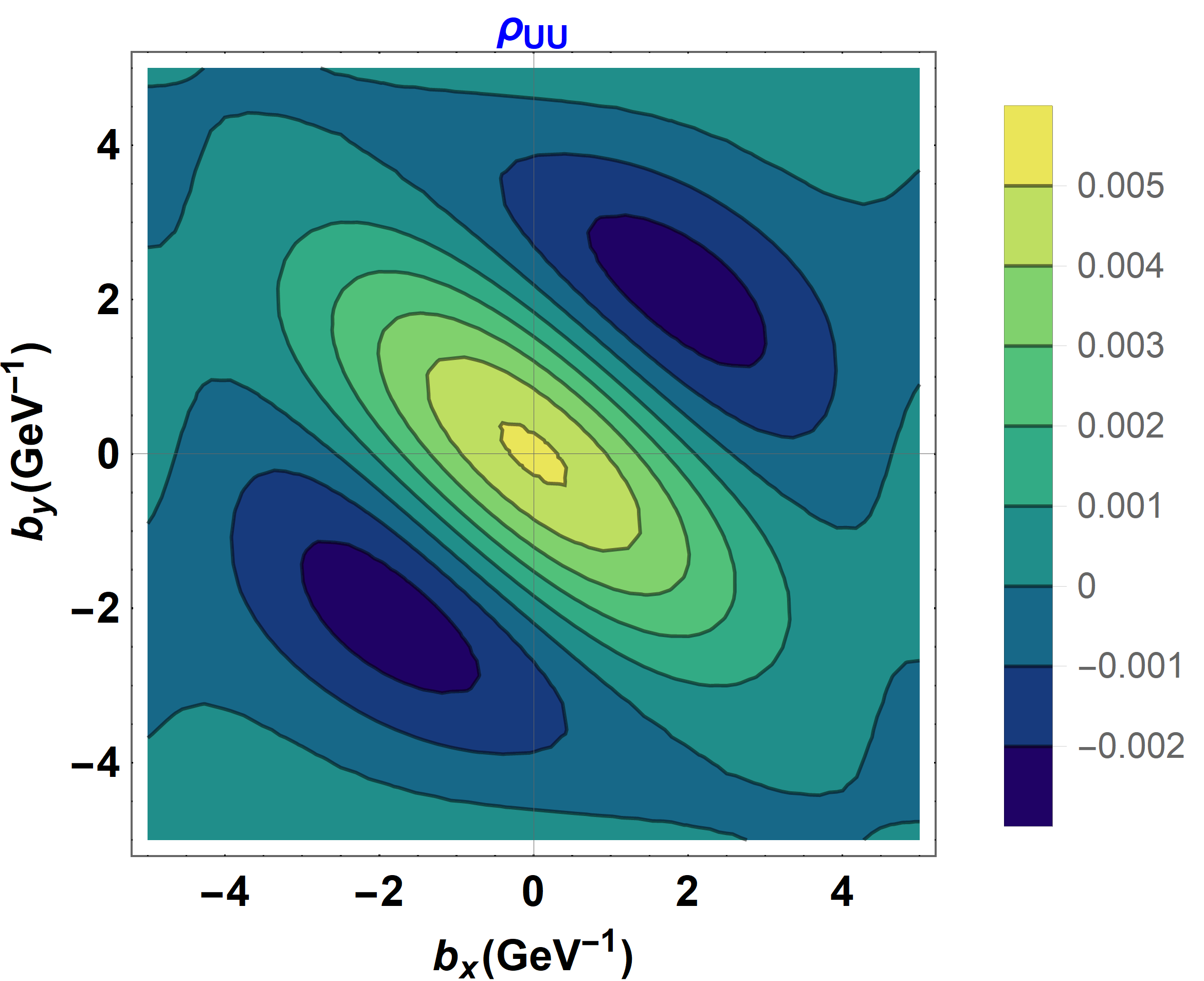}\\
\small{(d)}\includegraphics[width=5cm,height=4cm,clip]{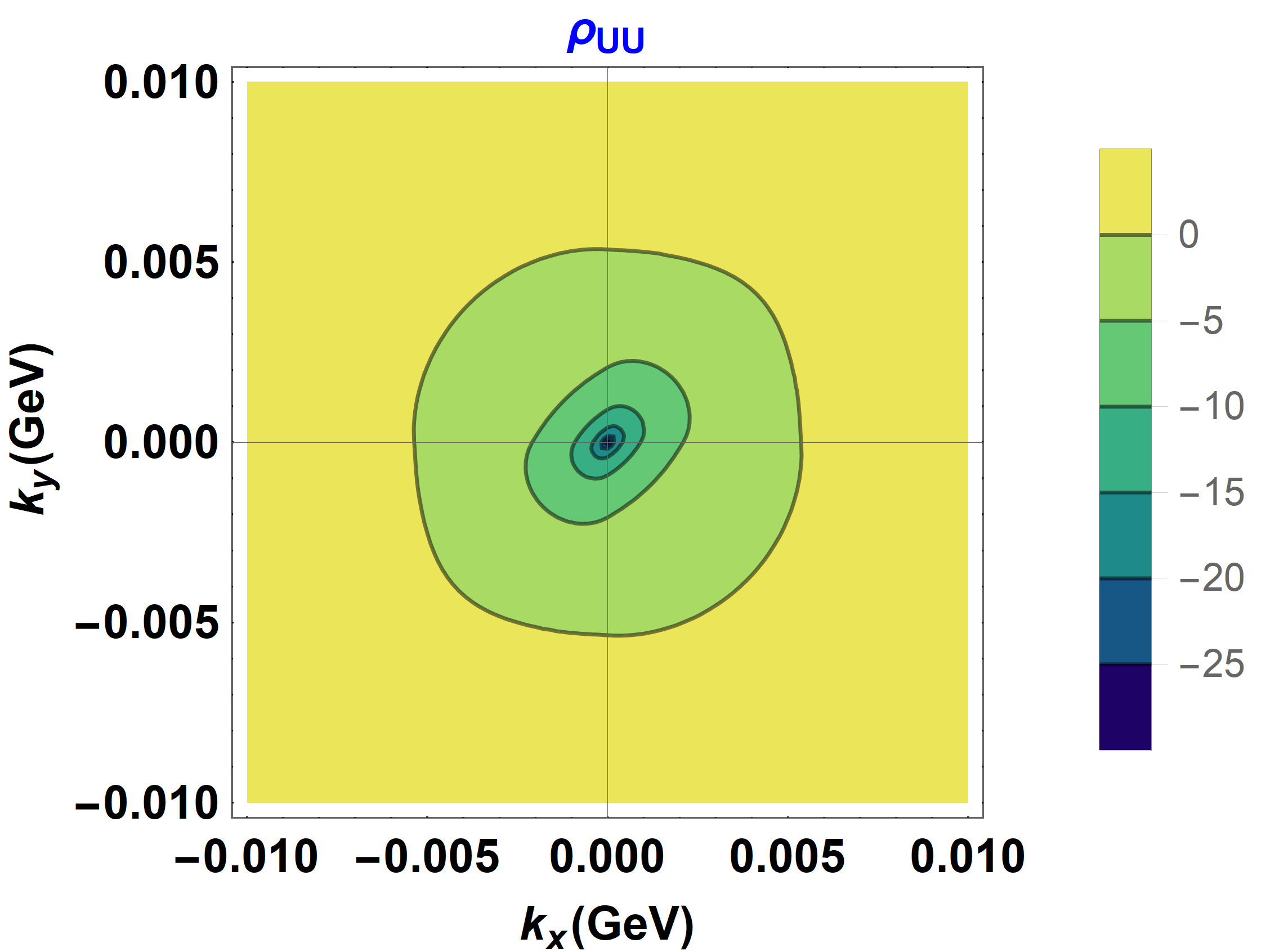} 
\small{(e)}\includegraphics[width=5cm,height=4cm,clip]{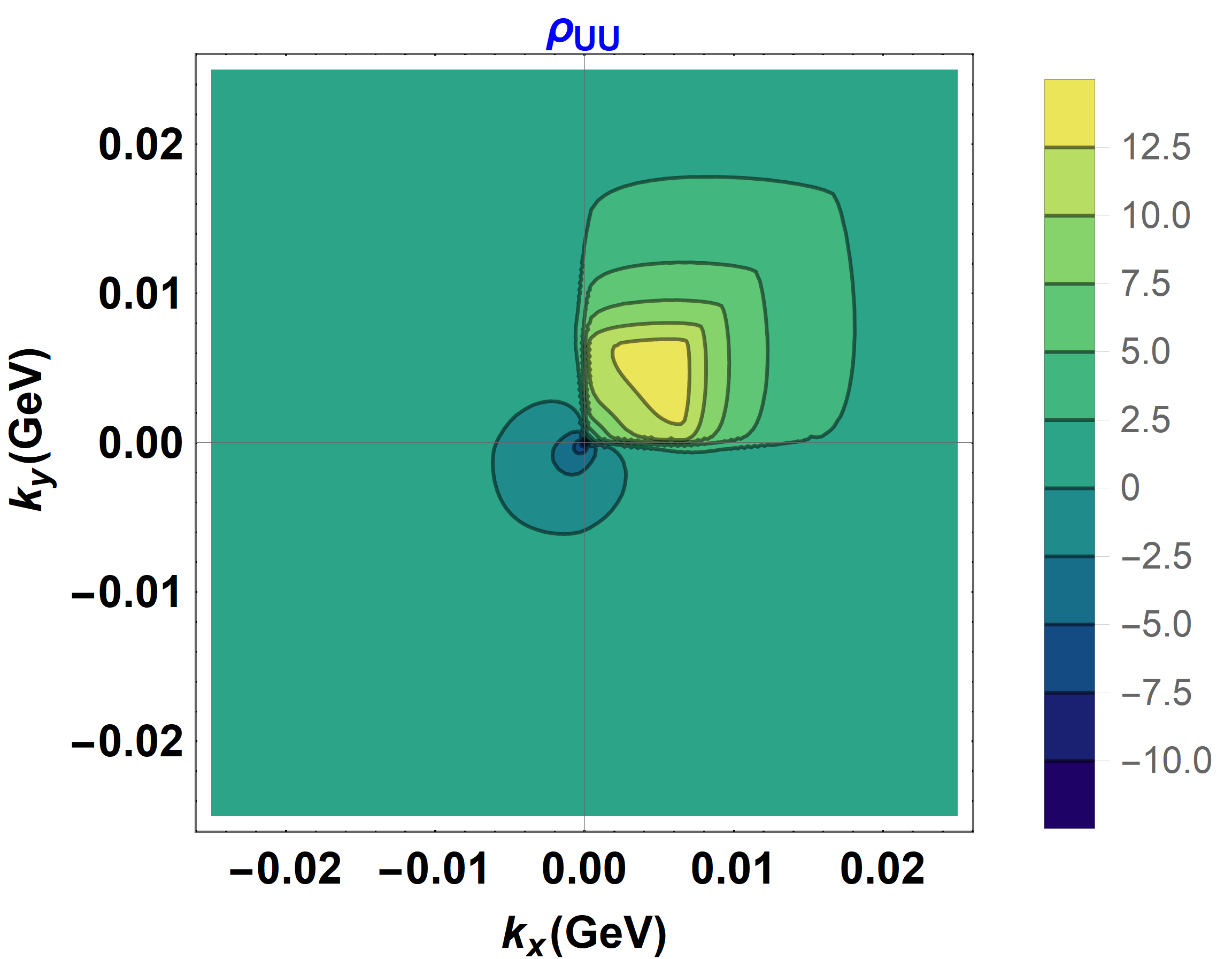}
\small{(f)}\includegraphics[width=5cm,height=4cm,clip]{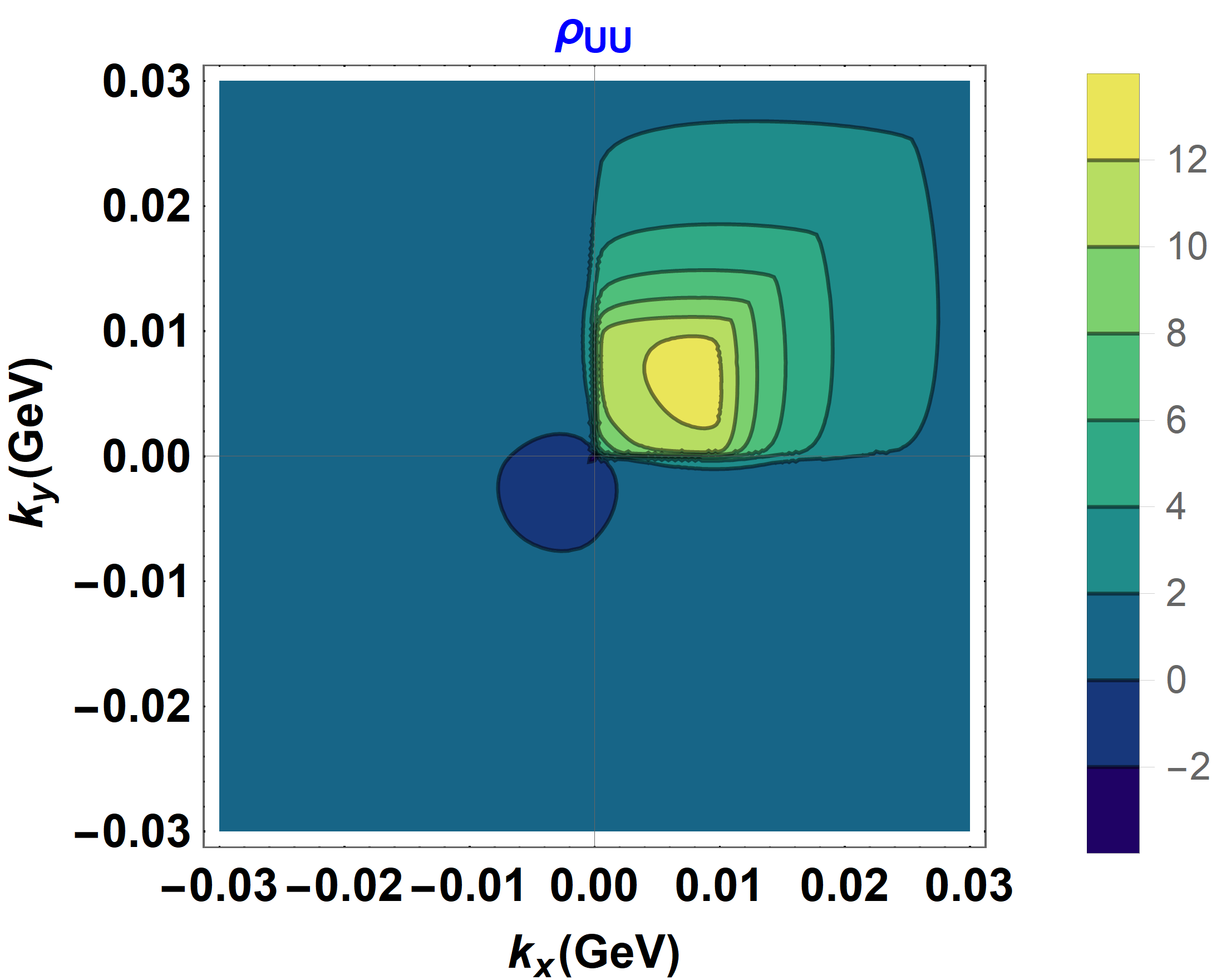} \\
\small{(g)}\includegraphics[width=5cm,height=4cm,clip]{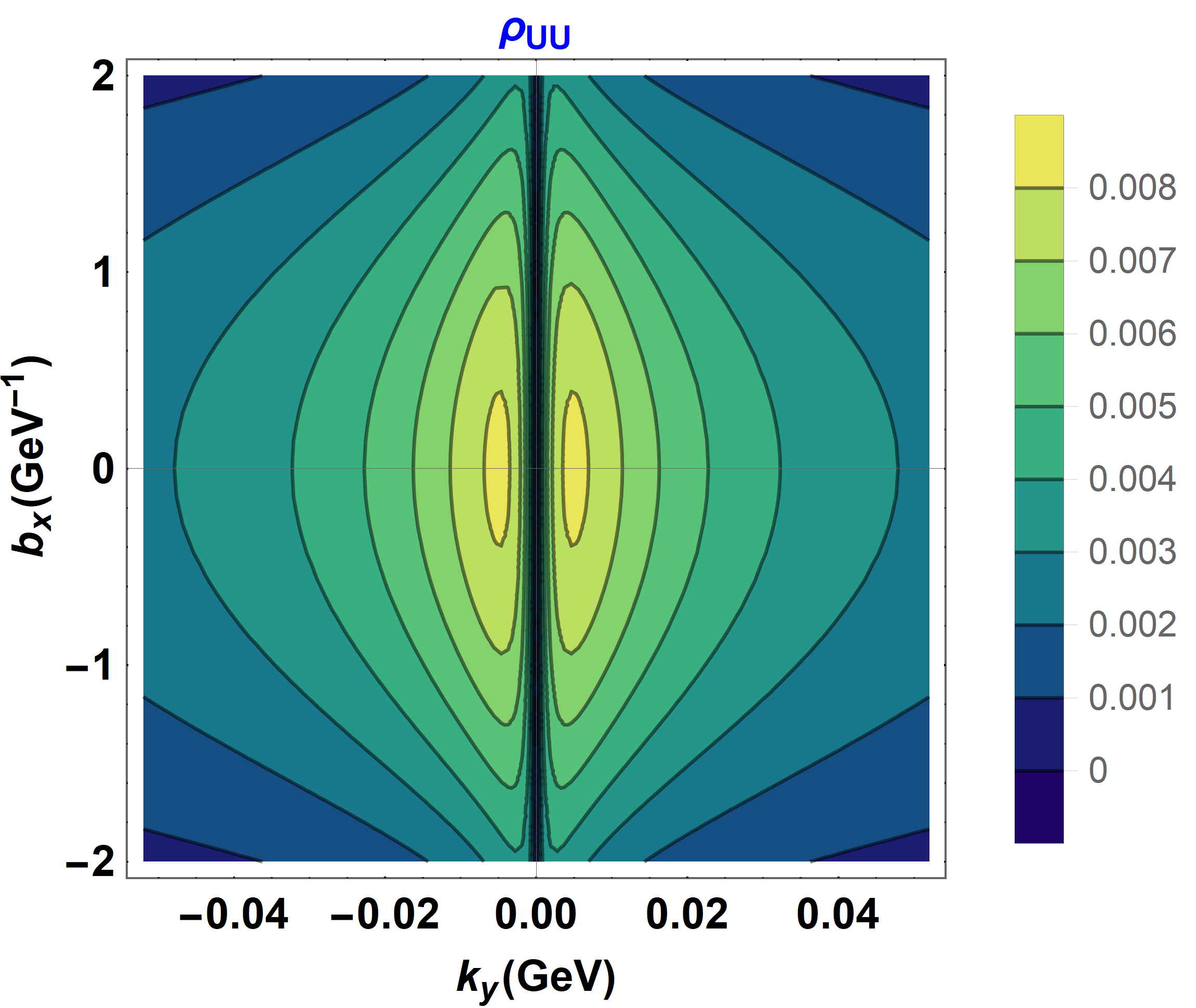} 
\small{(h)}\includegraphics[width=5cm,height=4cm,clip]{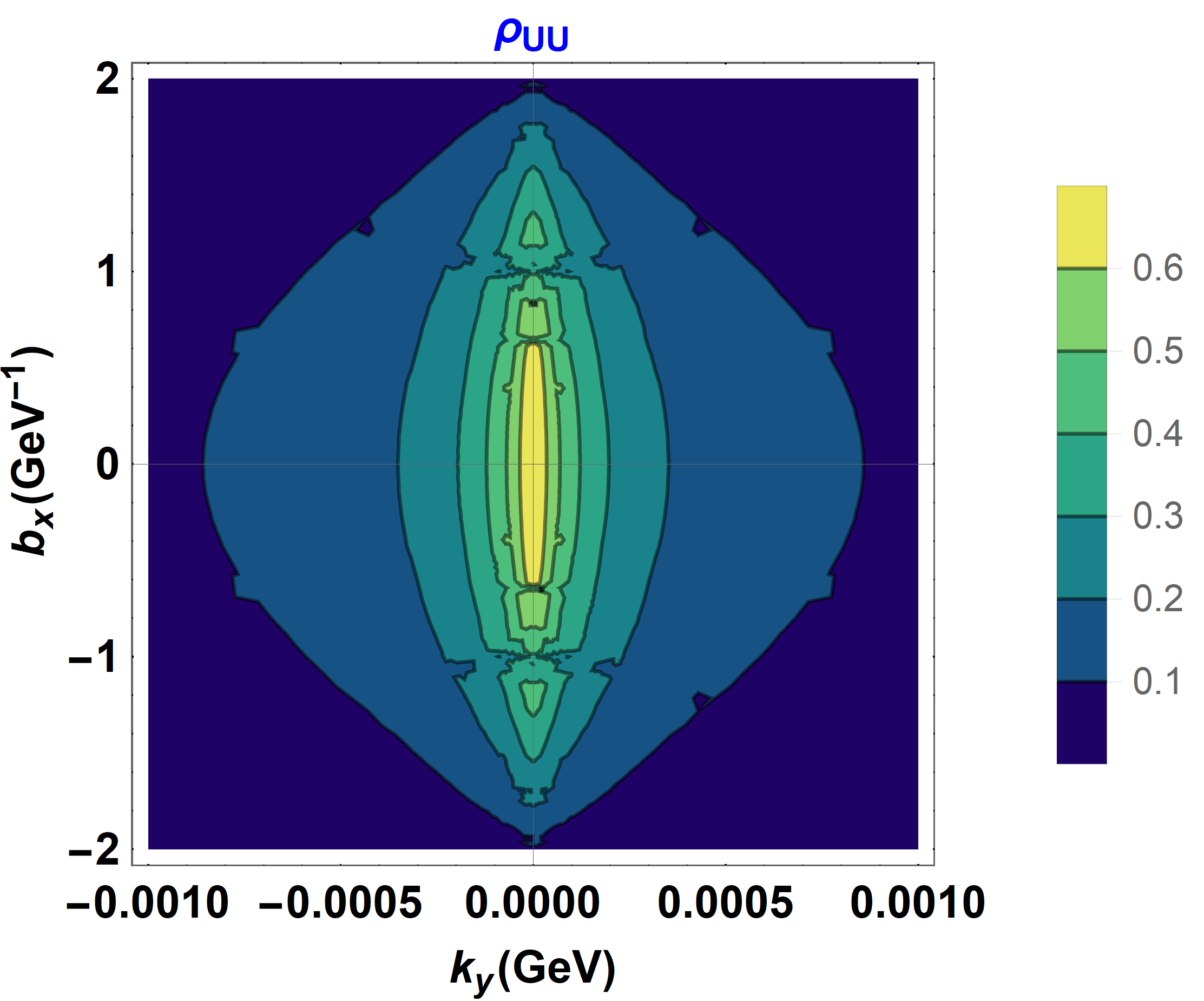}
\small{(i)}\includegraphics[width=5cm,height=4cm,clip]{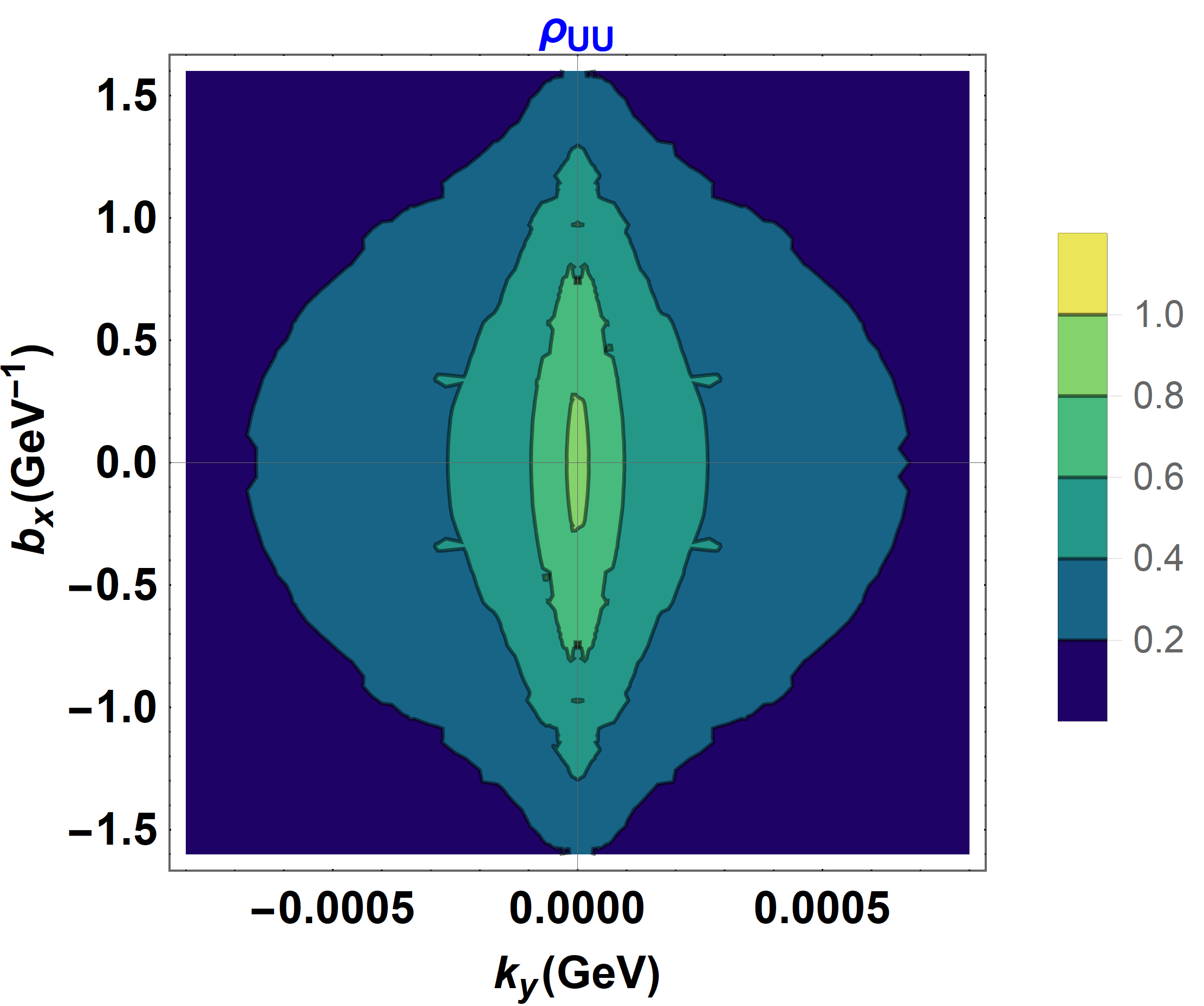} \\
\end{minipage}
\caption{\label{rho_UU} The quark Wigner distribution $\rho_{UU}$ in the transverse impact parameter plane, the transverse momentum plane, and the mixed plane. The left, middle, and right panels show the results for $\xi=0,\;\xi=0.25$, and $\xi=0.5$ respectively.}
\end{figure}

In this subsection, we present and analyze the Wigner distribution $\rho_{UU}$, corresponding to the unpolarized quark in the unpolarized target. The plots are shown in Fig.~\ref{rho_UU}, where the left, middle, and right columns correspond to skewness values $\xi = 0$, $\xi = 0.25$, and $\xi = 0.5$, respectively.

In the top row, we display $\rho_{UU}$ in the transverse impact parameter space, with the quark transverse momentum fixed at $\vec{k}_\perp = 0.4\,\hat{y}$ GeV. For $\xi = 0$, the distribution exhibits approximate rotational symmetry and a  quadruple-like structure, reflecting the spatial profile of the unpolarized quark in a forward kinematic configuration. As $\xi$ increases, the distribution becomes increasingly distorted, exhibiting a more asymmetric and deformed shape. This deformation arises due to the introduction of nonzero longitudinal momentum transfer, which induces a transverse shift in the quark distribution and breaks the symmetry between initial and final hadron states.

The middle row of Fig.~\ref{rho_UU} shows the distribution in transverse momentum space, with the transverse impact parameter fixed at $\vec{b}_\perp = 0.4\,\hat{y}$ GeV$^{-1}$. For $\xi = 0$, the distribution is nearly circularly symmetric and centered at the origin, as expected from the symmetry of the forward matrix element. However, for nonzero $\xi$, the symmetry is broken, and the distribution exhibits a clear shift along the transverse momentum directions. In particular, negative regions begin to appear, indicative of quantum interference effects between light-front wave function components with differing orbital angular momentum. These features reflect the nontrivial momentum-space correlations that emerge due to the off-forward kinematics.

In the bottom row, we show the Wigner distribution in the mixed phase-space $(b_x, k_y)$, with $b_y$ and $k_x$ integrated over the range $[0,\,0.4]$. This representation reveals the correlation between position and momentum in orthogonal transverse directions. For all three values of $\xi$, the distribution is sharply peaked along $b_x = 0$ and $k_y = 0$, indicating that the quark is most likely to be found near the transverse center with small transverse momentum. As $\xi$ increases, the distribution becomes increasingly localized, with the phase-space support narrowing significantly. This localization arises due to the reduction in overlap between initial and final states under larger longitudinal momentum transfer, leading to a kinematic squeezing of the available phase space.

%----------------------- subsection --------

\subsection{Evolution of $\rho_{UU}$ in momentum space}
\begin{figure}[htp!]
\centering
\begin{subfigure}[b]{0.30\textwidth}
\centering
\includegraphics[width=\textwidth]{Plots/rhoUUkspacecontour.png}
\caption{$\xi=0.00$}
\end{subfigure}
\centering
\begin{subfigure}[b]{0.30\textwidth}
\centering
\includegraphics[width=\textwidth]{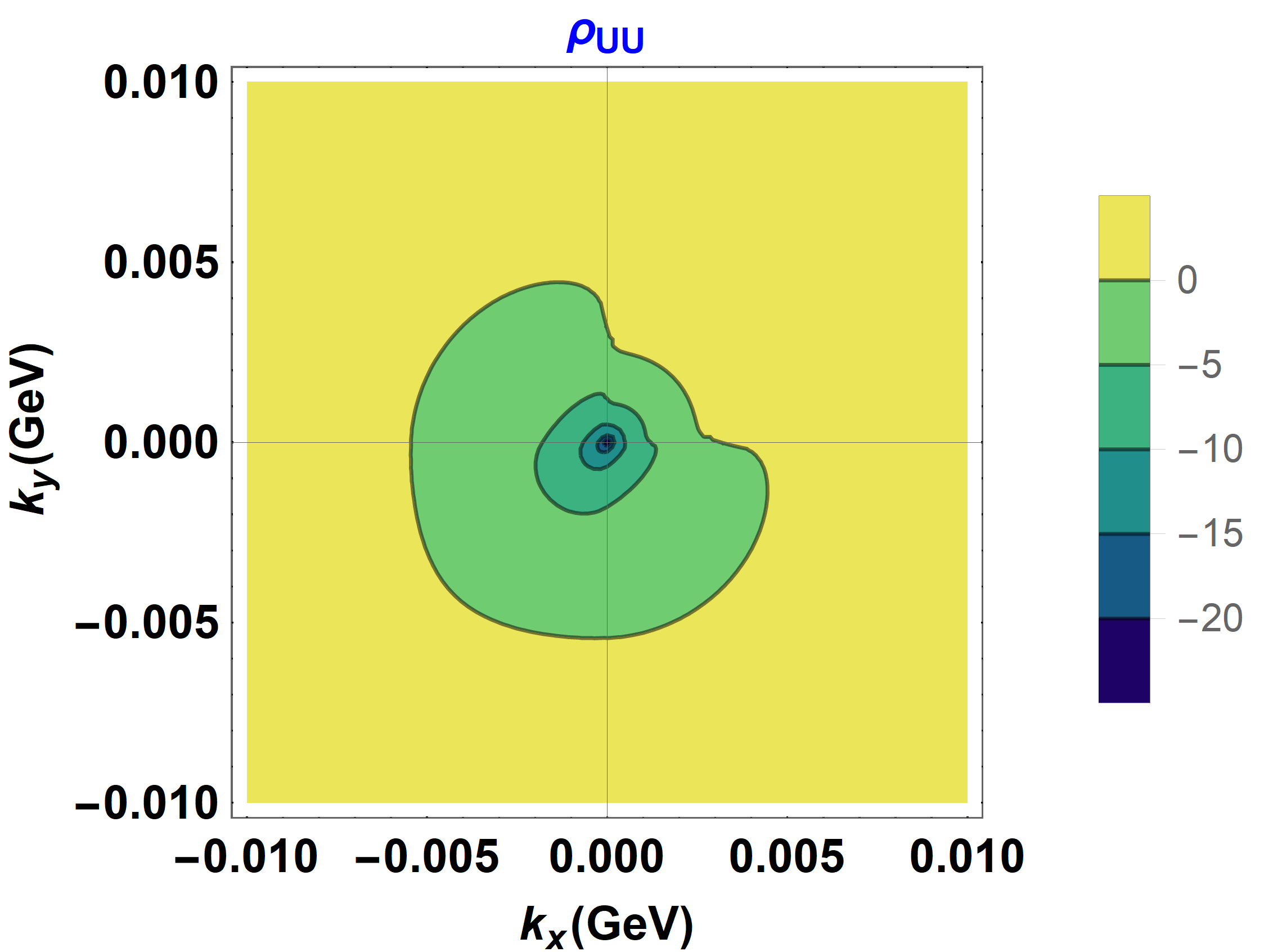}
\caption{$\xi=0.05$}
\end{subfigure}
\begin{subfigure}[b]{0.30\textwidth}
\centering
\includegraphics[width=\textwidth]{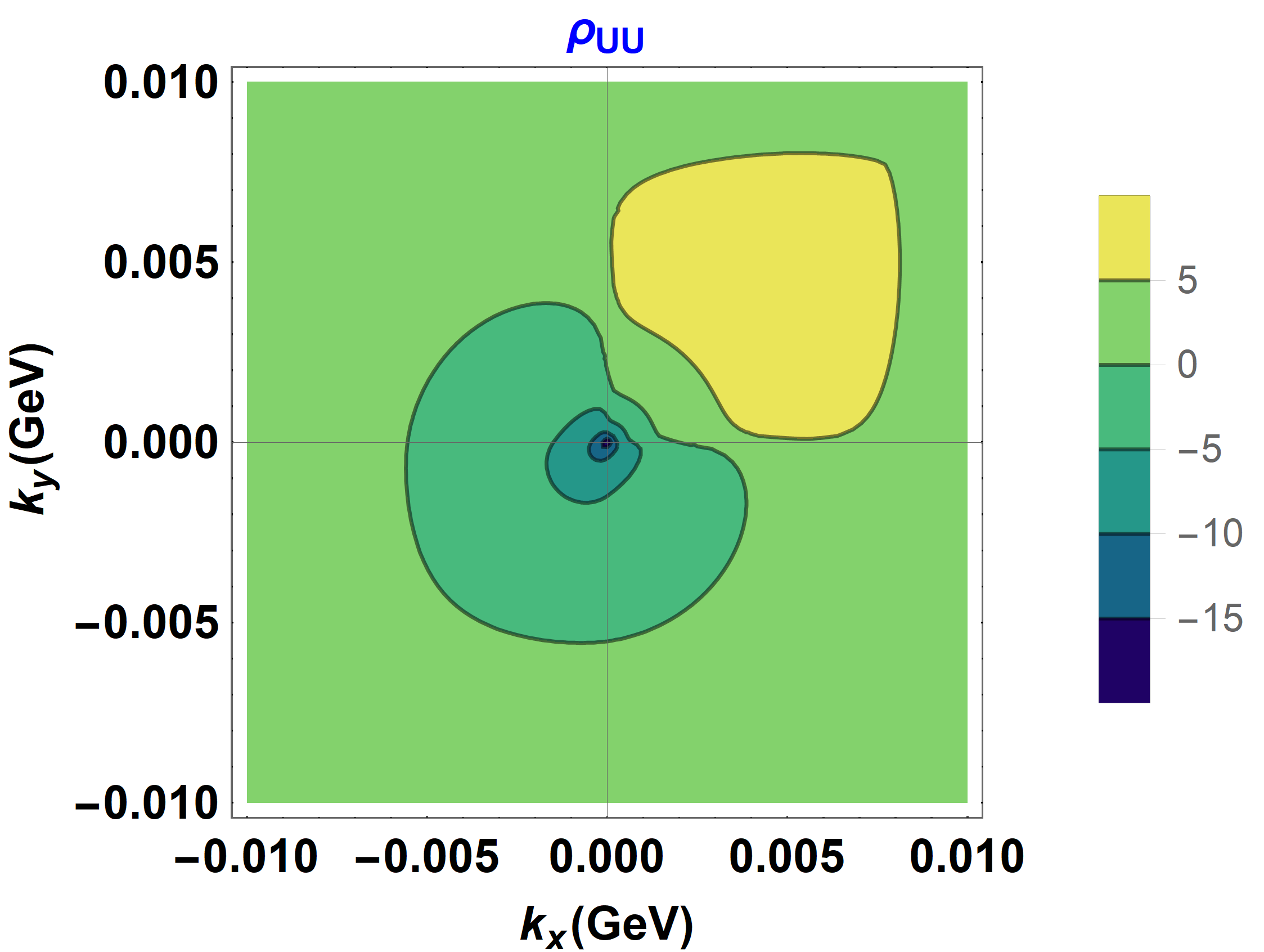}
\caption{$\xi=0.10$}
\end{subfigure}
\hfill
\centering
\begin{subfigure}[b]{0.30\textwidth}
\centering
\includegraphics[width=\textwidth]{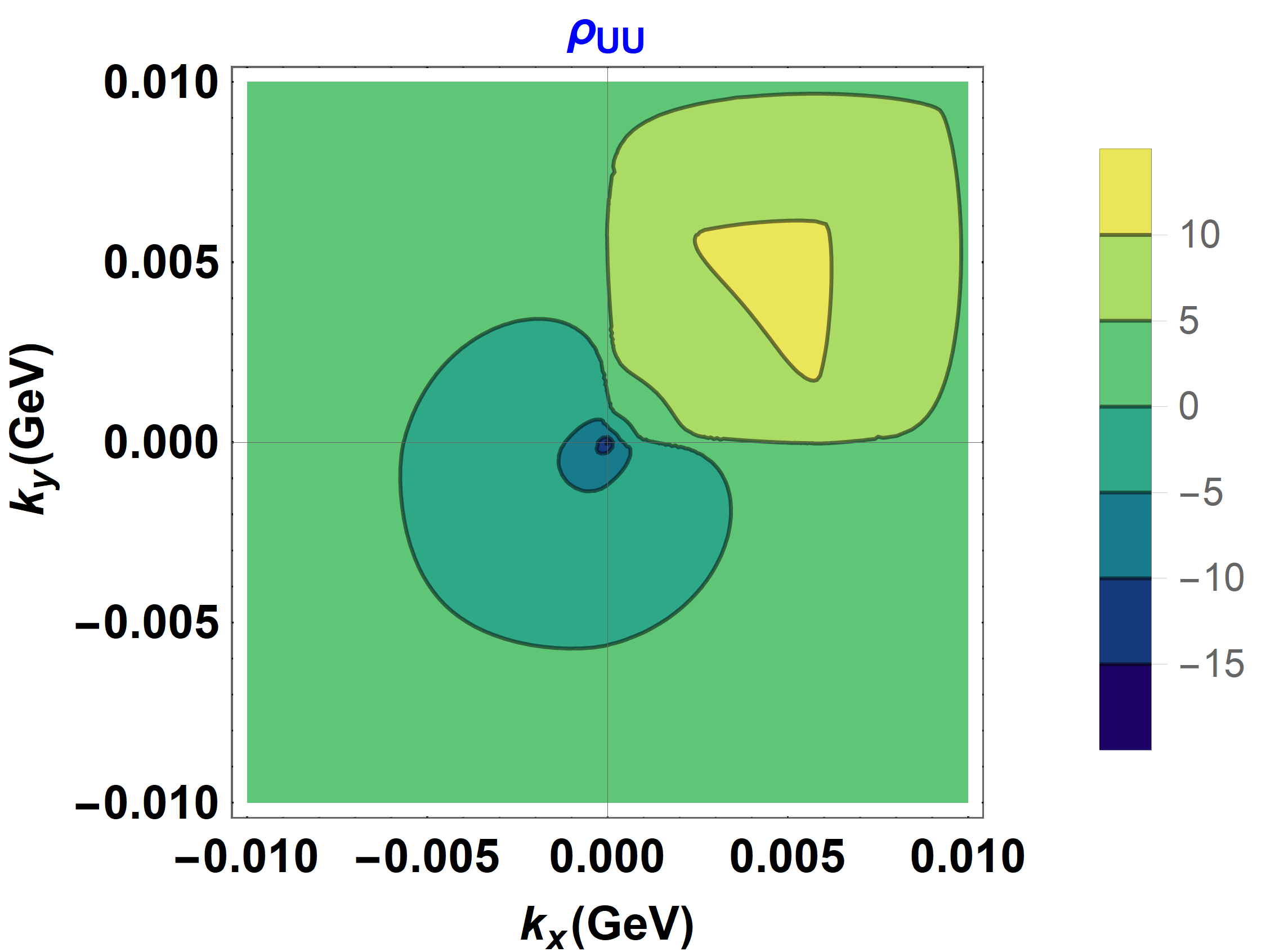}
\caption{$\xi=0.15$}
\end{subfigure}
\centering
\begin{subfigure}[b]{0.30\textwidth}
\centering
\includegraphics[width=\textwidth]{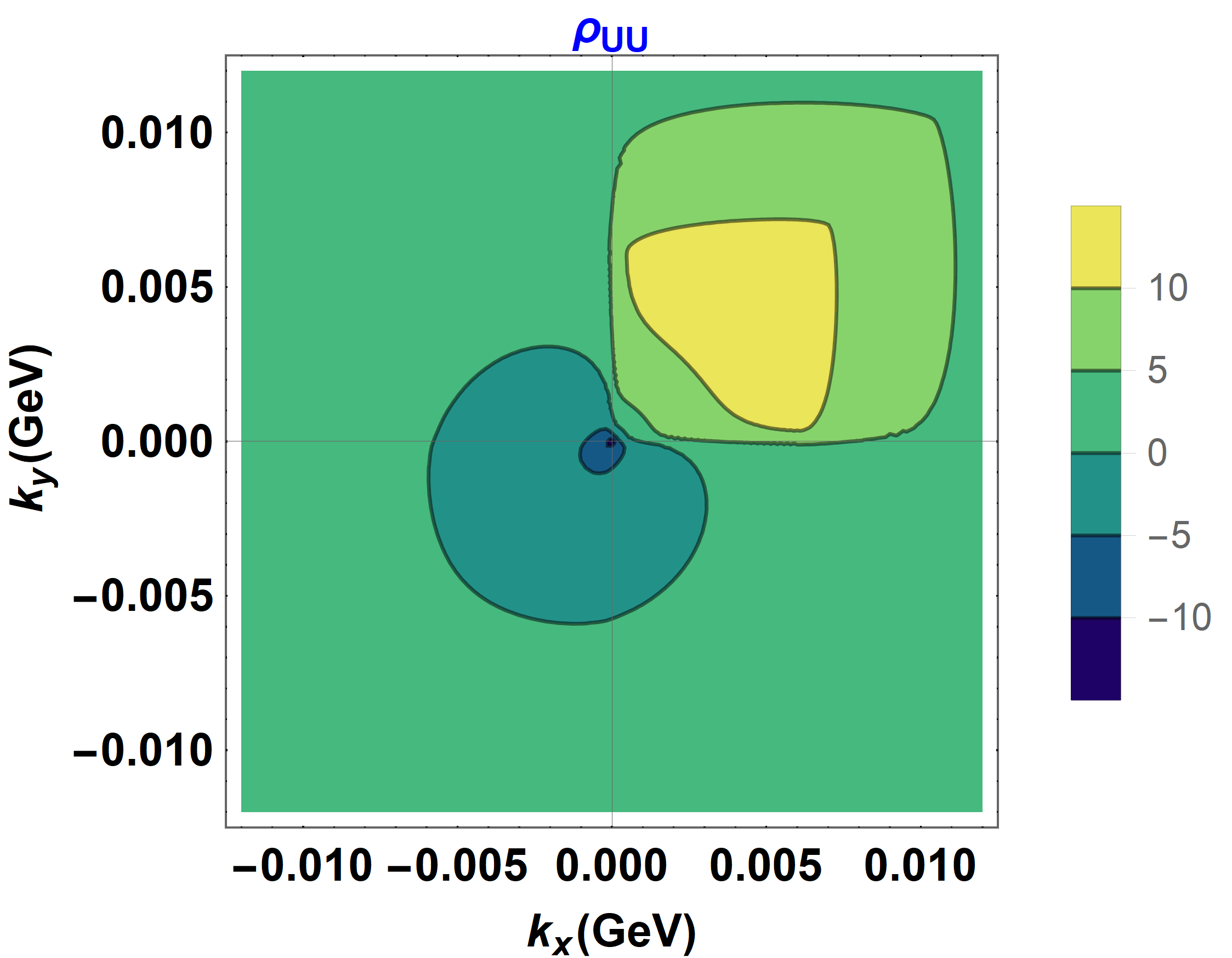}
\caption{$\xi=0.20$}
\end{subfigure}
\centering
\begin{subfigure}[b]{0.30\textwidth}
\centering
\includegraphics[width=\textwidth]{Plots/rhoUUkspacecontour5.png}
\caption{$\xi=0.25$}
\end{subfigure}
\centering
\begin{subfigure}[b]{0.30\textwidth}
\centering
\includegraphics[width=\textwidth]{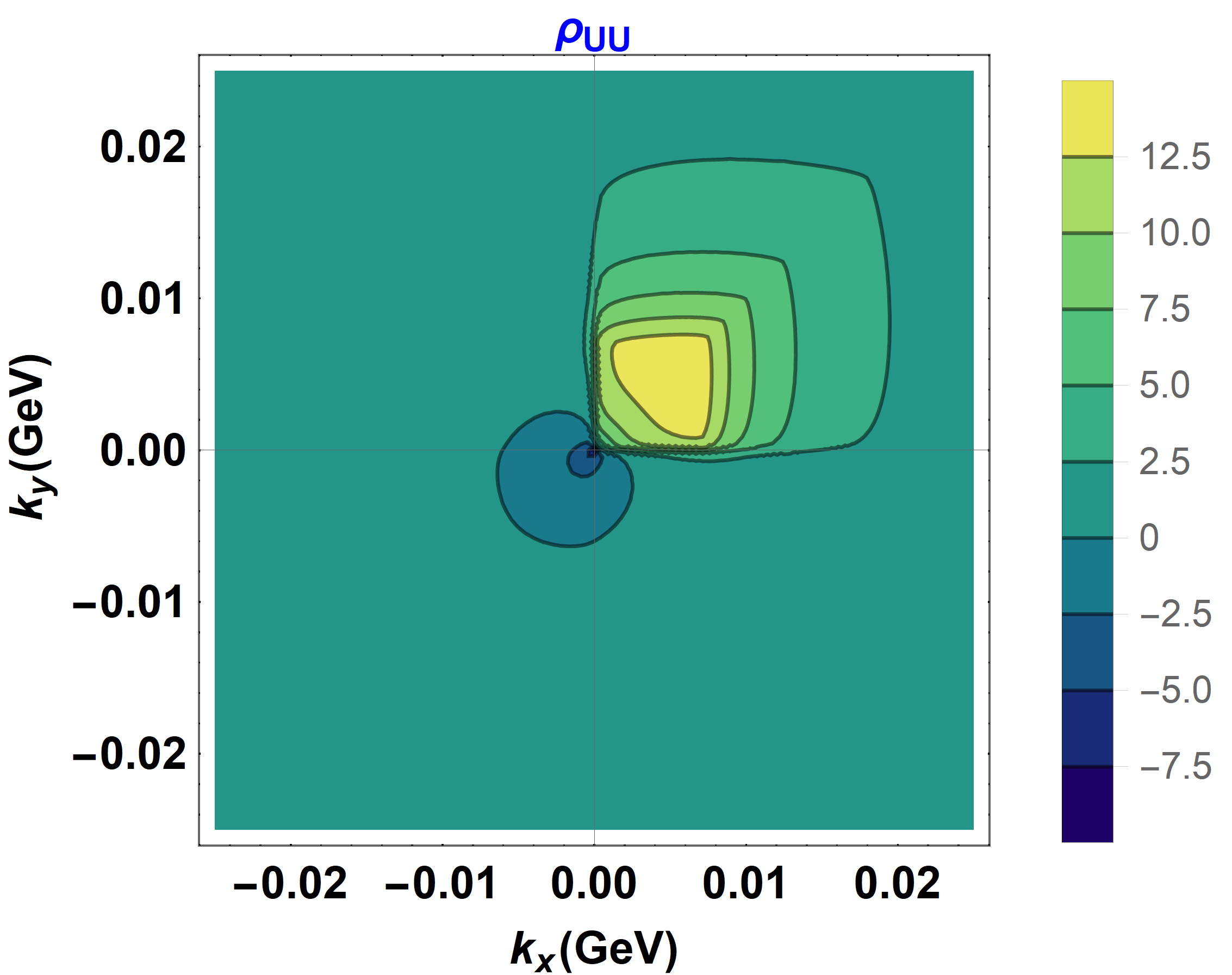}
\caption{$\xi=0.30$}
\end{subfigure}
\centering
\begin{subfigure}[b]{0.30\textwidth}
\centering
\includegraphics[width=\textwidth]{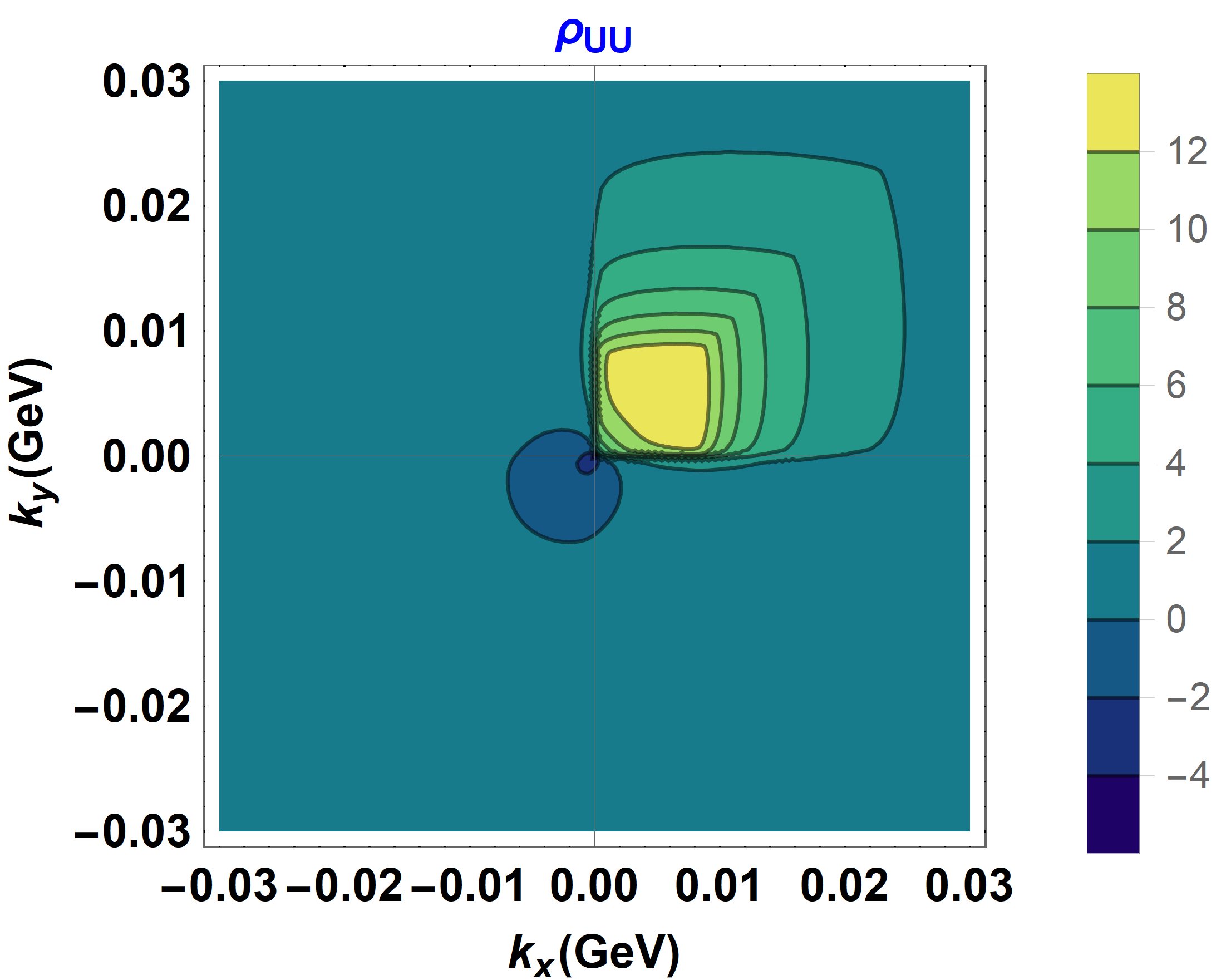}
\caption{$\xi=0.40$}
\end{subfigure}
\centering
\begin{subfigure}[b]{0.30\textwidth}
\centering
\includegraphics[width=\textwidth]{Plots/rhoUUkspacecontour10.png}
\caption{$\xi=0.50$}
\end{subfigure}
\caption{\label{rho_UUk} The evolution of the quark Wigner distribution $\rho_{UU}$ with skewness in the transverse momentum plane.  }
\end{figure}

Distinct structural patterns emerge in momentum space upon examining the evolution of $\rho_{UU}$. Figure~\ref{rho_UUk} illustrates the evolution of the unpolarized quark Wigner distribution $\rho_{UU}$ in the transverse momentum plane $(k_x, k_y)$ for increasing values of skewness $\xi$, ranging from $0$ to $0.5$. The impact parameter is fixed at $\vec{b}_\perp = 0.4\,\hat{y}$ GeV$^{-1}$ in all panels. These plots provide insight into how the quark momentum-space distribution is modified under longitudinal momentum transfer between the initial and final hadron states.

%At $\xi = 0$ (Fig.~\ref{rho_UUk}(a)), the distribution is nearly circularly symmetric and centered at the origin. This symmetry reflects the forward nature of the matrix element in the absence of skewness, where the initial and final states are indistinguishable in momentum space. A central negative region is observed, surrounded by a positive ring, which is characteristic of quantum interference patterns arising from the overlap of light-front wave functions.

As skewness increases, a systematic deformation of the distribution emerges. From $\xi = 0.05$ to $\xi = 0.15$ (Figs.~\ref{rho_UUk}(b)--(d)), the center of the distribution gradually shifts towards the positive $k_x$ and $k_y$ directions, accompanied by a breaking of rotational symmetry. This shift indicates that the quark acquires a net transverse momentum as a result of the nonzero longitudinal momentum transfer. The spatial displacement also implies the onset of dynamical recoil effects that are absent in the forward limit.

At intermediate skewness values $\xi = 0.20$ to $\xi = 0.30$ (Figs.~\ref{rho_UUk}(e)--(g)), the distribution develops a well-defined lobe structure, with positive and negative regions clearly separated in the transverse momentum plane. This structure reflects enhanced interference between different orbital components in the quark wave function. The transition from a symmetric to an asymmetric profile is a direct manifestation of the underlying non-forwardness and the modification of the quark's internal kinematics.

For large skewness values, $\xi = 0.40$ and $\xi = 0.50$ (Figs.~\ref{rho_UUk}(h)--(i)), the distortion becomes more pronounced. The positive density region becomes sharply localized in the first quadrant, while the negative region remains in the third quadrant. This pronounced asymmetry suggests that the quark momentum is increasingly aligned with the net momentum transfer direction. Moreover, the emergence of these localized lobes signals a squeezing of the available phase space in momentum space, likely due to the reduced overlap between the initial and final hadronic states at high $\xi$.

\subsection{Longitudinally Polarized quark in unpolarized target}
\begin{figure}[!htp]
\begin{minipage}[c]{1\textwidth}
\small{(a)}\includegraphics[width=5cm,height=4cm,clip]{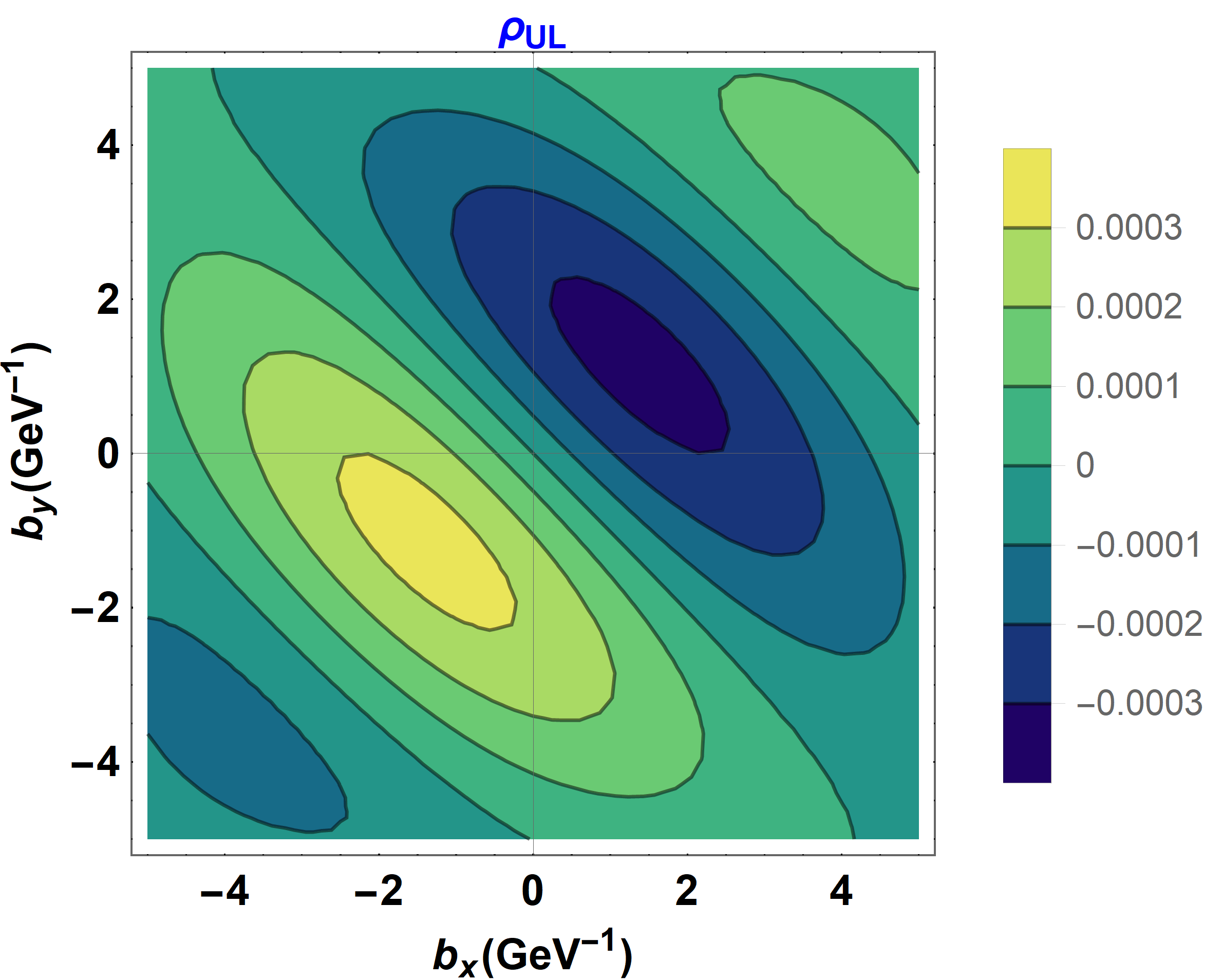}
\small{(b)}\includegraphics[width=5cm,height=4cm,clip]{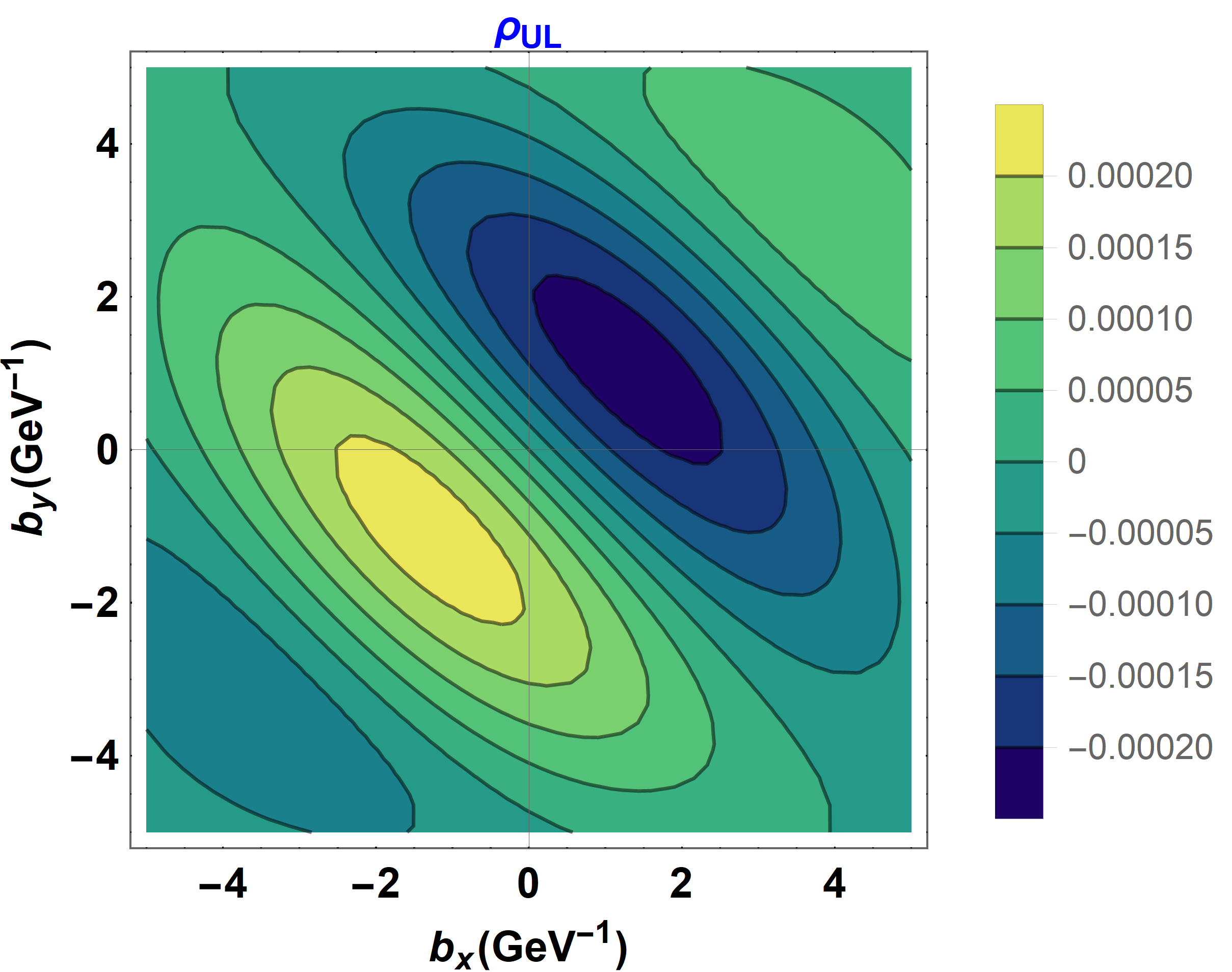} 
\small{(c)}\includegraphics[width=5cm,height=4cm,clip]{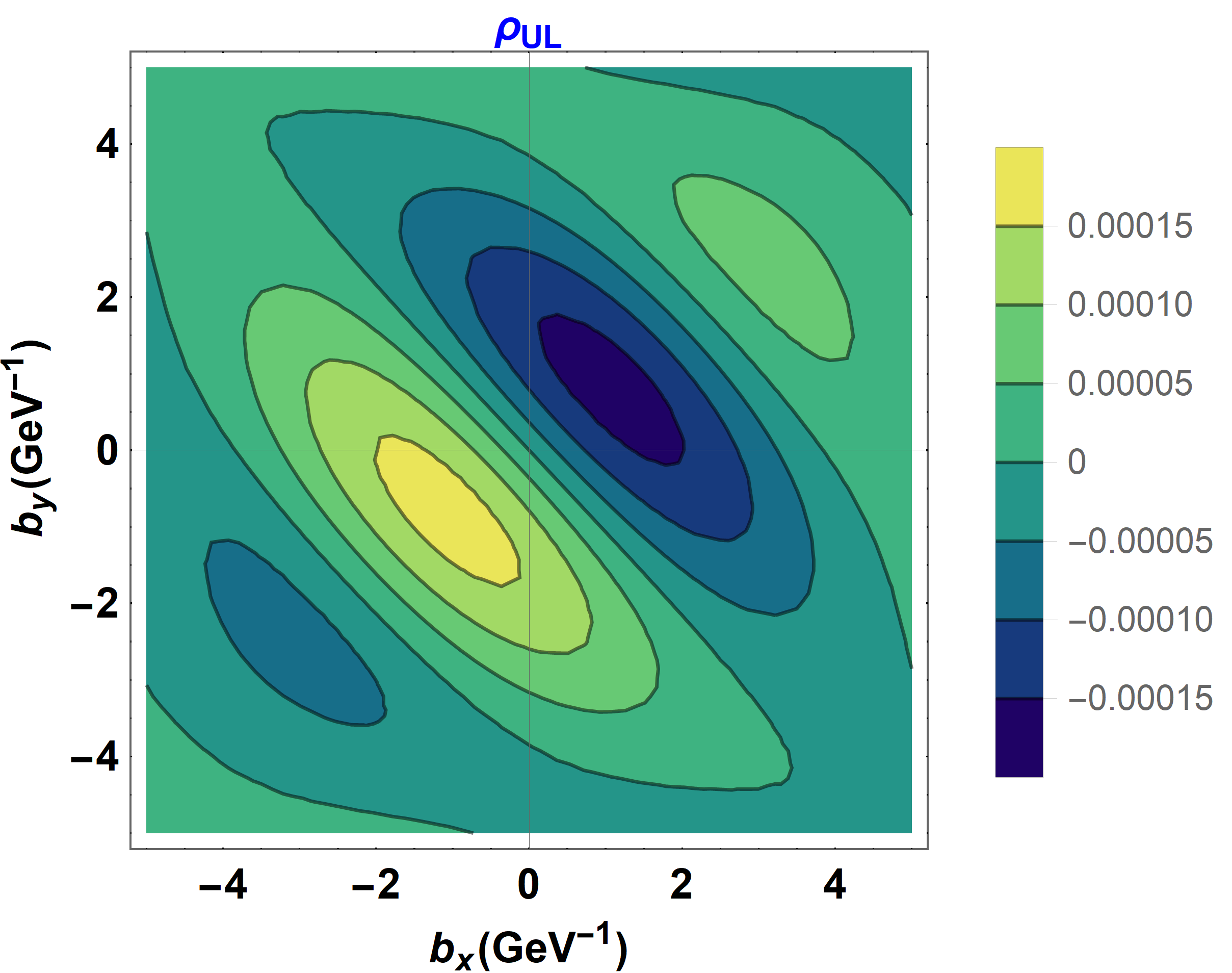}\\
\small{(d)}\includegraphics[width=5cm,height=4cm,clip]{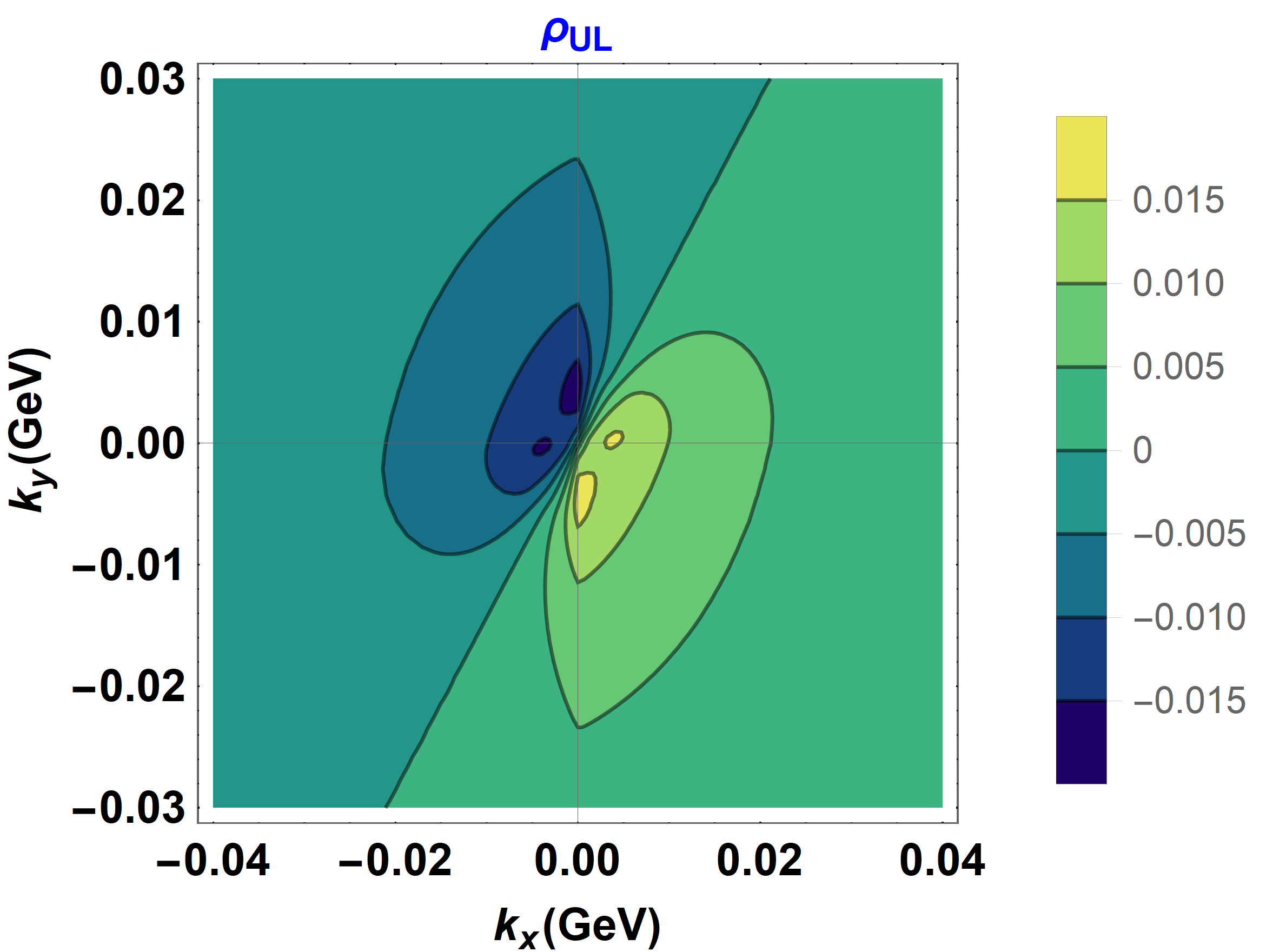} 
\small{(e)}\includegraphics[width=5cm,height=4cm,clip]{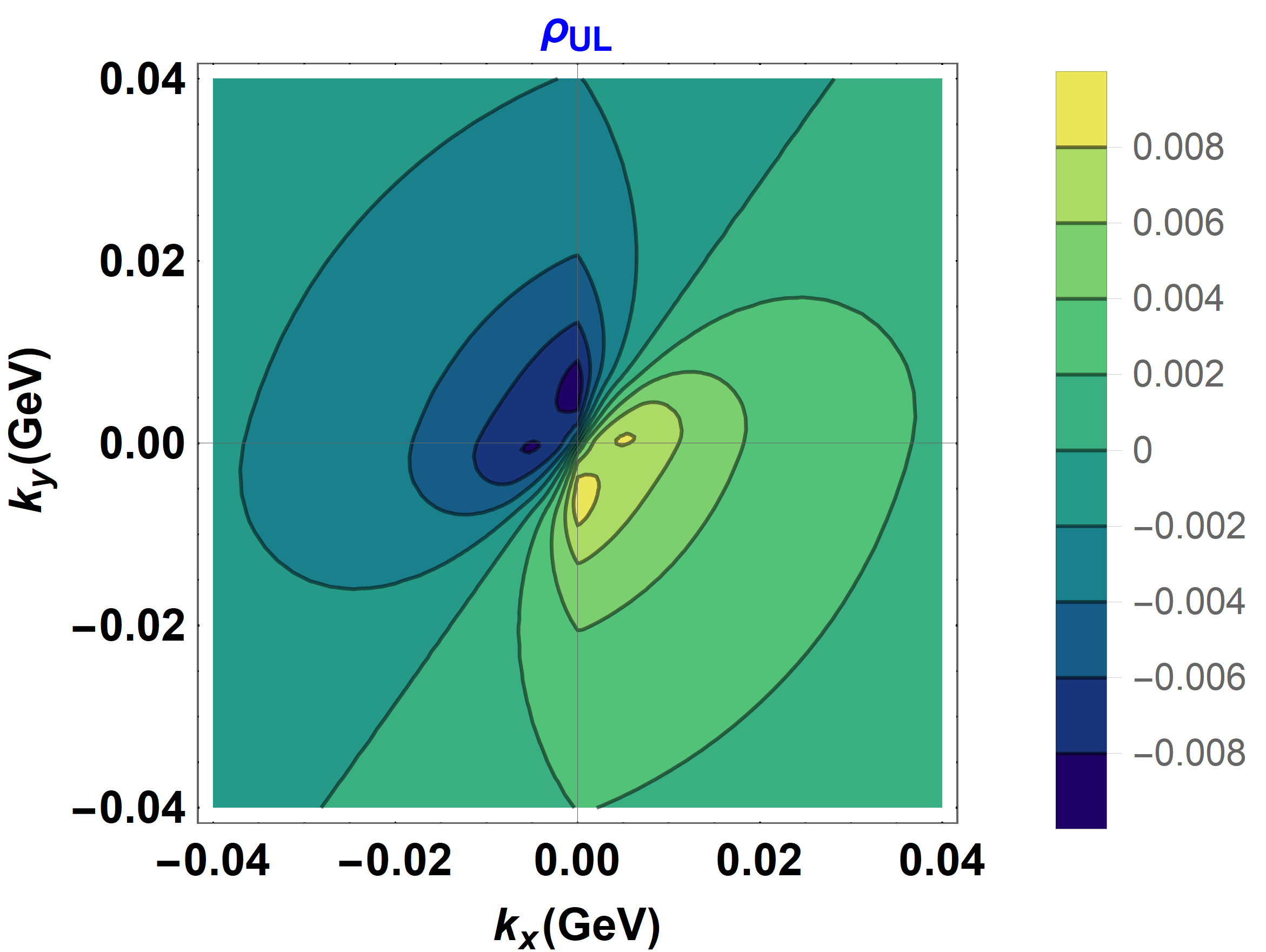}
\small{(f)}\includegraphics[width=5cm,height=4cm,clip]{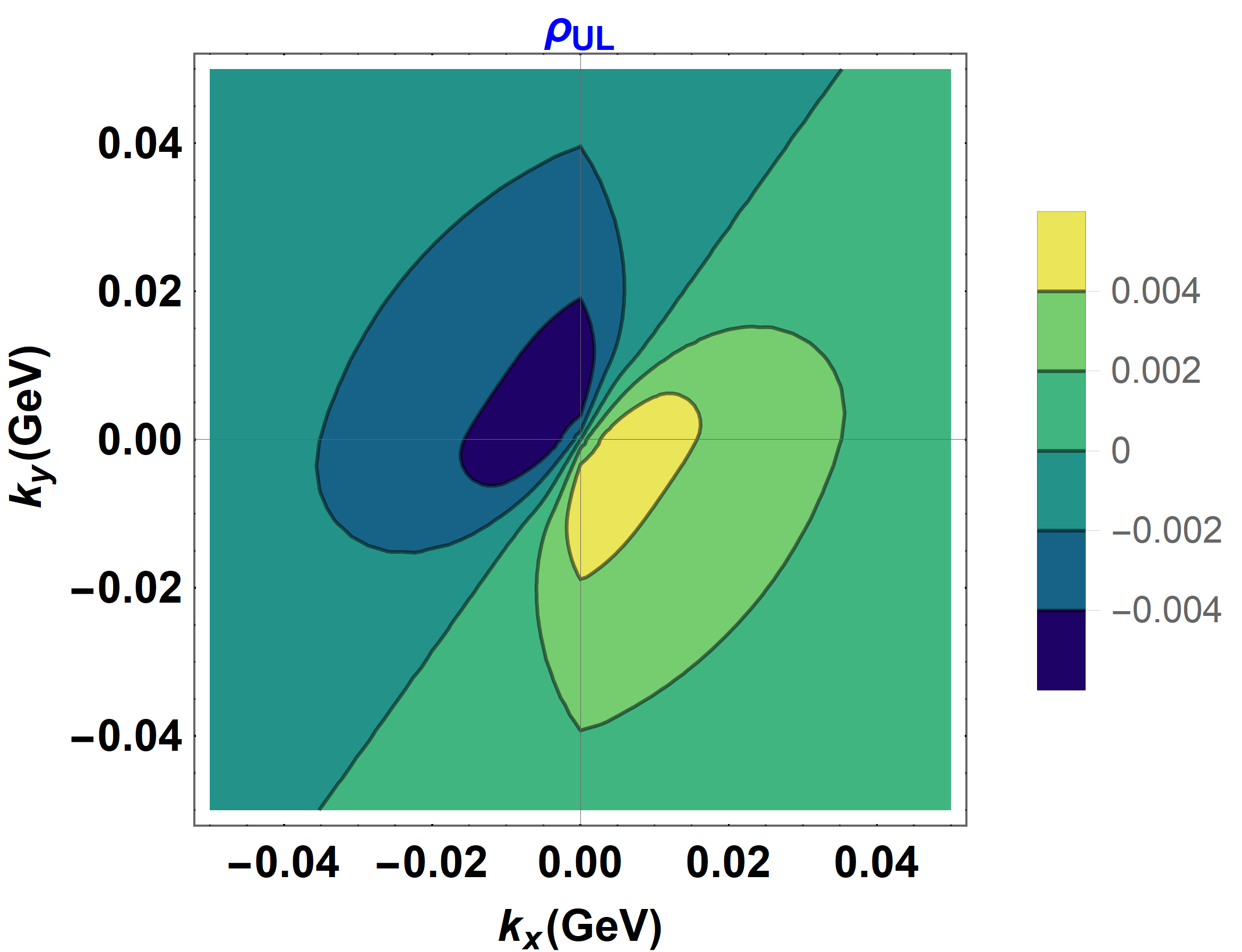} \\
\small{(g)}\includegraphics[width=5cm,height=4cm,clip]{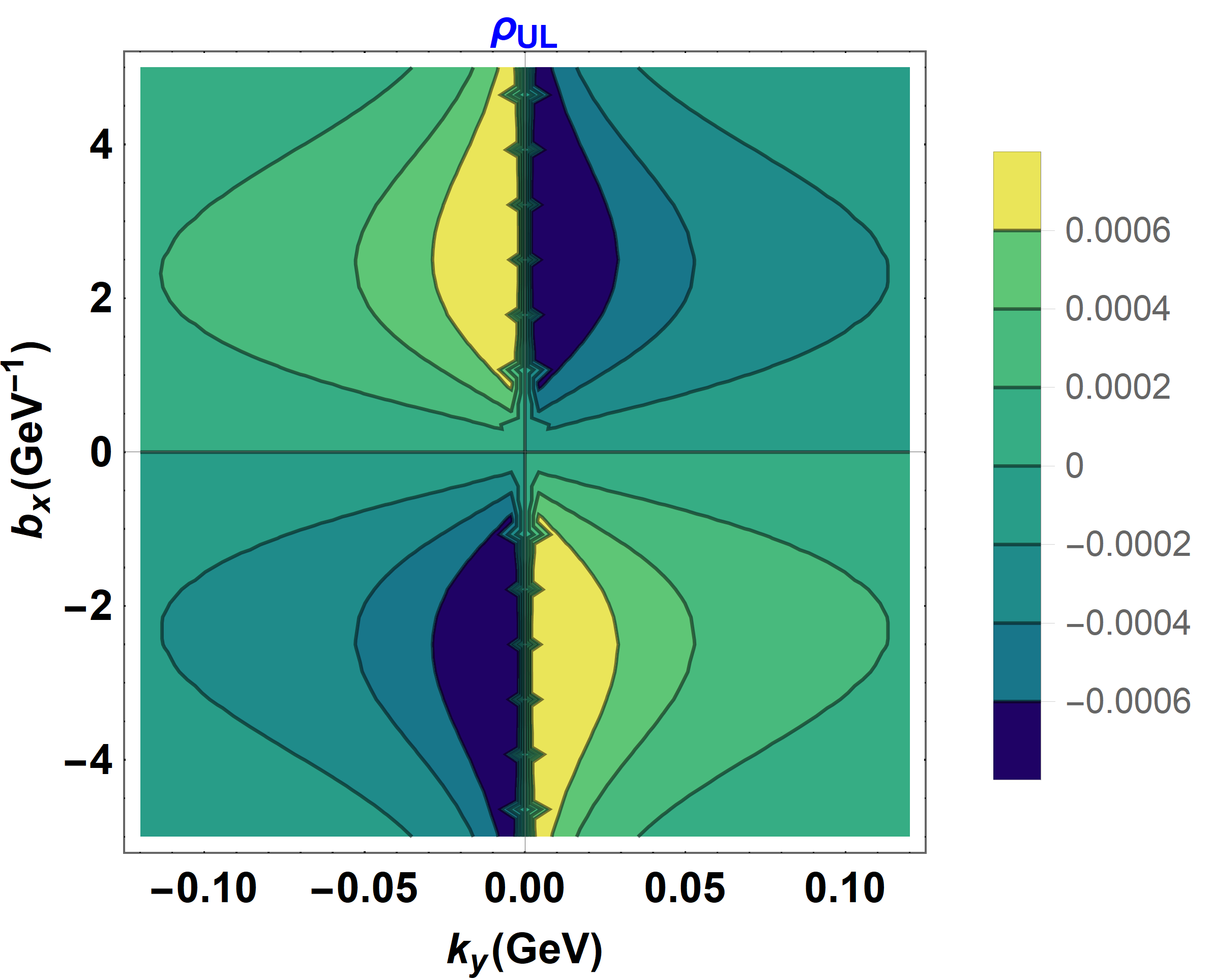} 
\small{(h)}\includegraphics[width=5cm,height=4cm,clip]{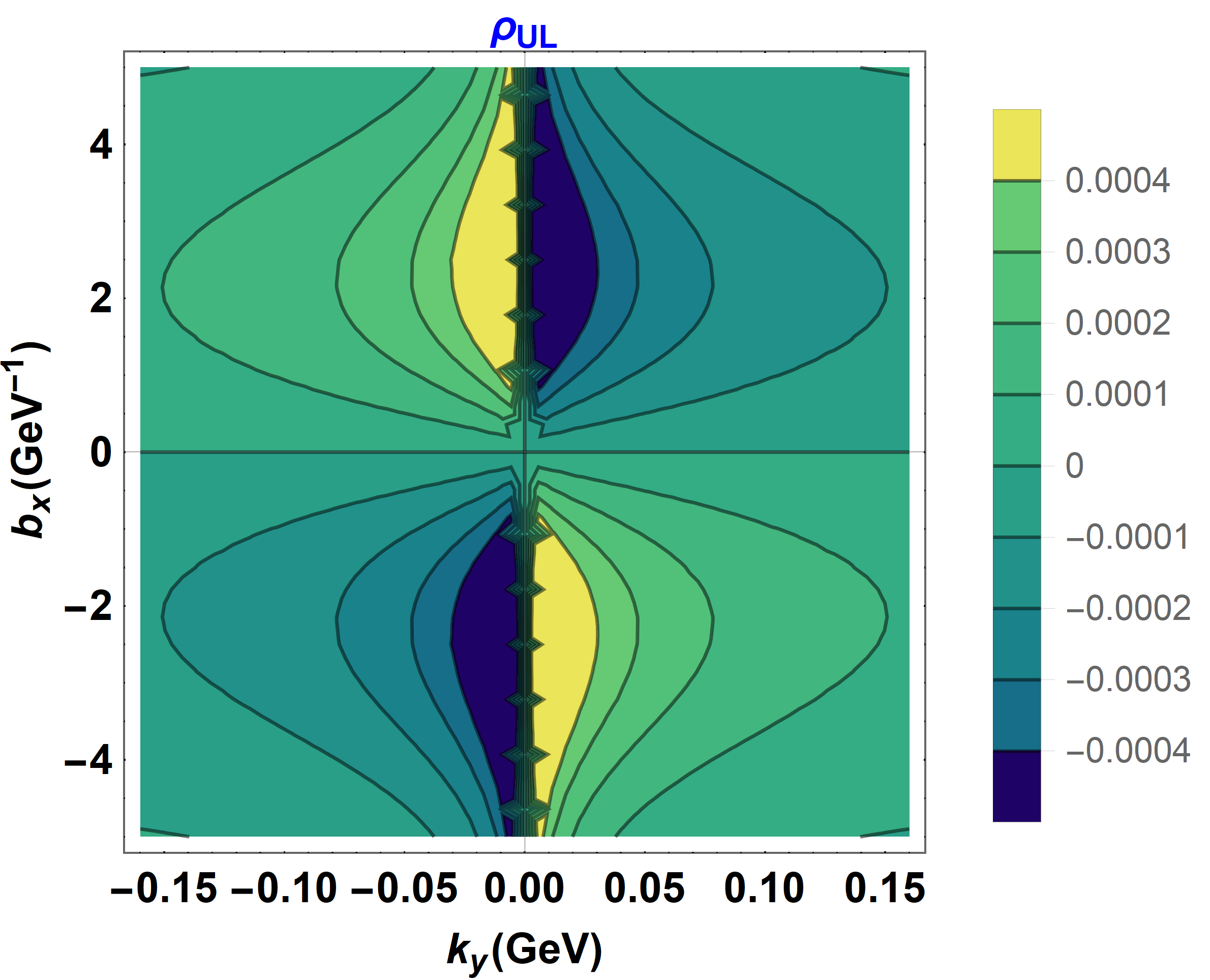}
\small{(i)}\includegraphics[width=5cm,height=4cm,clip]{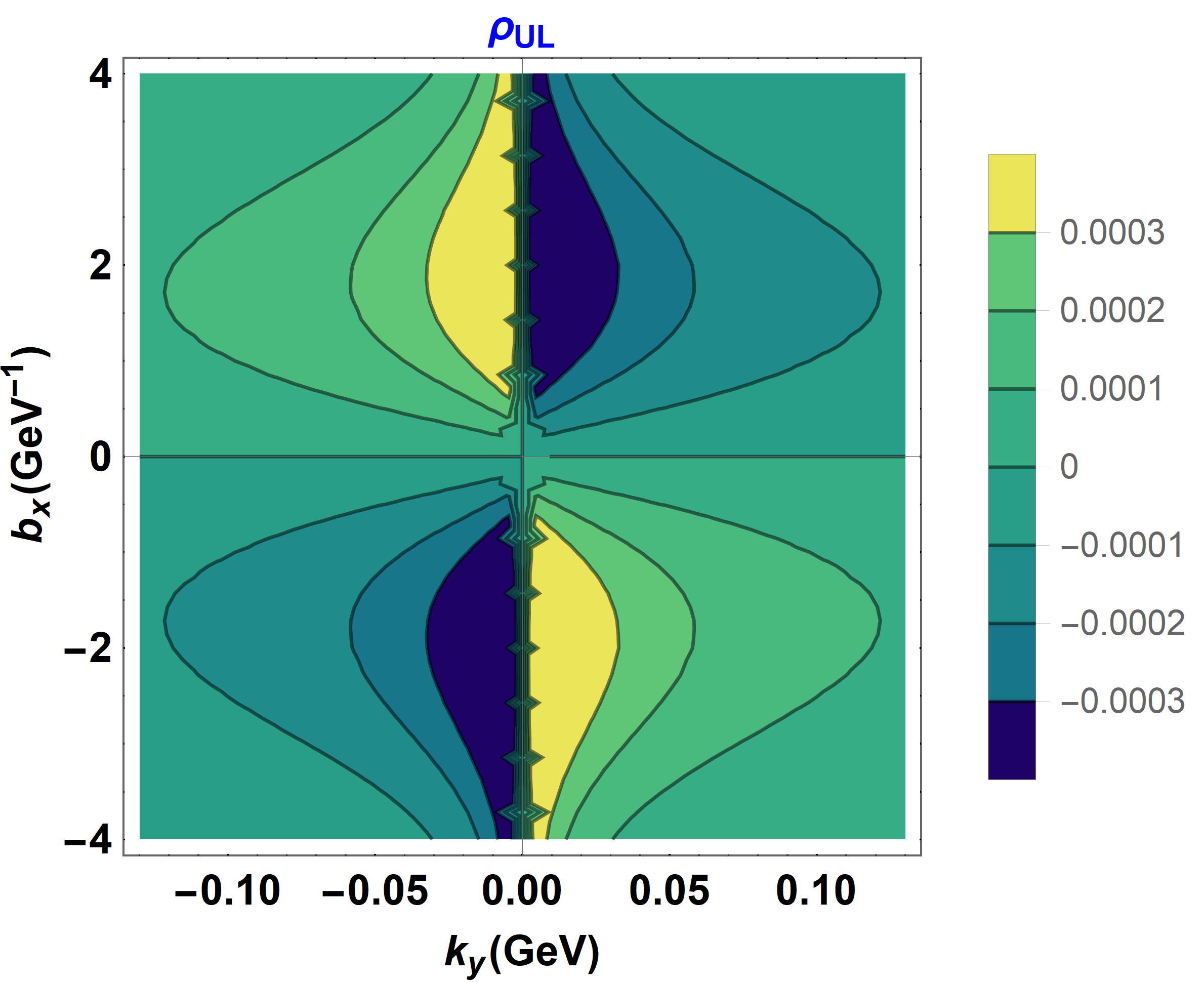} \\
\end{minipage}
\caption{\label{rho_UL}The quark Wigner distribution $\rho_{UL}$ in the transverse impact parameter plane, the transverse momentum plane, and the mixed plane. The left, middle, and right panels show the results for $\xi=0,\;\xi=0.25$, and $\xi=0.5$ respectively. }
\end{figure}

Figure~\ref{rho_UL} presents the Wigner distribution $\rho_{UL}$, corresponding to a longitudinally polarized quark in an unpolarized target. %The distribution is shown in three different representations: the transverse impact parameter plane $(b_x, b_y)$, the transverse momentum plane $(k_x, k_y)$, and the mixed phase-space $(b_x, k_y)$ . The columns from left to right correspond to skewness values $\xi = 0$, $\xi = 0.25$, and $\xi = 0.5$, respectively.
In the impact parameter space (Figs.~\ref{rho_UL}(a)--(c)), with fixed transverse momentum $\vec{k}_\perp = 0.4\,\hat{y}$ GeV, the distribution $\rho_{UL}(b_x, b_y)$ exhibits a characteristic dipole pattern. For $\xi = 0$, the dipole is oriented along the diagonal axis, and its sign structure indicates a spatial shift of the quark distribution depending on its helicity. This asymmetry is a hallmark of orbital angular momentum contributions associated with spin-orbit correlations. As $\xi$ increases, the dipole pattern persists but becomes more distorted and asymmetric, reflecting the influence of nonzero longitudinal momentum transfer on the spatial localization of the polarized quark.

In the transverse momentum space (Figs.~\ref{rho_UL}(d)--(f)), with $\vec{b}_\perp = 0.4\,\hat{y}$ GeV$^{-1}$ fixed, the distribution $\rho_{UL}(k_x, k_y)$ also displays a prominent dipole structure. At $\xi = 0$, the distribution shows an anti-symmetric pattern about the $k_y$ axis, with positive and negative lobes indicating a transverse momentum asymmetry induced by the quark helicity. This behavior signals the presence of spin-momentum correlations - specifically, a correlation between the quark's longitudinal spin and its transverse motion in the hadron. As $\xi$ increases, the distribution becomes increasingly asymmetric, and the lobes shift further apart, suggesting enhanced sensitivity to recoil effects and changes in parton orbital dynamics due to the off-forward nature of the matrix element.

The mixed-space representation (Figs.~\ref{rho_UL}(g)--(i)), where $k_x$ and $b_y$ are integrated over $[0,\,0.4]$, reveals further insights into the coupling between spatial and momentum degrees of freedom. At $\xi = 0$, the distribution is odd in $b_x$ and $k_y$, and is sharply peaked around $b_x = 0$, forming a vertical ridge structure. This pattern represents a strong correlation between the quark’s transverse position and its momentum, modulated by the quark helicity. As $\xi$ increases, the central structure becomes narrower and more localized, indicating a contraction in the phase-space overlap. The anti-symmetric profile in $b_x$ persists across all $\xi$ values, consistent with the helicity-sensitive nature of $\rho_{UL}$.

Overall, the behavior of $\rho_{UL}$ reflects nontrivial spin-orbit correlations and the presence of quark orbital angular momentum in the longitudinal polarization sector. The distributions exhibit clear dipole structures in both position and momentum space, which deform under increasing skewness due to kinematic shifts in the quark momentum and position.

\subsection{Transversely Polarized quark in unpolarized target}
\begin{figure}[!htp]
\begin{minipage}[c]{1\textwidth}
\small{(a)}\includegraphics[width=5cm,height=4cm,clip]{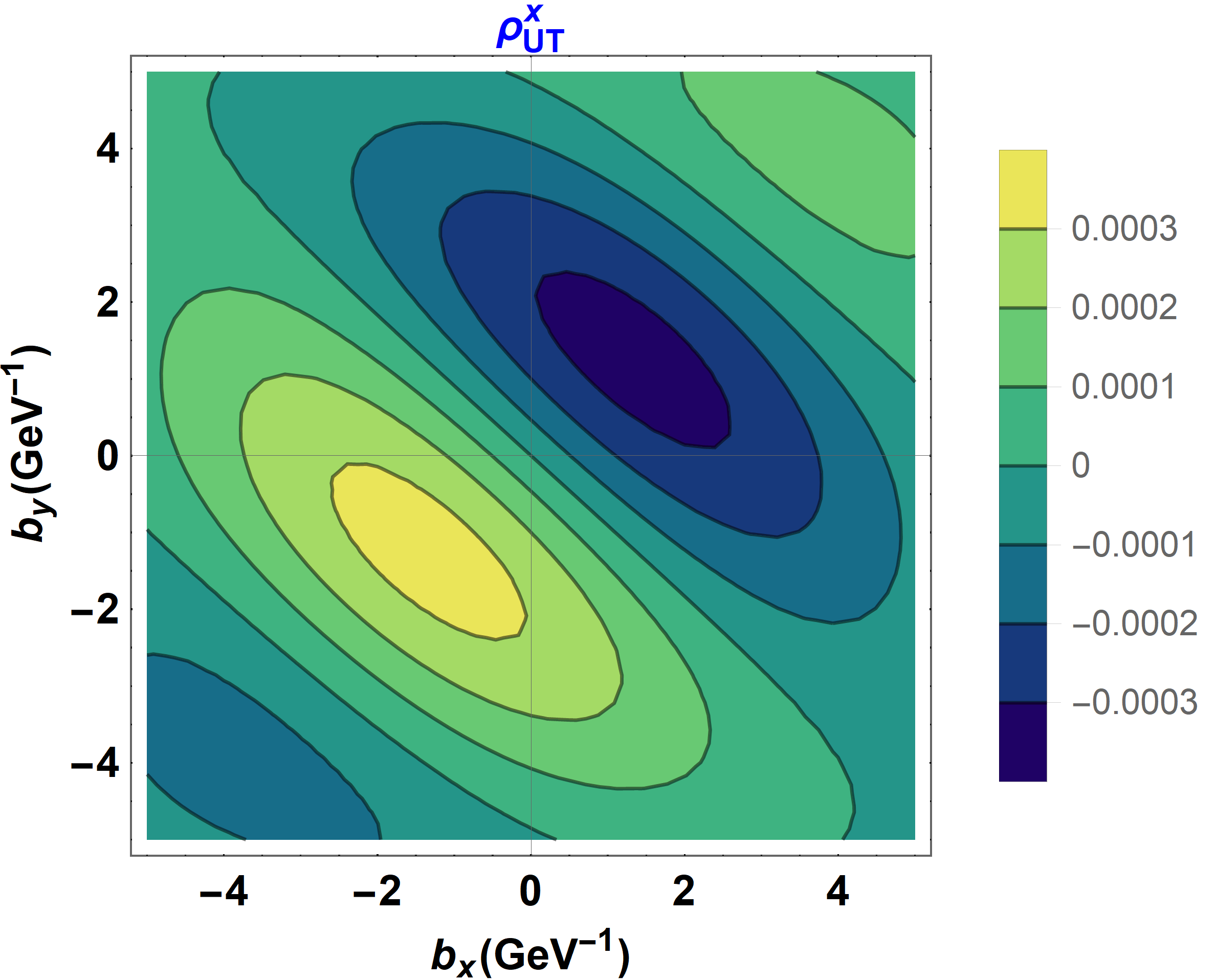}
\small{(b)}\includegraphics[width=5cm,height=4cm,clip]{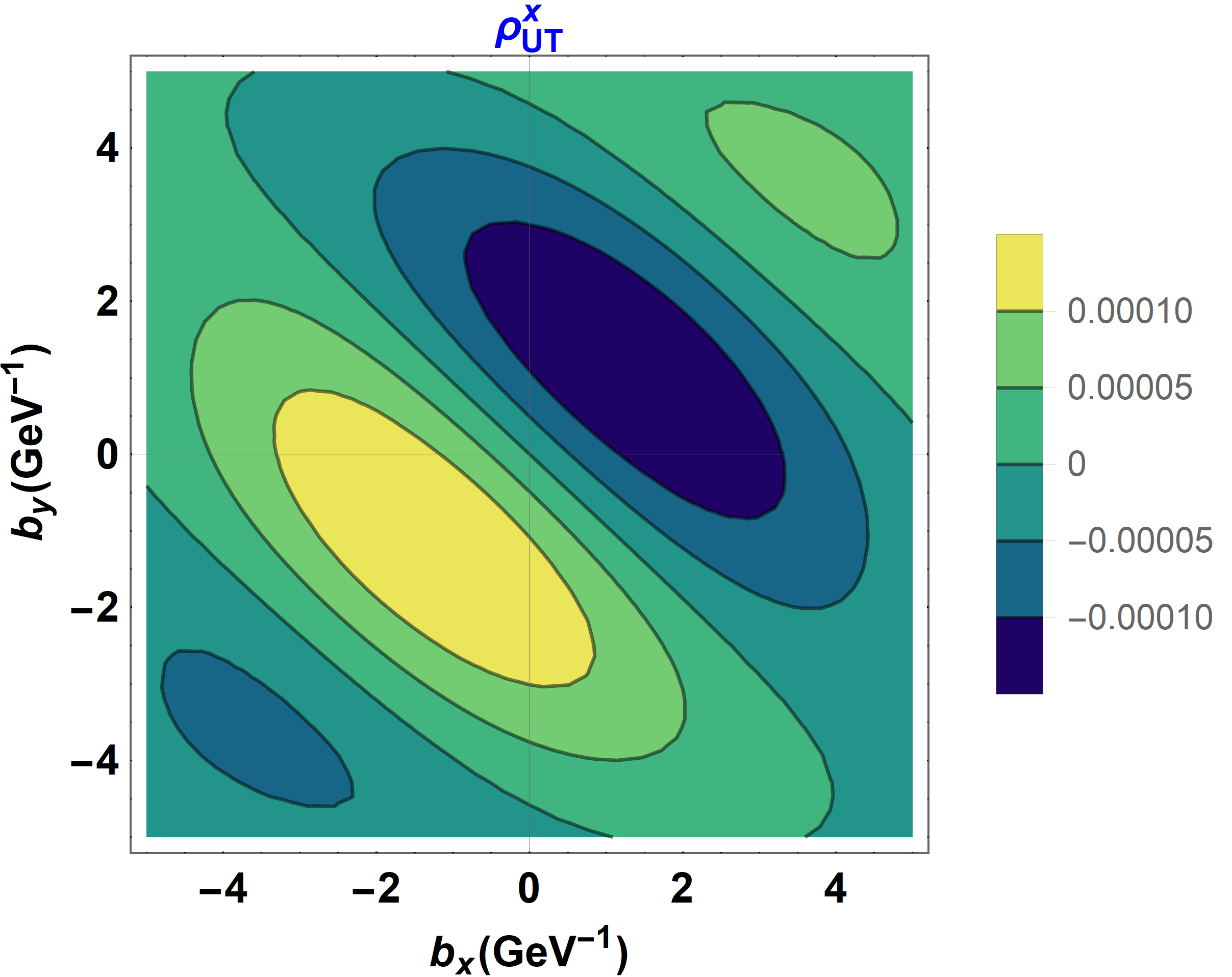} 
\small{(c)}\includegraphics[width=5cm,height=4cm,clip]{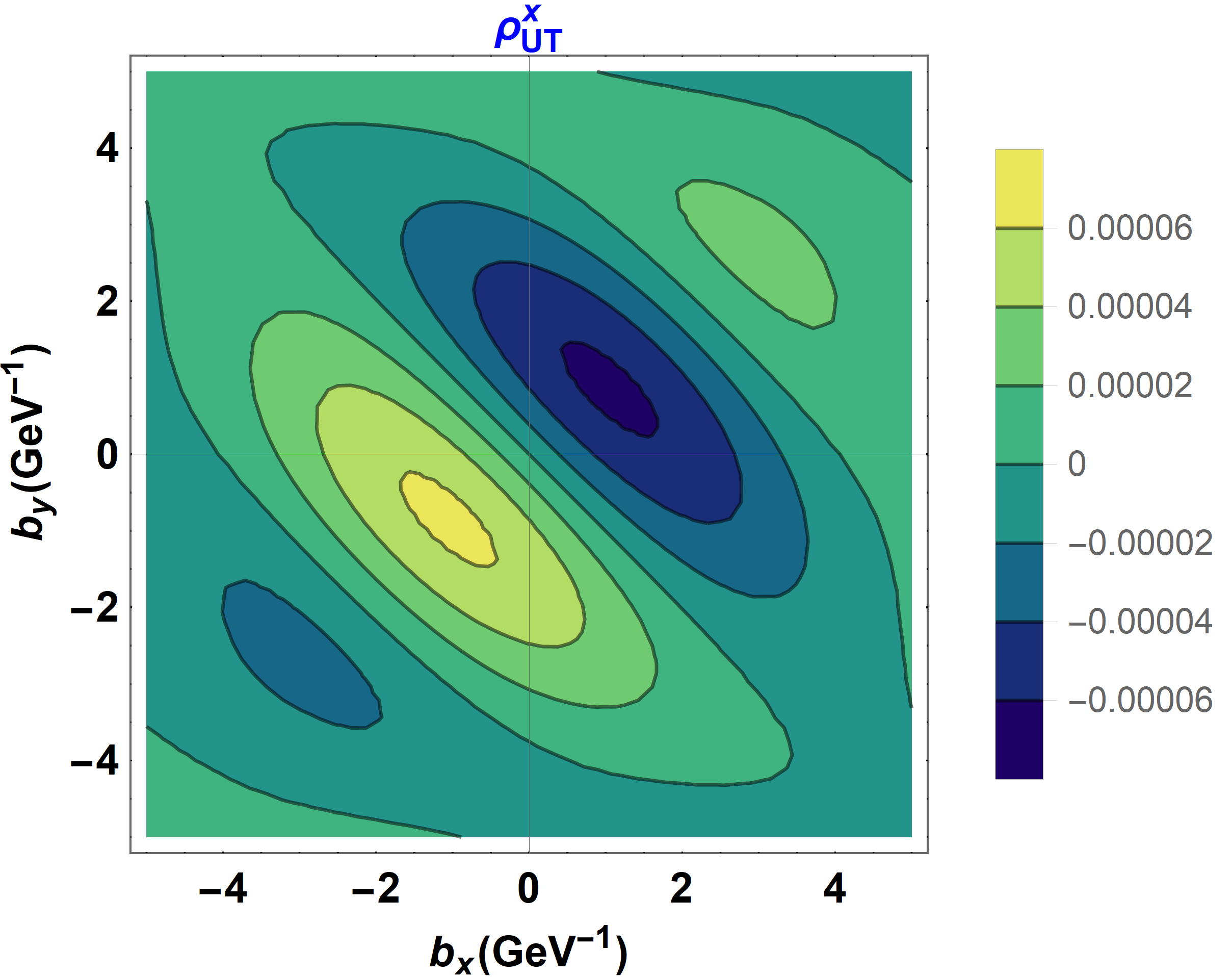}\\
\small{(d)}\includegraphics[width=5cm,height=4cm,clip]{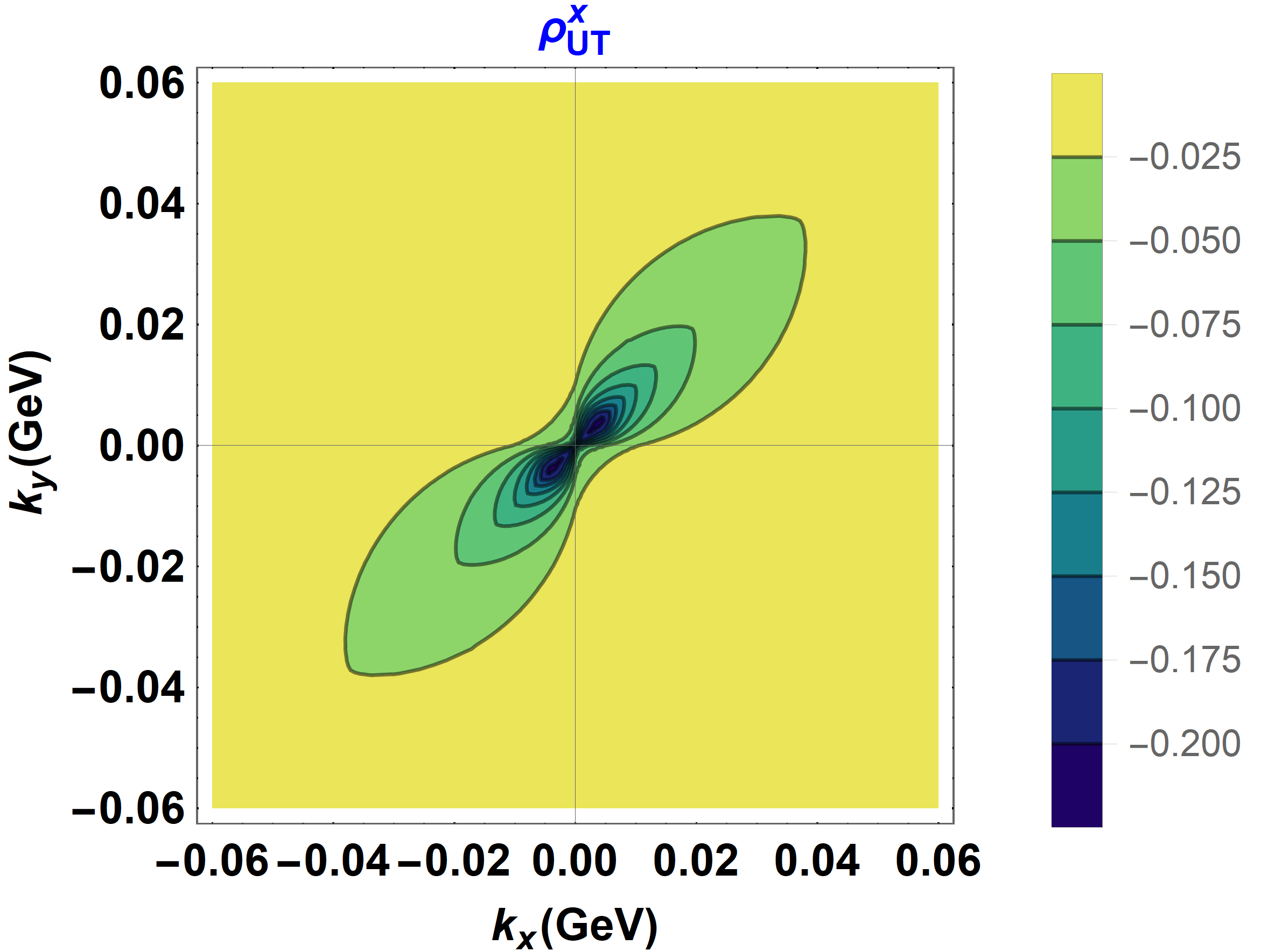} 
\small{(e)}\includegraphics[width=5cm,height=4cm,clip]{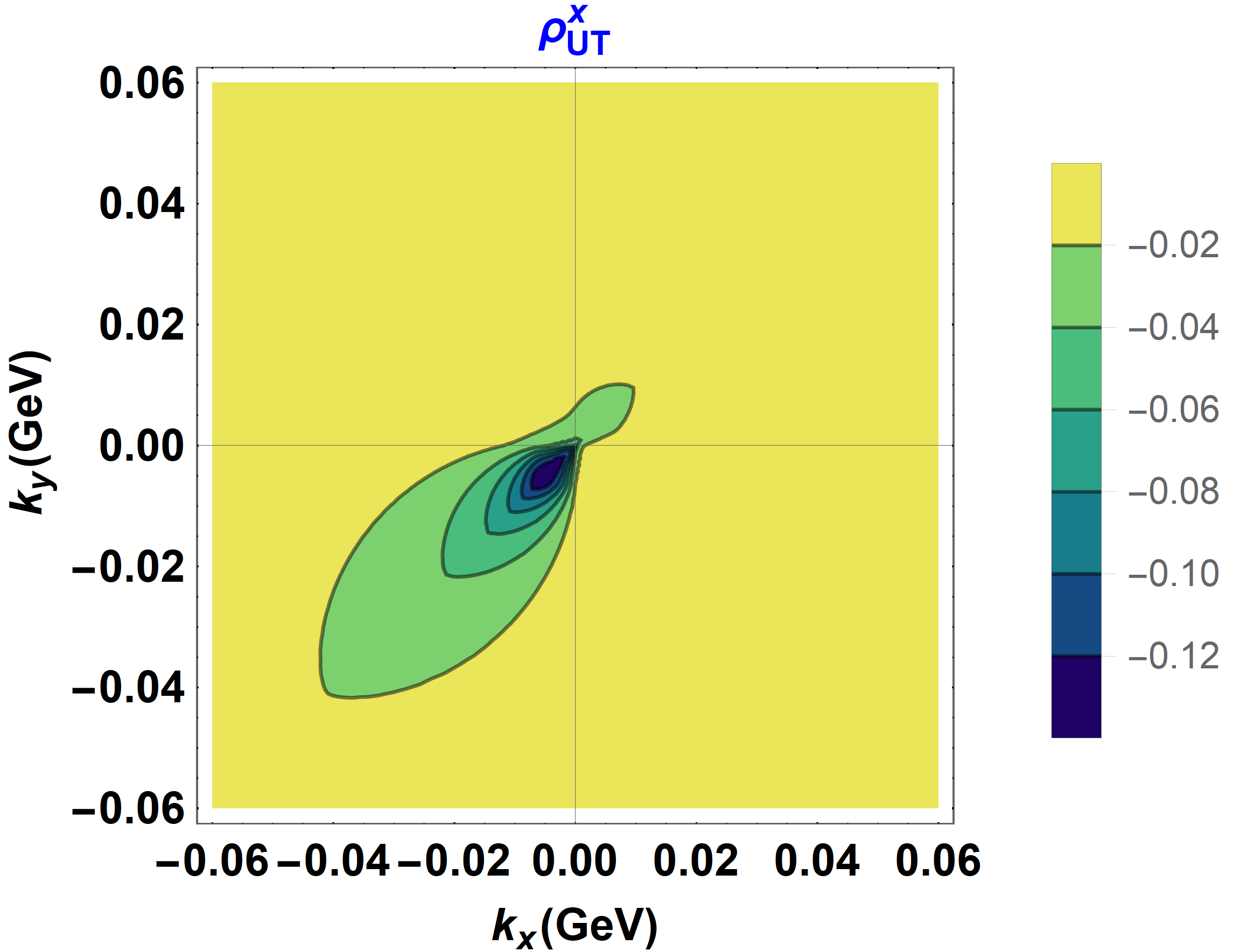}
\small{(f)}\includegraphics[width=5cm,height=4cm,clip]{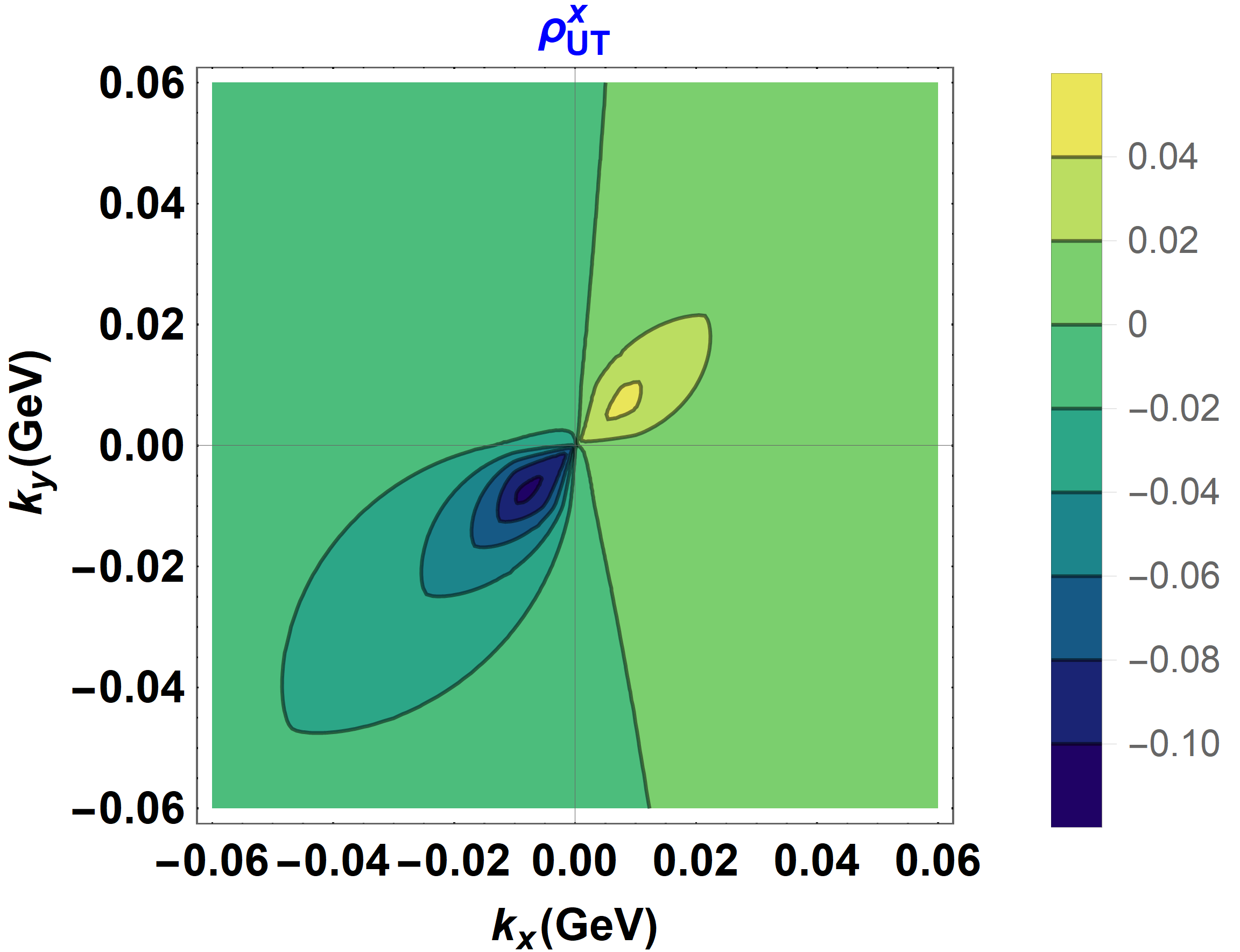} \\
\small{(g)}\includegraphics[width=5cm,height=4cm,clip]{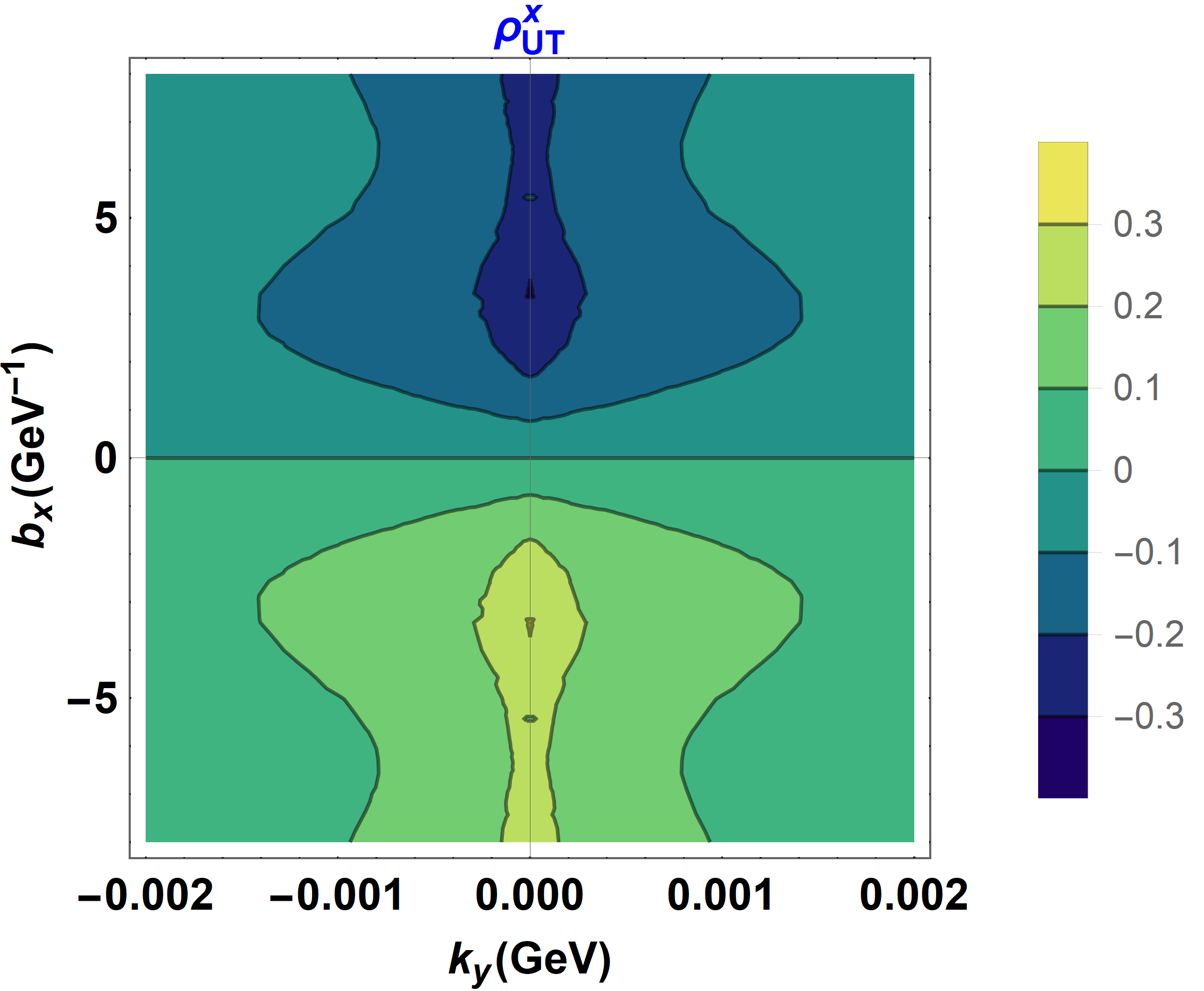} 
\small{(h)}\includegraphics[width=5cm,height=4cm,clip]{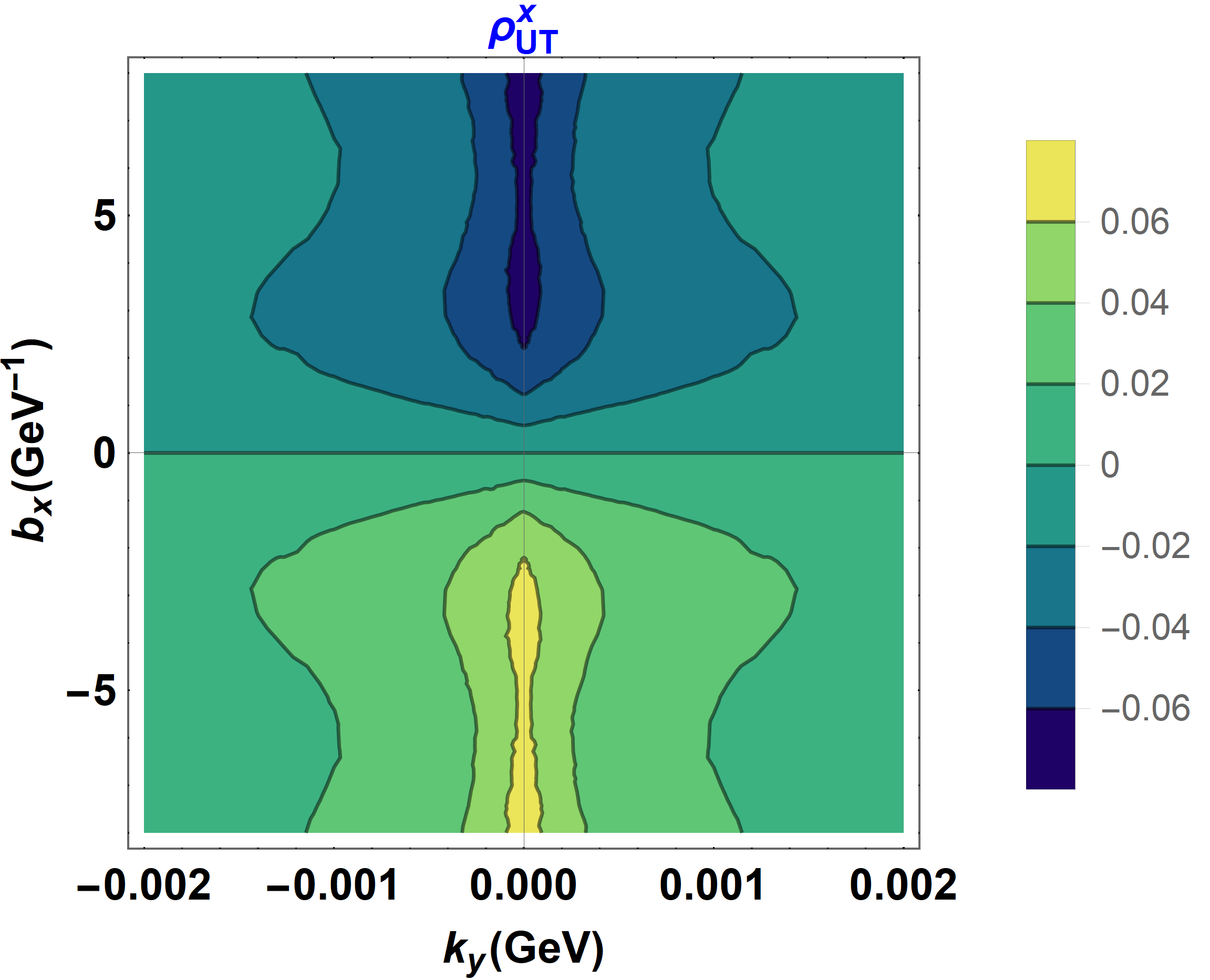}
\small{(i)}\includegraphics[width=5cm,height=4cm,clip]{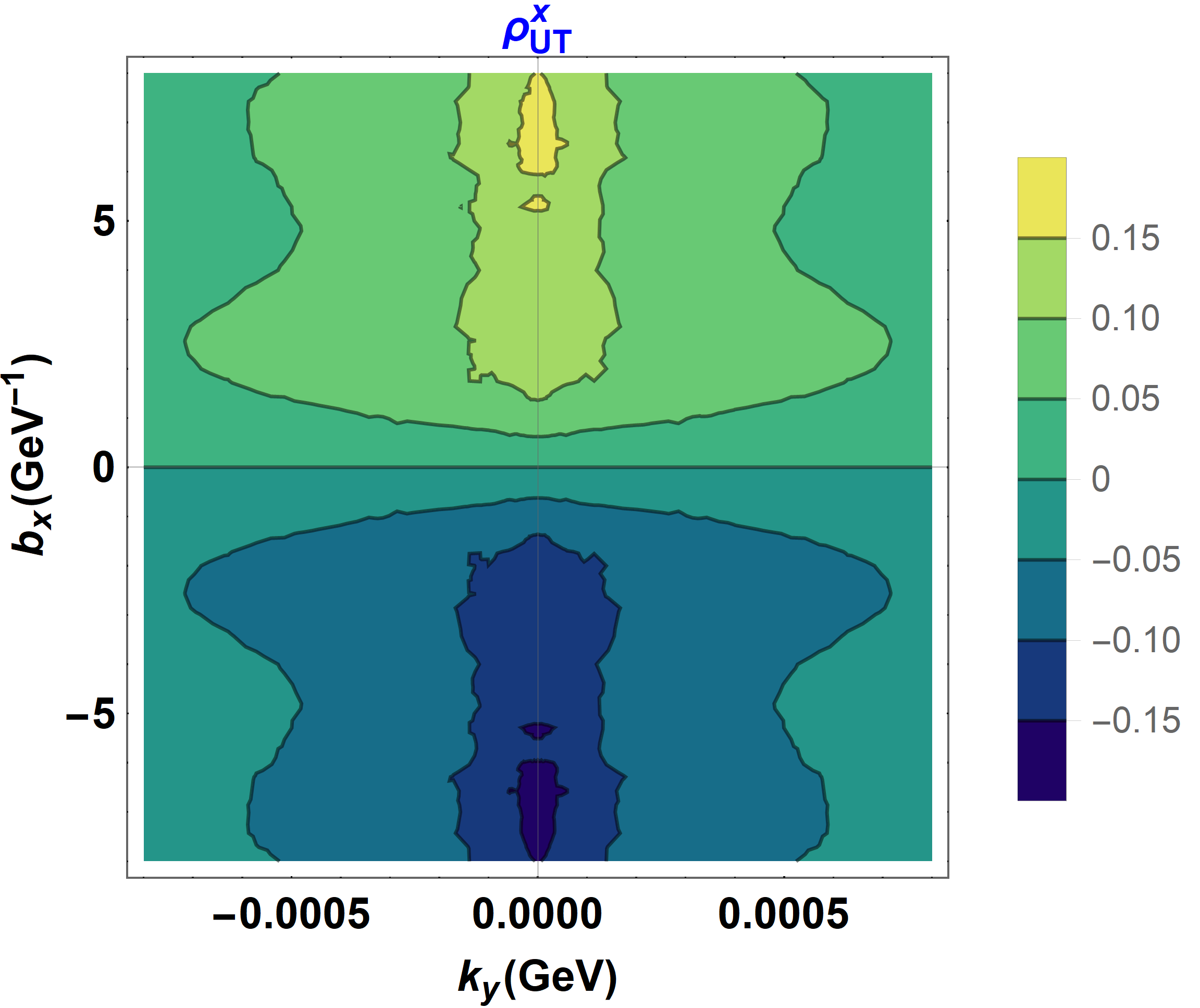} \\
\end{minipage}
\caption{\label{rho_UT}The quark Wigner distribution $\rho_{UT}^x$ in the transverse impact parameter plane, the transverse momentum plane, and the mixed plane. The left, middle, and right panels display the results for $\xi = 0$, $\xi = 0.25$, and $\xi = 0.5$, respectively. }
\end{figure}

Figure~\ref{rho_UT} displays the Wigner distribution $\rho_{UT}^x$, which corresponds to a transversely polarized quark (polarized along the $\hat{x}$ direction) in an unpolarized target. % The distribution is presented in the transverse impact parameter space (top row), the transverse momentum space (middle row), and the mixed phase-space (bottom row). The left, middle, and right columns show the results for skewness values $\xi = 0$, $0.25$, and $0.5$, respectively.
In the impact parameter space (Figs.~\ref{rho_UT}(a)--(c)), with quark transverse momentum fixed at $\vec{k}_\perp = 0.4\,\hat{y}$ GeV, the distribution exhibits a clear dipole structure oriented along the $b_y$ axis. For $\xi = 0$, this dipole is symmetric and its sign flips across the $b_y = 0$ axis. This pattern is a direct manifestation of the spin-orbit correlation between the quark’s transverse spin and its transverse spatial position. As skewness increases, the dipole becomes distorted and loses its mirror symmetry, with the positive and negative regions shifting asymmetrically. This distortion indicates the sensitivity of the spatial structure to longitudinal momentum transfer, which induces transverse deformation in the quark distribution.

The transverse momentum space representation (Figs.~\ref{rho_UT}(d)--(f)), with fixed transverse position $\vec{b}_\perp = 0.4\,\hat{y}$ GeV$^{-1}$, reveals a pronounced dipole structure in the $k_x$ direction for $\xi = 0$. This asymmetry indicates a nonzero average transverse momentum generated by the transverse spin of the quark, consistent with the behavior expected from T-odd distributions like the Sivers function in the TMD framework. As $\xi$ increases, the dipole weakens and compresses toward the origin, with the transverse momentum support narrowing significantly. This reflects a suppression of spin-momentum correlations at higher skewness, likely due to the decreasing overlap between initial and final wavefunction components.

The mixed-space distribution (Figs.~\ref{rho_UT}(g)--(i)), with $k_x$ and $b_y$ integrated over a fixed range, further emphasizes the interplay between transverse position and momentum for the transversely polarized quark. For $\xi = 0$, the distribution is anti-symmetric in $b_x$ and exhibits a sharp vertical ridge centered at $k_y = 0$, indicating strong correlations between the transverse spin and the quark’s position and momentum. As $\xi$ increases, the ridge becomes broader and less pronounced, and the overall magnitude of the distribution decreases. This behavior signals a loss of coherence in the spin-dependent phase-space correlations under increasing longitudinal momentum transfer.

\subsection{Longitudinally Polarized quark in longitudinally polarized target}
\begin{figure}[!htp]
\begin{minipage}[c]{1\textwidth}
\small{(a)}\includegraphics[width=5cm,height=4cm,clip]{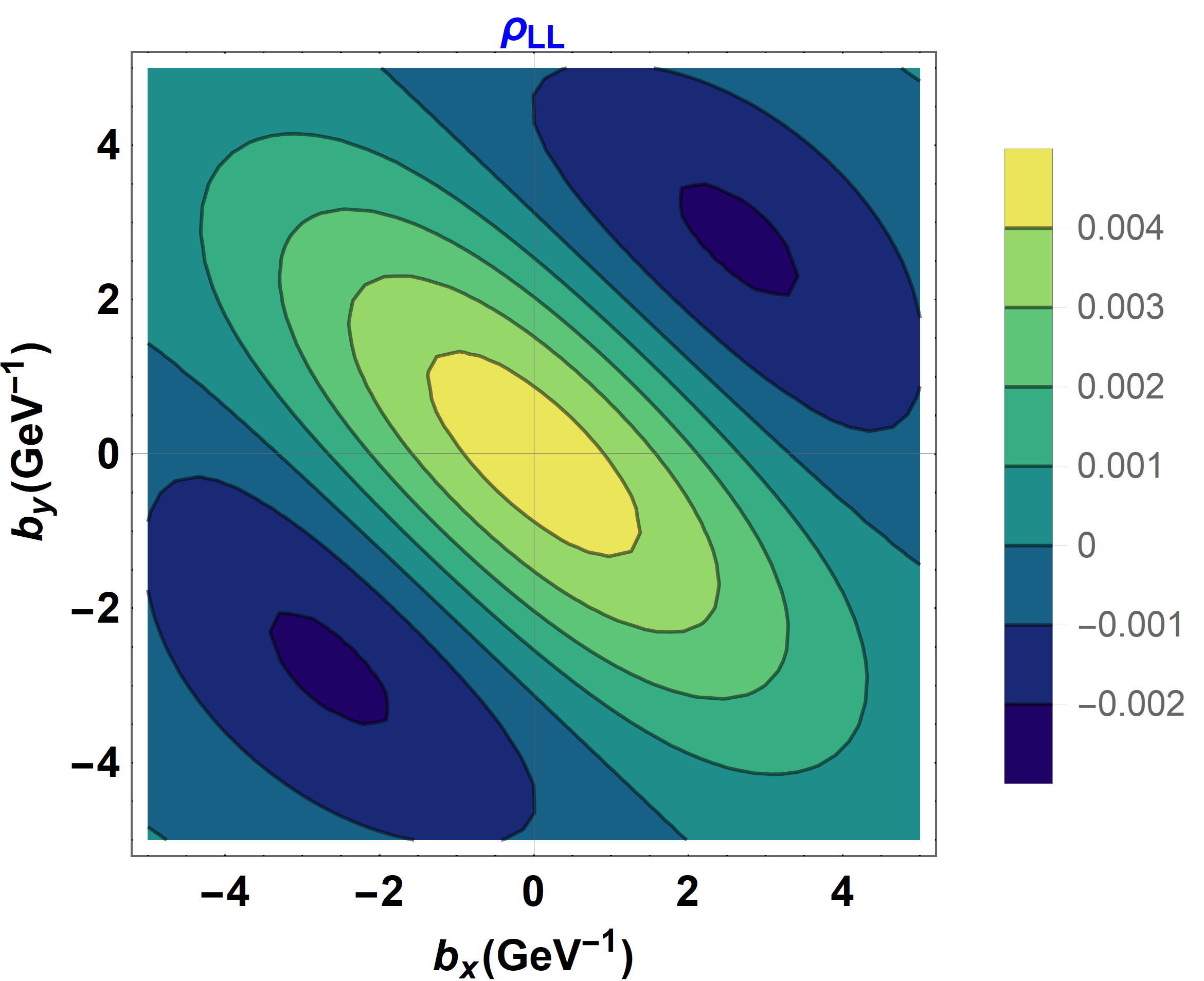}
\small{(b)}\includegraphics[width=5cm,height=4cm,clip]{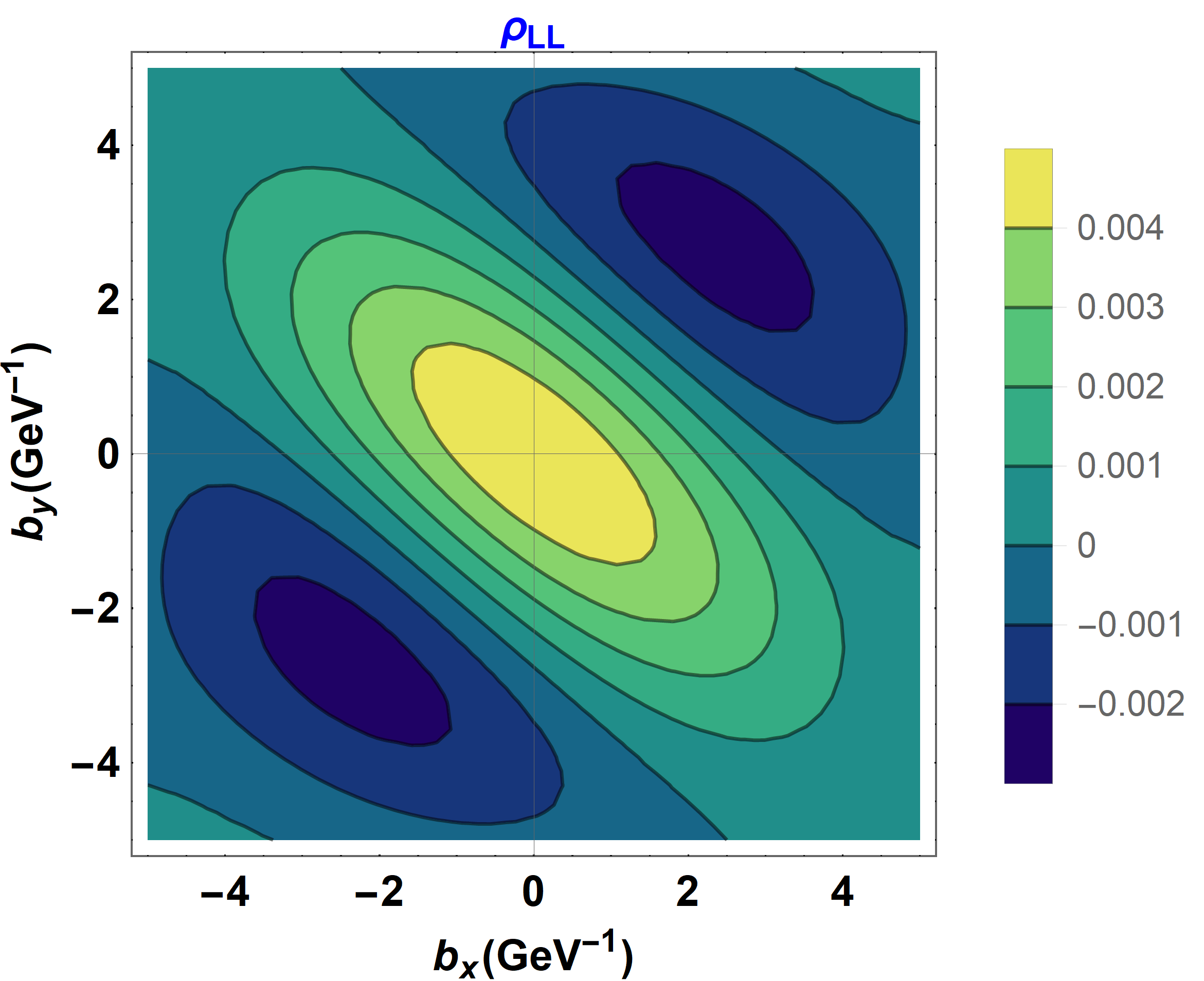} 
\small{(c)}\includegraphics[width=5cm,height=4cm,clip]{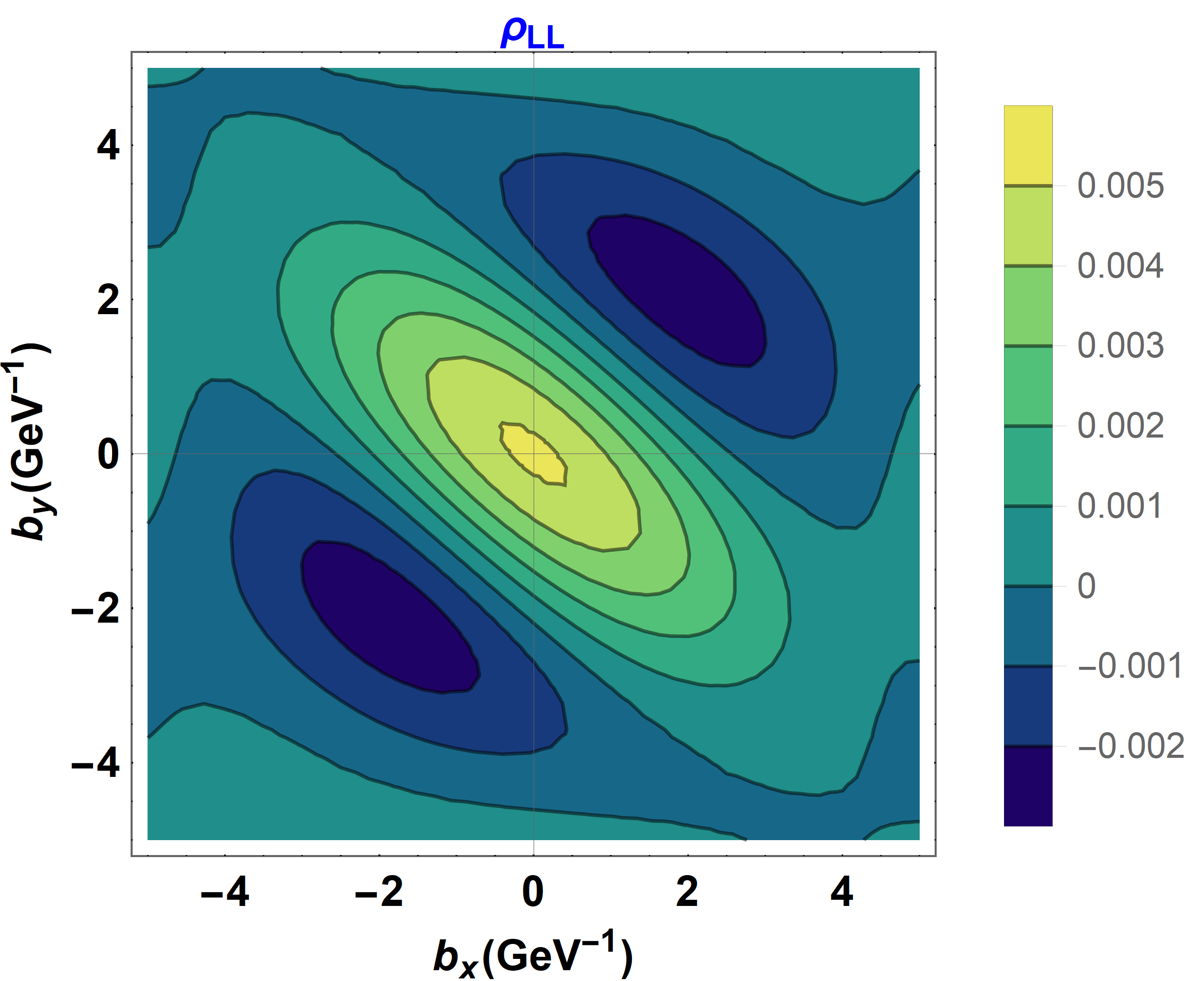}\\
\small{(d)}\includegraphics[width=5cm,height=4cm,clip]{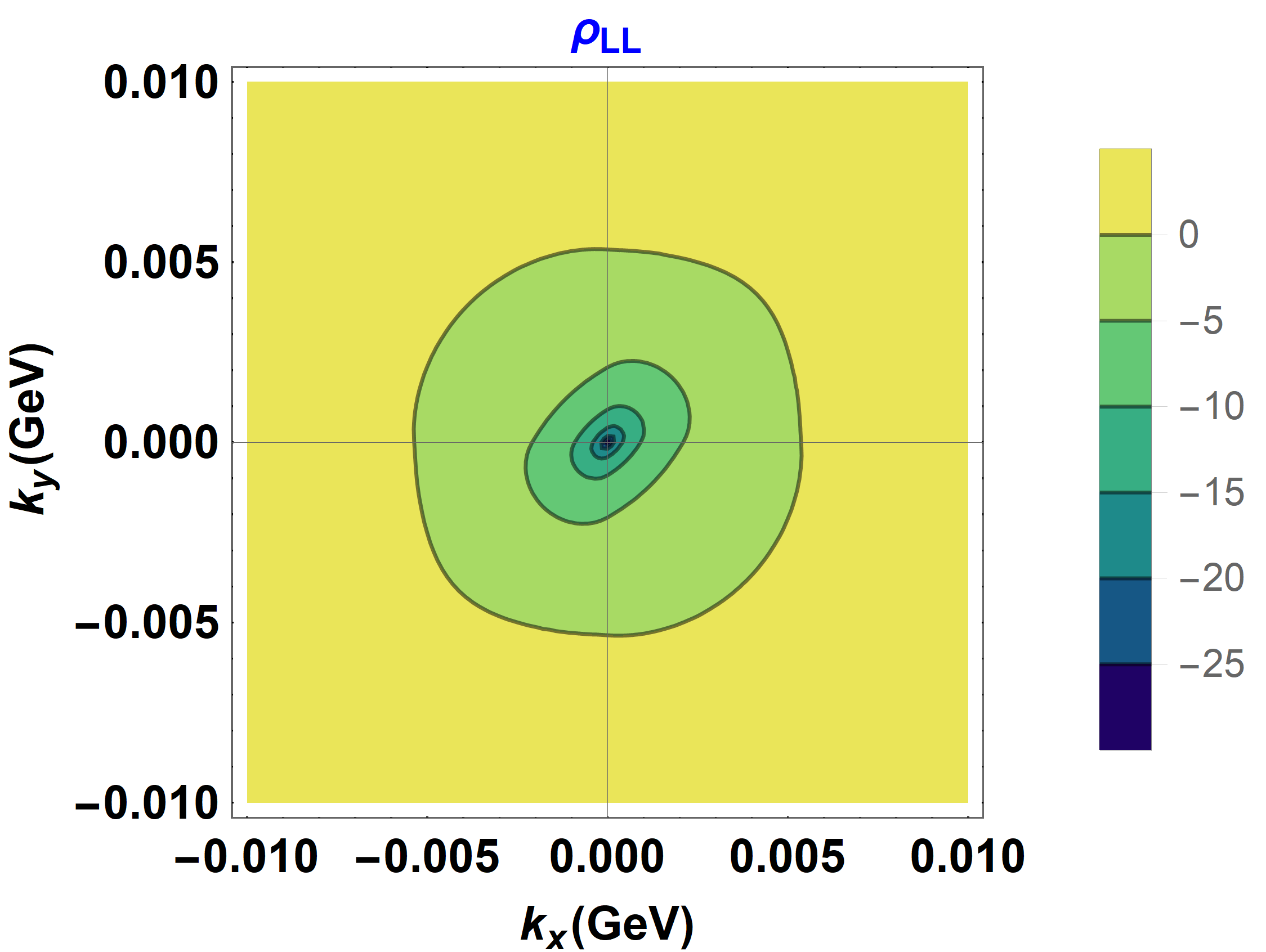} 
\small{(e)}\includegraphics[width=5cm,height=4cm,clip]{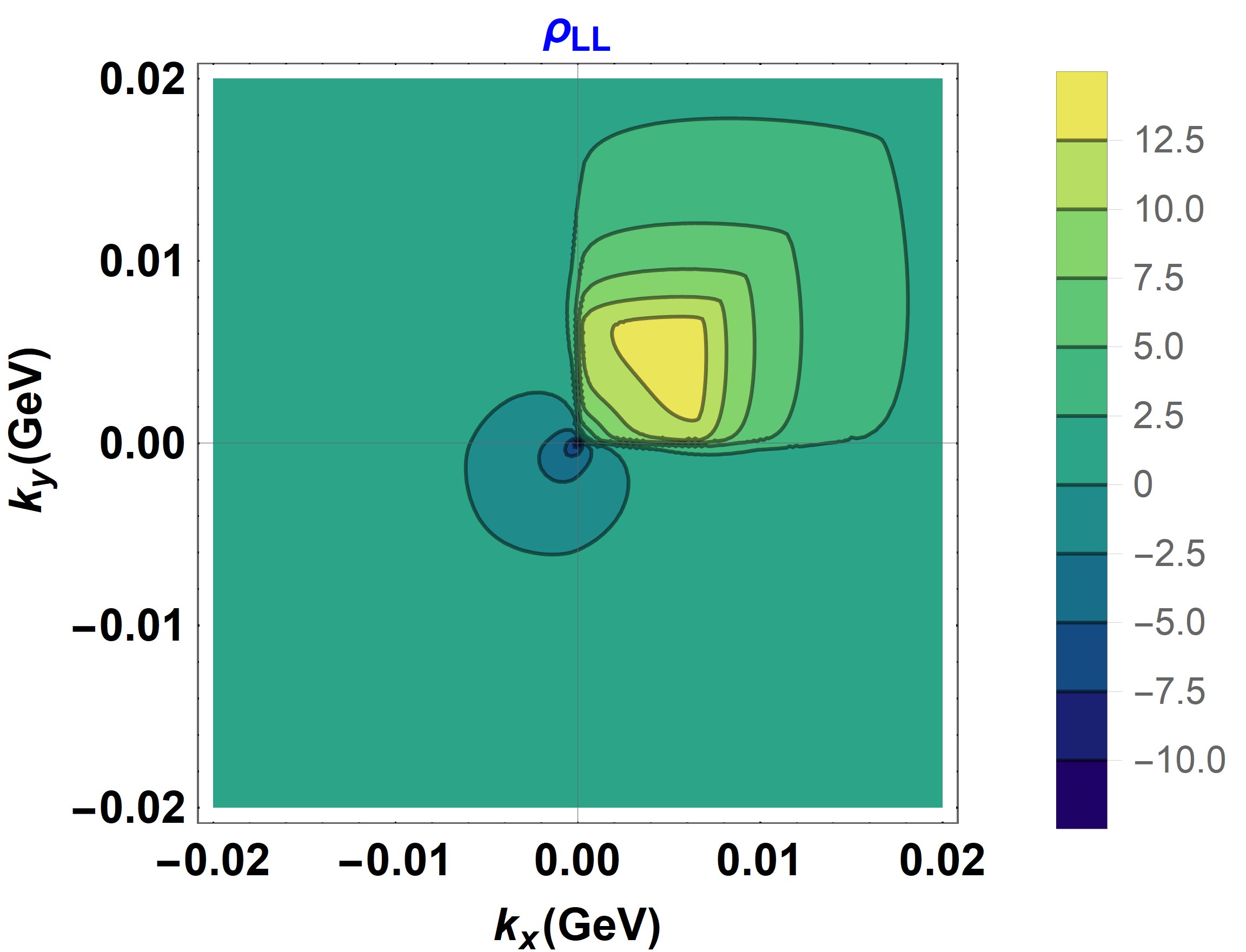}
\small{(f)}\includegraphics[width=5cm,height=4cm,clip]{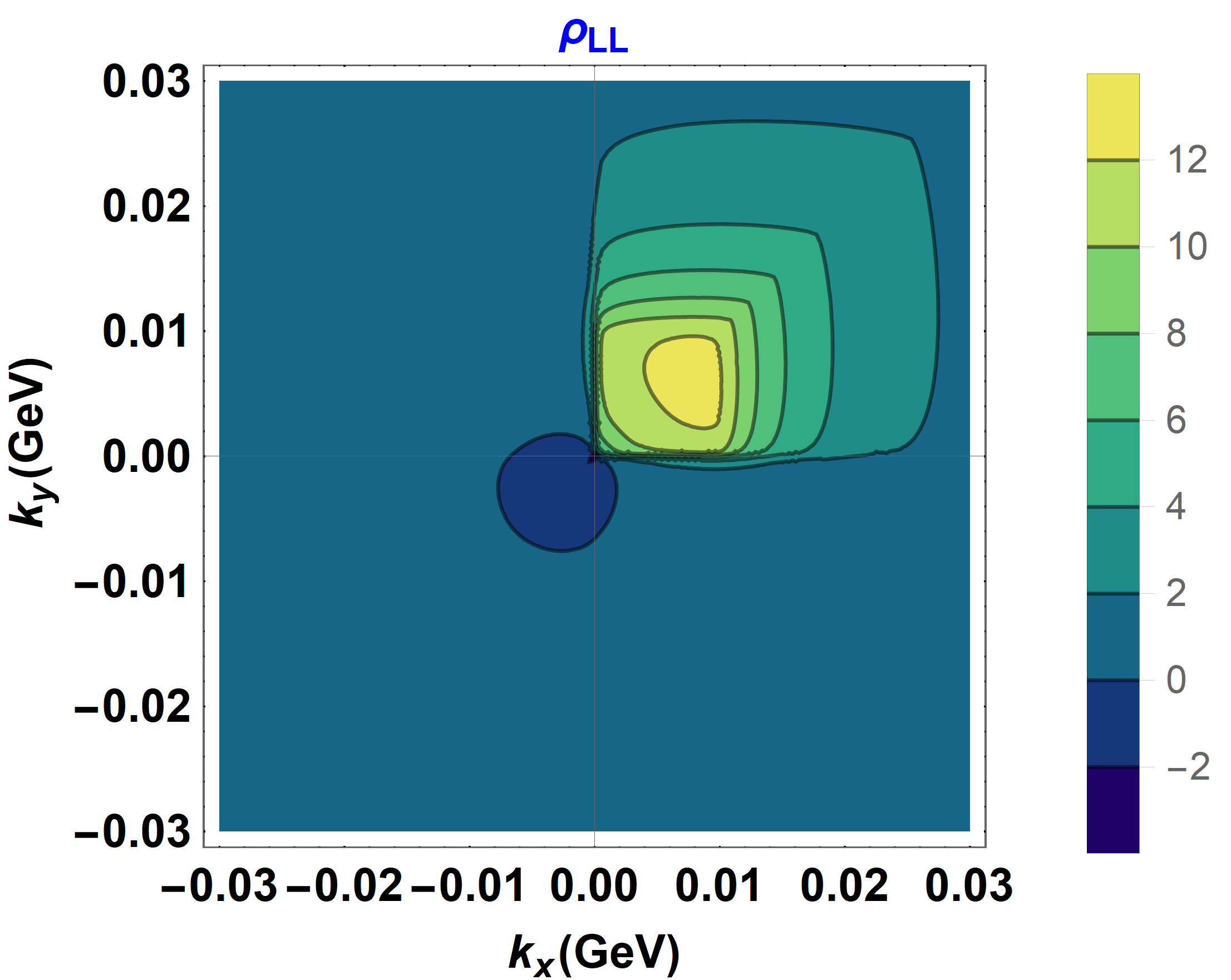} \\
\small{(g)}\includegraphics[width=5cm,height=4cm,clip]{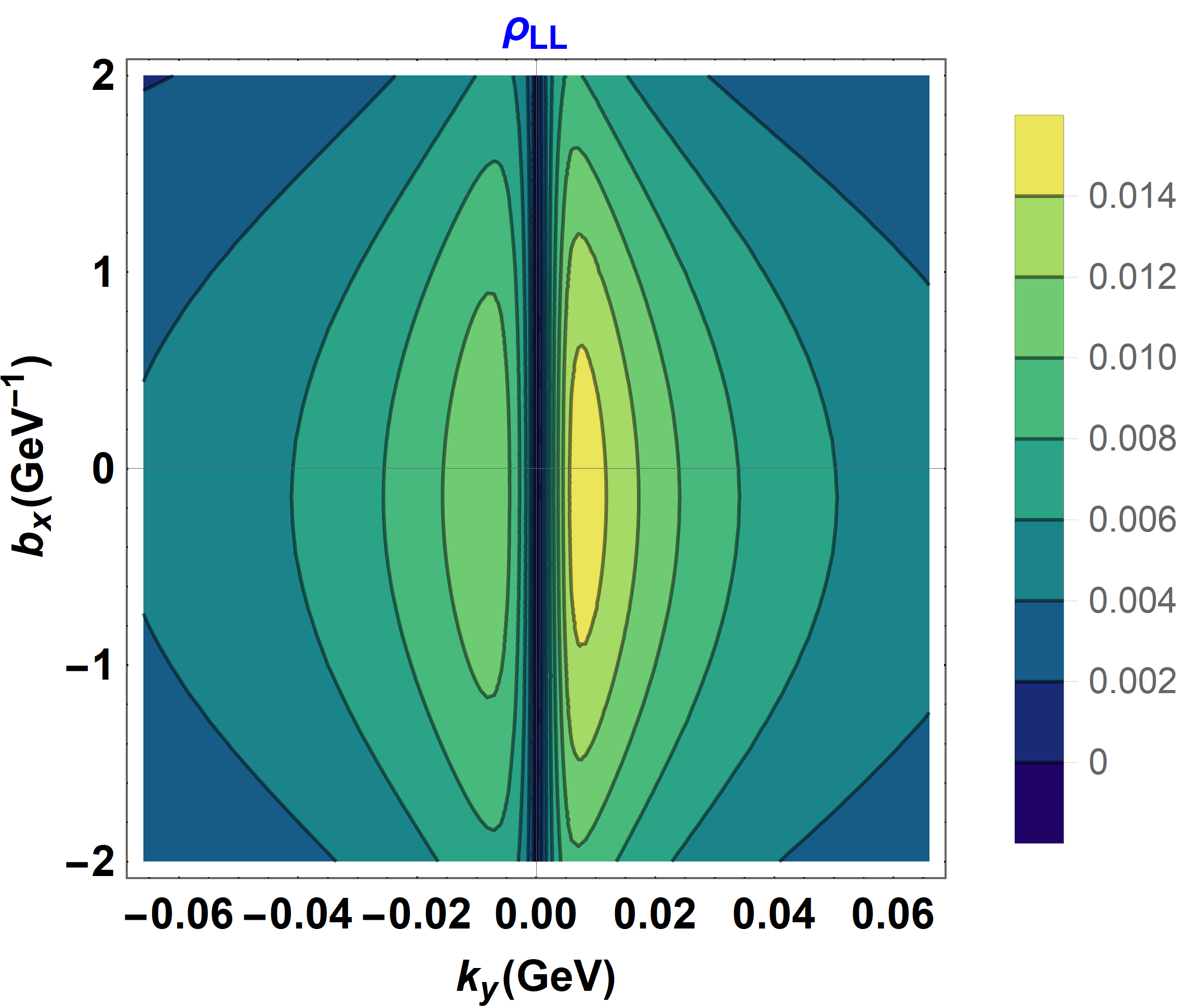} 
\small{(h)}\includegraphics[width=5cm,height=4cm,clip]{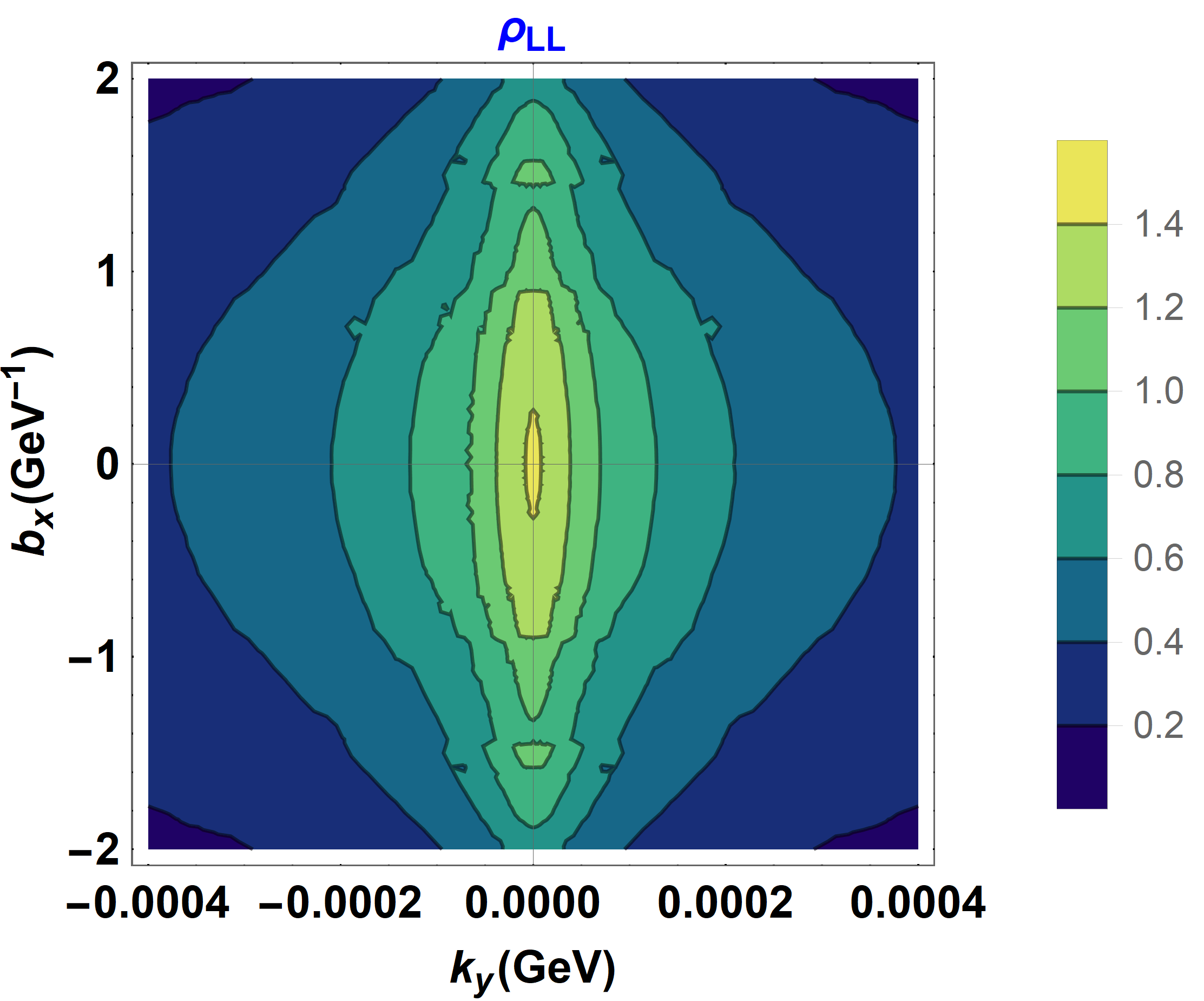}
\small{(i)}\includegraphics[width=5cm,height=4cm,clip]{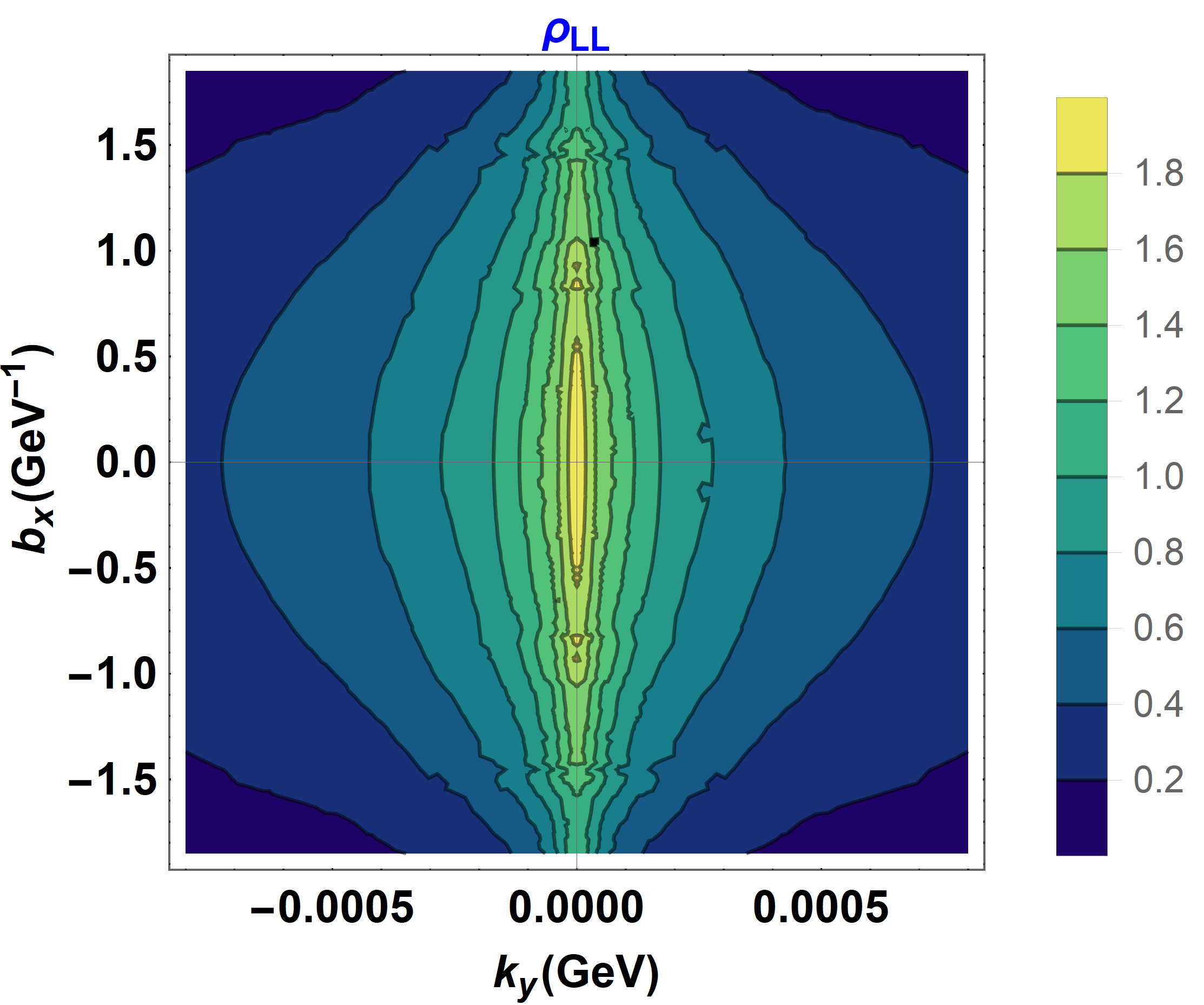} \\
\end{minipage}
\caption{\label{rho_LL}The quark Wigner distribution $\rho_{LL}$ are displayed in the transverse impact parameter plane, the transverse momentum plane, and the mixed plane. The left, middle, and right panels present the results for $\xi = 0$, $\xi = 0.25$, and $\xi = 0.5$, respectively. }
\end{figure}

Figure~\ref{rho_LL} presents the Wigner distribution $\rho_{LL}$ corresponding to a longitudinally polarized quark in a longitudinally polarized target. %shown in three different kinematic projections: the transverse impact parameter space $(b_x, b_y)$, the transverse momentum space $(k_x, k_y)$, and the mixed space $(b_x, k_y)$, for three values of skewness $\xi = 0$, $0.25$, and $0.5$.
In the impact parameter space (Figs.~\ref{rho_LL}(a)-(c)), the distribution exhibits a clear quadrupole structure, characterized by four distinct lobes with alternating signs. This pattern reflects spin-orbit correlations that arise from the interference between LFWF components differing by two units of orbital angular momentum ($\Delta L_z = 2$). The distributions are symmetric under parity in the transverse plane and exhibit smooth localization around the origin. As the skewness increases from $\xi = 0$ to $\xi = 0.5$, the spatial distribution becomes increasingly compressed in the transverse direction, indicating that larger longitudinal momentum transfer results in reduced transverse spatial overlap between the initial and final LFWFs.

The momentum-space distributions (Figs.~\ref{rho_LL}(d)-(f)) show significant evolution with increasing skewness. At $\xi = 0$, the distribution is nearly circularly symmetric, peaked at the origin, and exhibits no directional preference. However, at nonzero skewness, the distribution becomes deformed, developing a dipole-like feature along the $k_y$ direction. This asymmetry, which results in the appearance of negative regions in the distribution, is a clear signature of quantum interference effects between LFWF components carrying different orbital angular momentum. These effects become more pronounced at larger $\xi$, where the positive lobe shifts upward in $k_y$, while a negative lobe emerges in the lower hemisphere. Such patterns are indicative of spin–momentum correlations sensitive to the off-forward kinematics encoded by the skewness parameter.

In the mixed phase space (Figs.~\ref{rho_LL}(g)-(i)), which shows the distribution in $(b_x, k_y)$, the structure is elongated along the $b_x$-axis and concentrated around $b_x = 0$. At $\xi = 0$, the distribution displays a ridge-like shape that is symmetric with respect to both axes. This configuration suggests strong correlations between the quark's transverse position and transverse momentum, modulated by the longitudinal spin alignment. As skewness increases, the distribution becomes increasingly localized and sharply peaked around the origin, further supporting the interpretation that large longitudinal momentum transfer restricts the available phase space for coherent quark configurations.
\subsection{Unpolarized quark in transversely polarized target}
\begin{figure}[!htp]
\begin{minipage}[c]{1\textwidth}
\small{(a)}\includegraphics[width=5cm,height=4cm,clip]{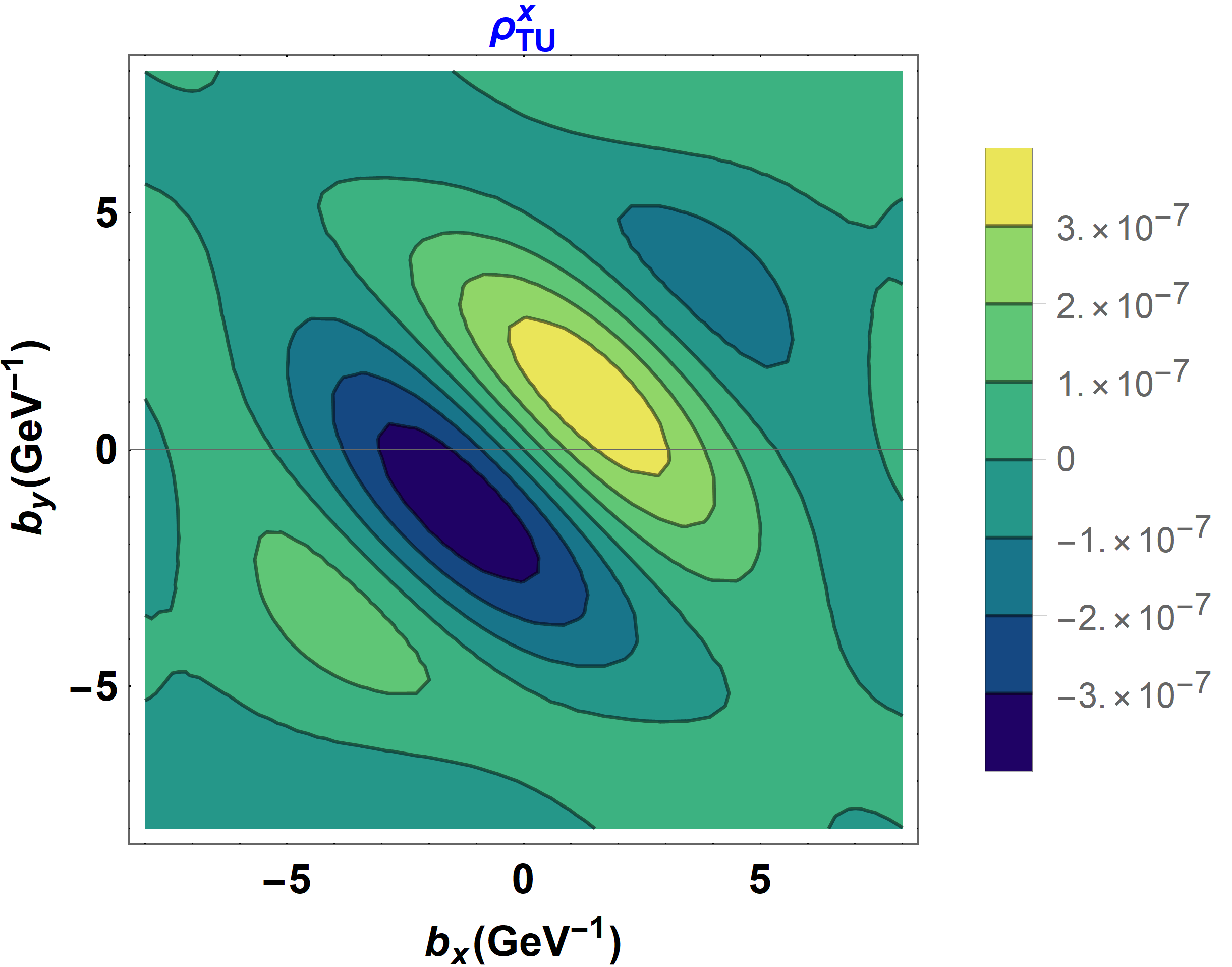}
\small{(b)}\includegraphics[width=5cm,height=4cm,clip]{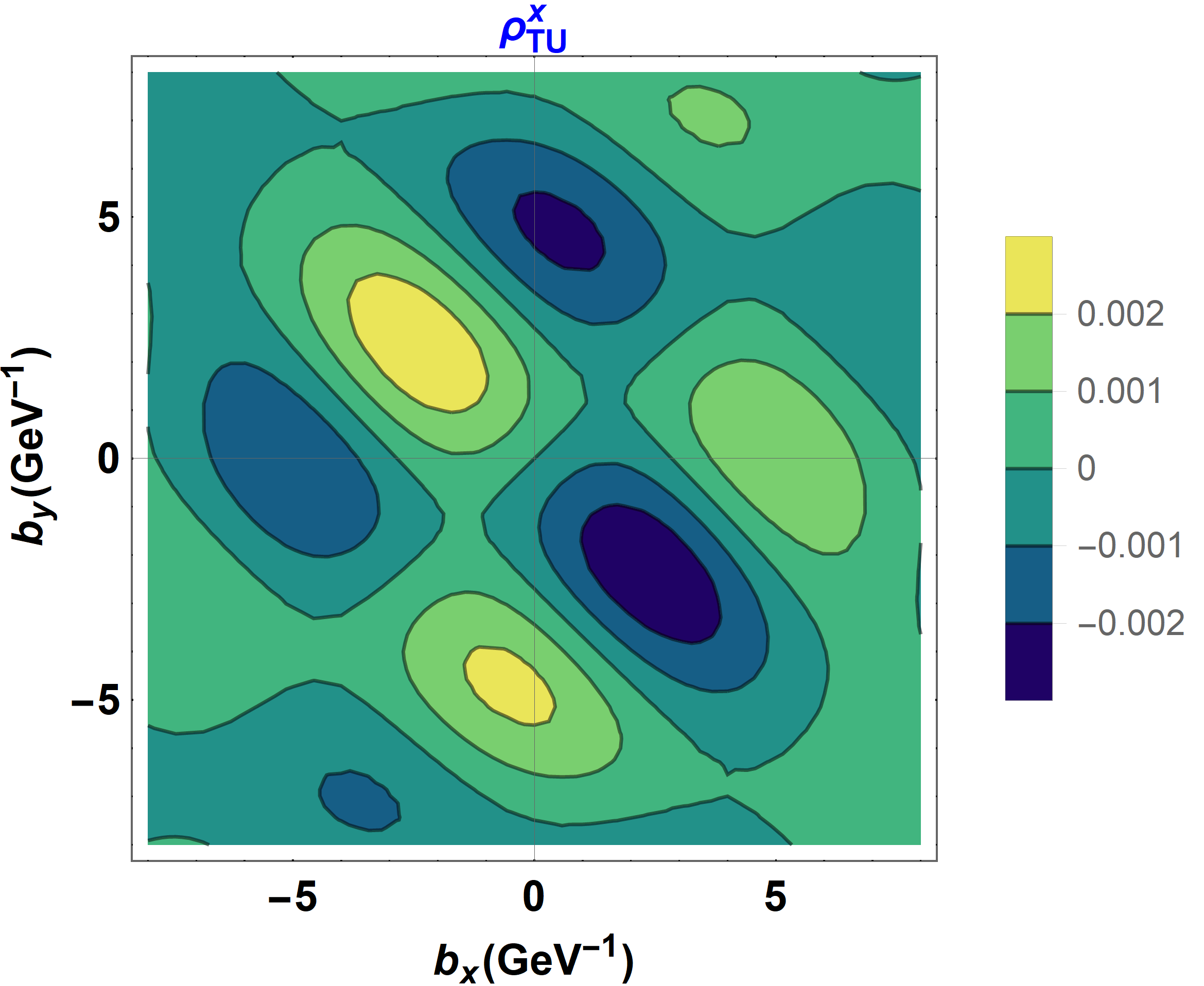} 
\small{(c)}\includegraphics[width=5cm,height=4cm,clip]{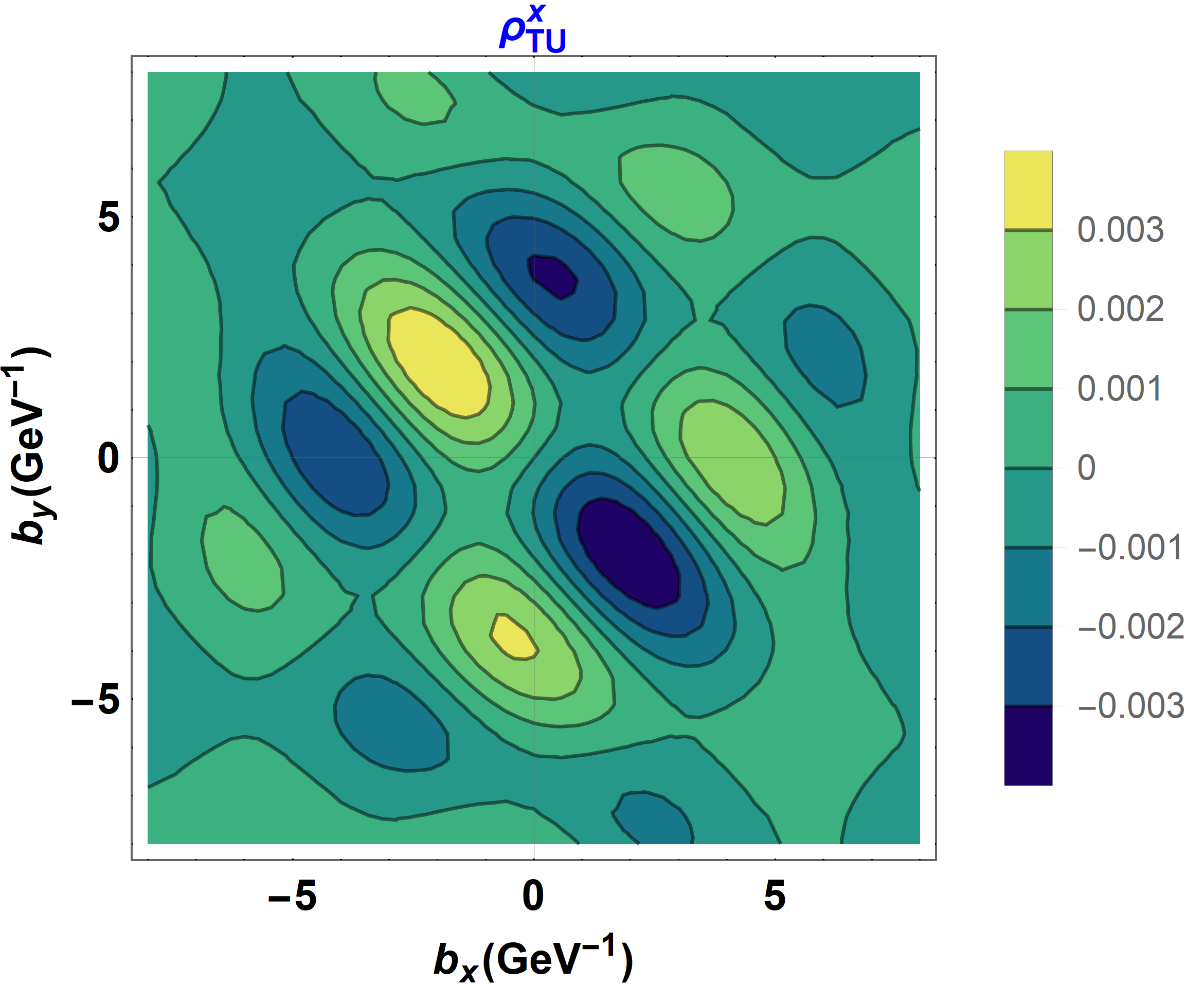}\\
\small{(d)}\includegraphics[width=5cm,height=4cm,clip]{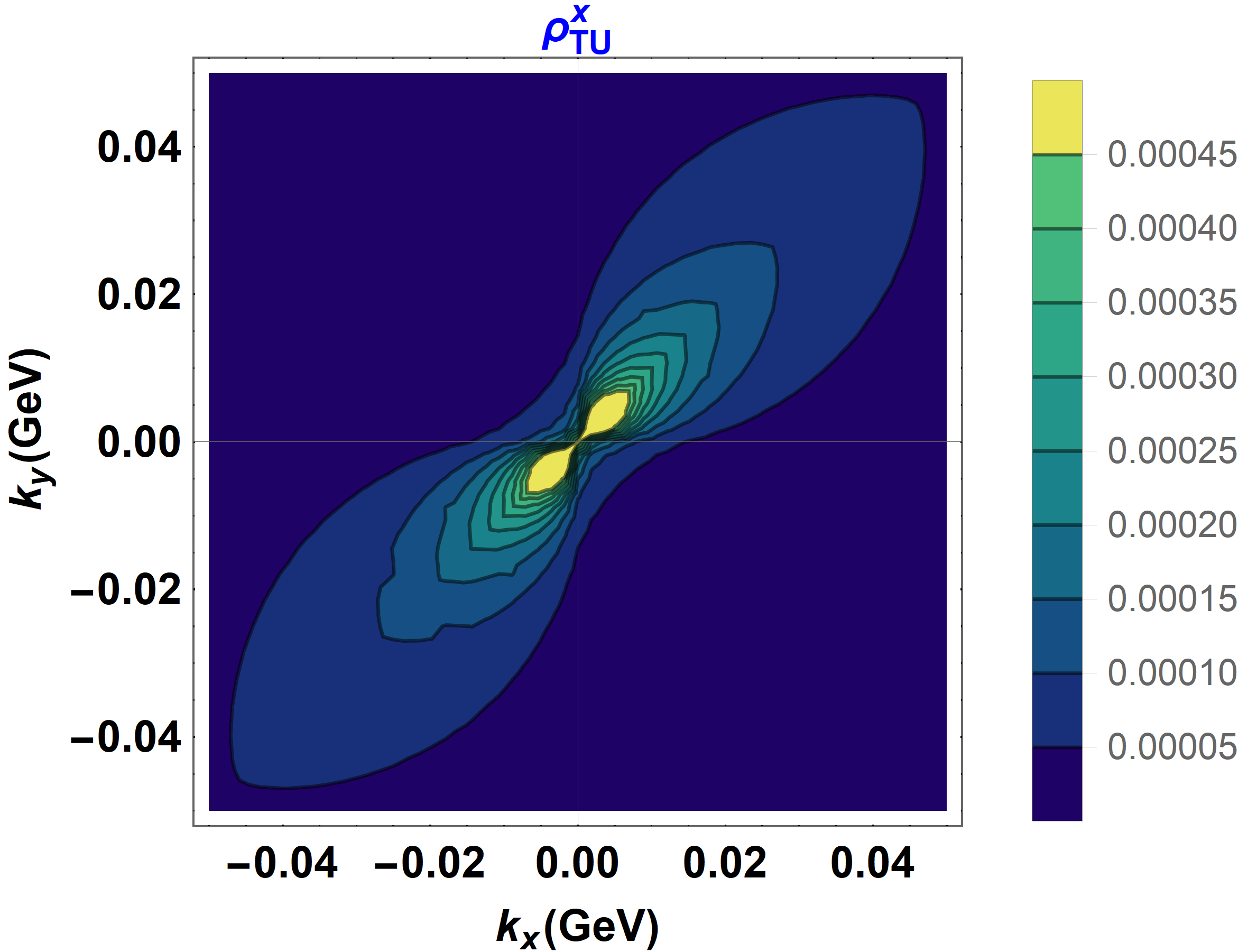} 
\small{(e)}\includegraphics[width=5cm,height=4cm,clip]{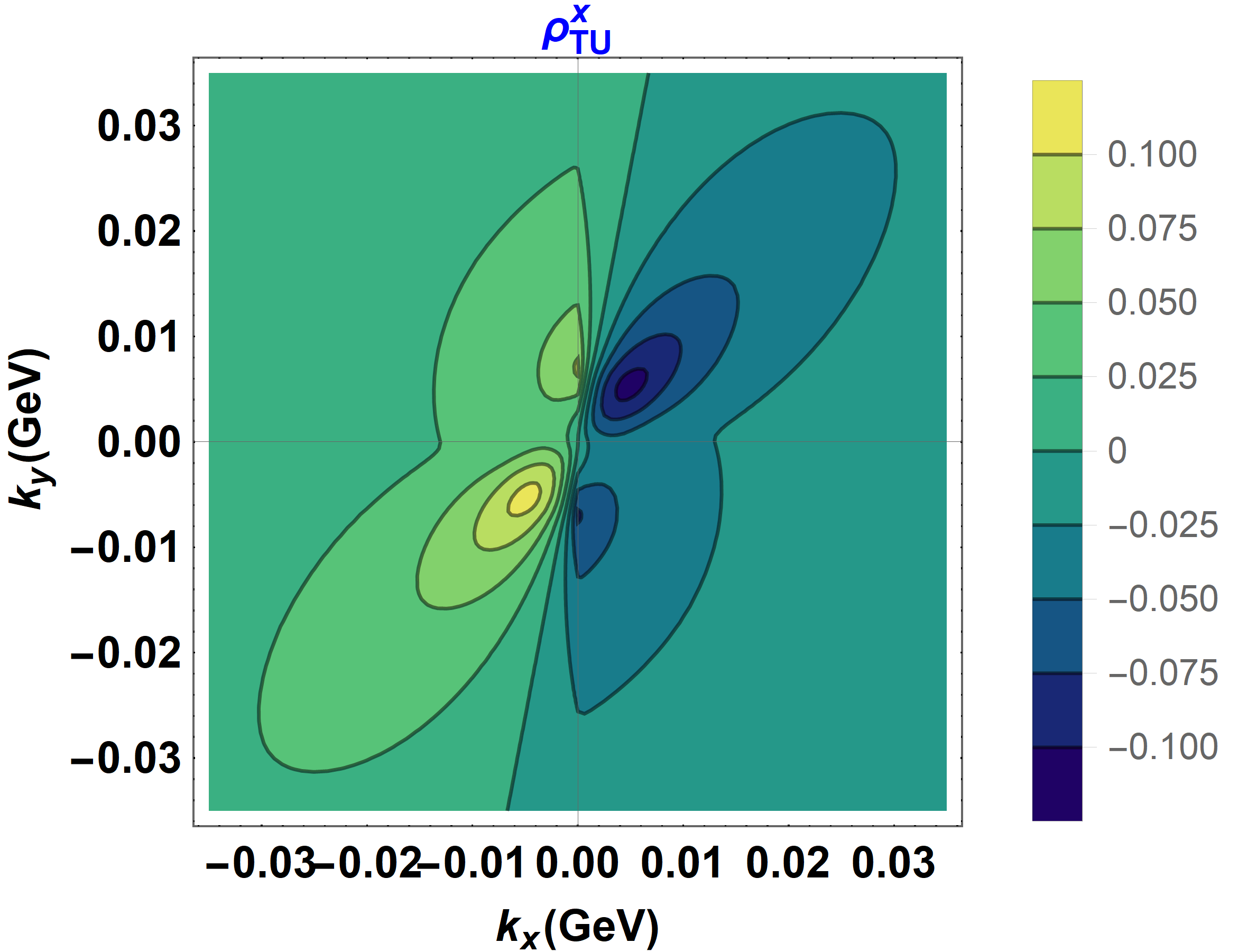}
\small{(f)}\includegraphics[width=5cm,height=4cm,clip]{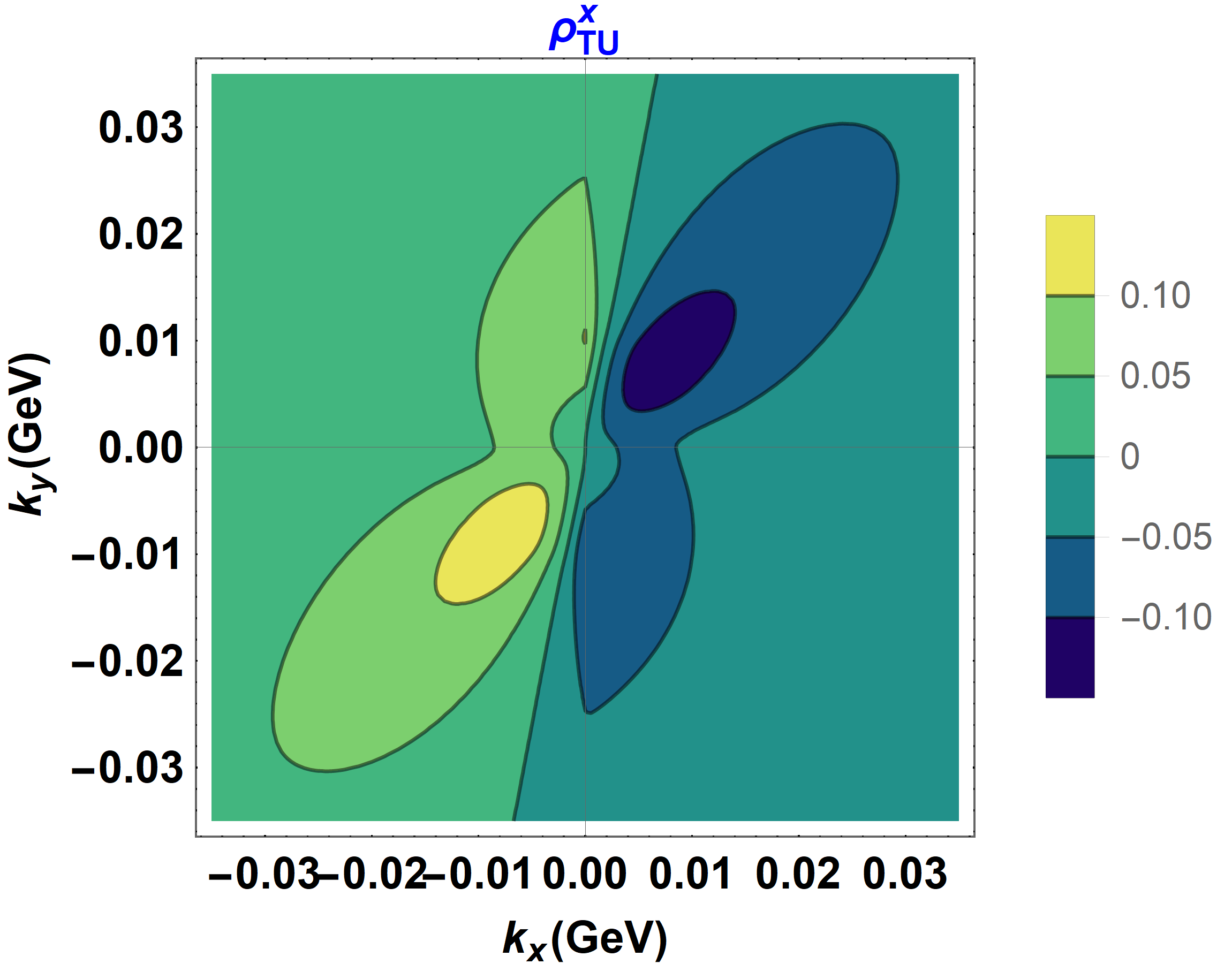} \\
\small{(g)}\includegraphics[width=5cm,height=4cm,clip]{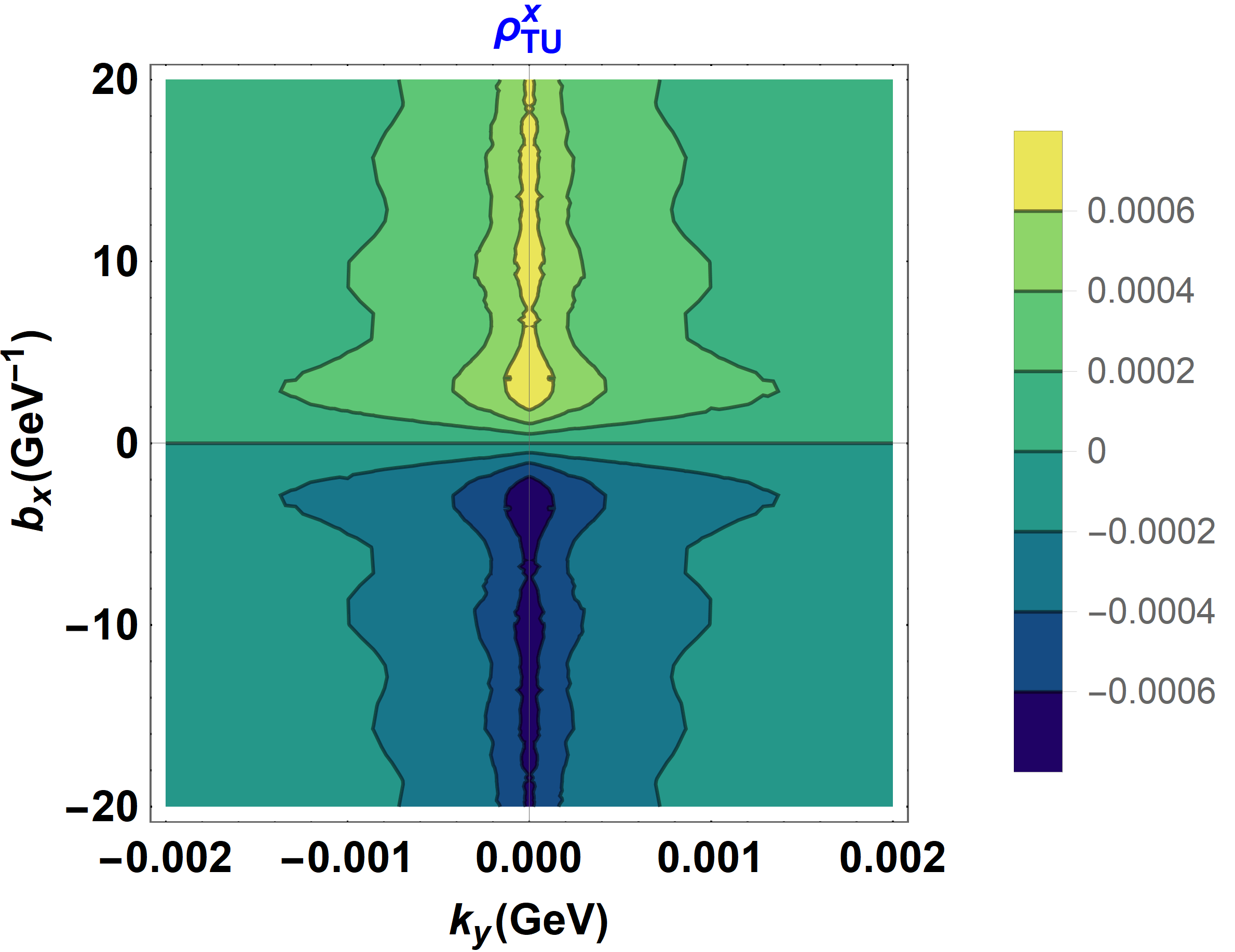} 
\small{(h)}\includegraphics[width=5cm,height=4cm,clip]{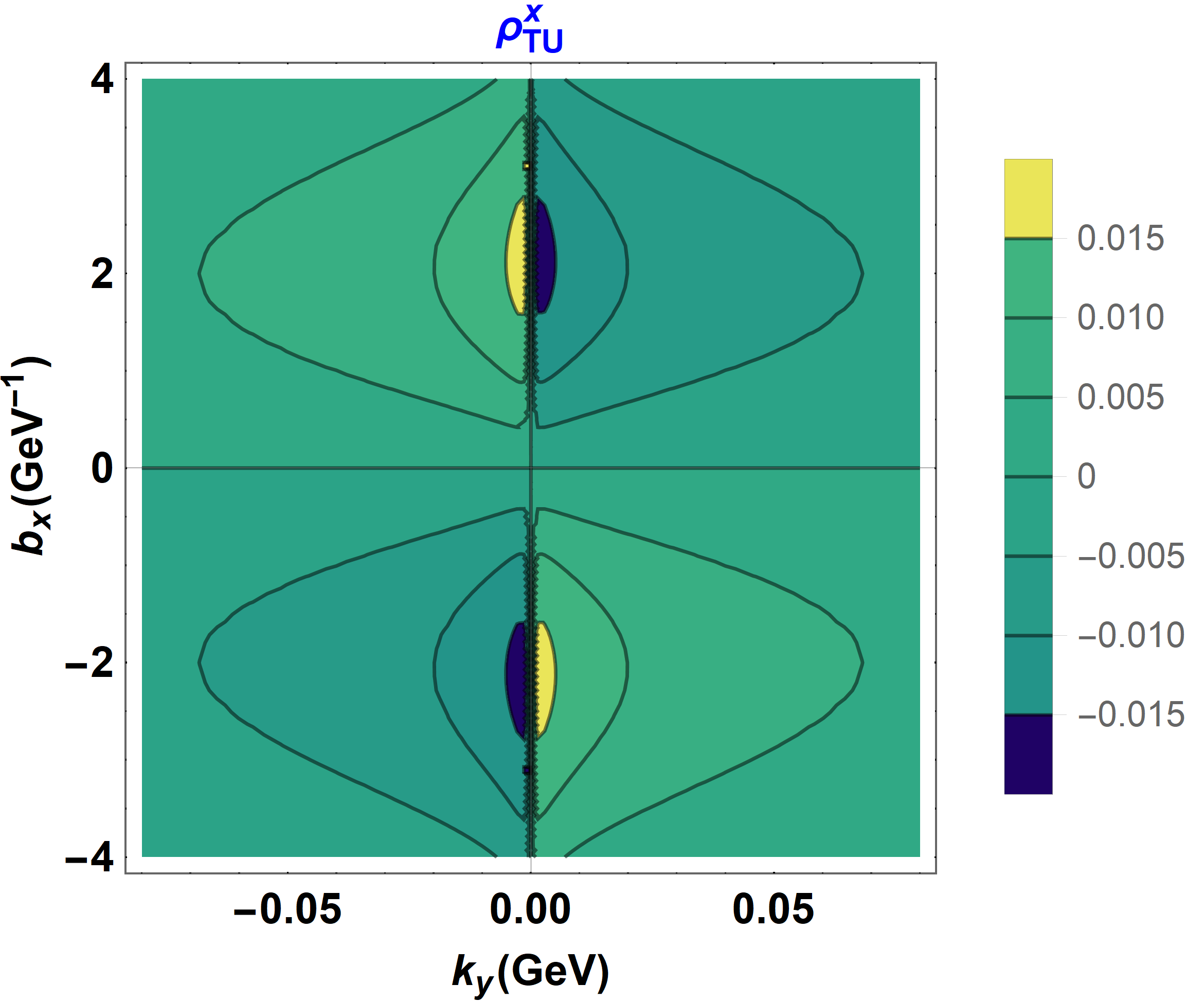}
\small{(i)}\includegraphics[width=5cm,height=4cm,clip]{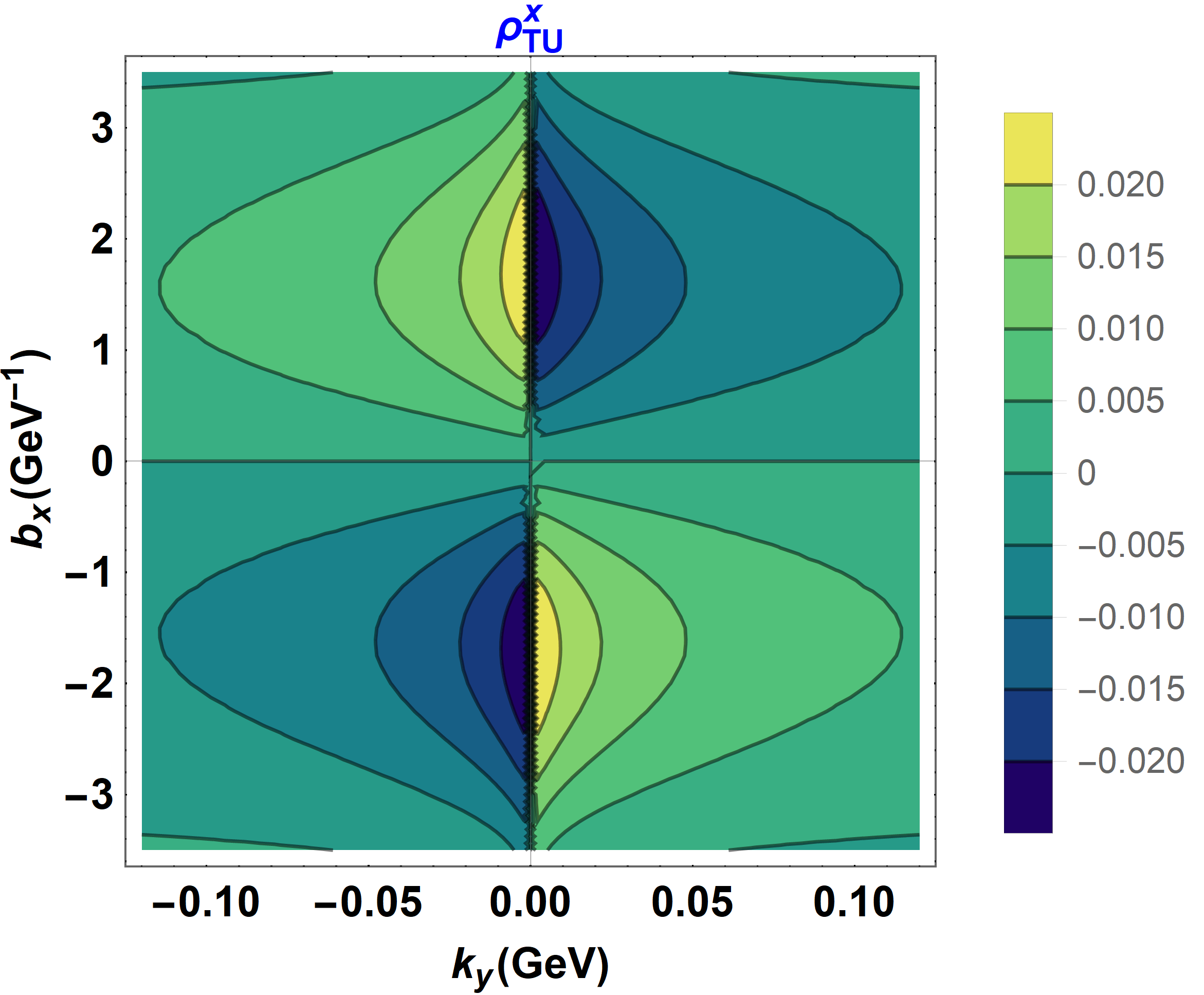} \\
\end{minipage}
\caption{\label{rho_TU}The quark Wigner distribution $\rho_{TU}^x$ are shown in the transverse impact parameter plane, the transverse momentum plane, and the mixed plane. The left, middle, and right panels present the results for $\xi = 0$, $\xi = 0.25$, and $\xi = 0.5$, respectively. }
\end{figure}

Figure~\ref{rho_TU} illustrates the Wigner distribution $\rho_{TU}^x$, which describes an unpolarized quark in the target that is transversely polarized along the $\hat{x}$ direction. %The distributions are presented in the transverse impact parameter plane $(b_x, b_y)$, the transverse momentum plane $(k_x, k_y)$, and the mixed plane $(b_y, k_y)$ for three representative values of skewness: $\xi = 0$, $0.25$, and $0.5$.
In the impact parameter space (Figs.~\ref{rho_TU}(a)-(c)), the Wigner distribution exhibits distinct azimuthal asymmetries, which become more intricate as skewness increases. For $\xi = 0$, the distribution is relatively smooth and localized, with a single central peak. As $\xi$ grows, the pattern evolves into a clear multipole structure, especially for $\xi = 0.5$, where a rich interference pattern appears, featuring multiple alternating positive and negative lobes. These patterns signal interference between light-front wave functions with different orbital angular momentum components and reflect strong spin–orbit correlations driven by the transverse polarization of the target.

The transverse momentum space distributions (Figs.~\ref{rho_TU}(d)-(f)) further emphasize the role of spin-momentum correlations. At zero skewness, the distribution in momentum space displays a dipolar structure oriented diagonally in the $k_x$-$k_y$ plane, indicating that the transverse polarization of the target induces a preferred direction in the quark’s transverse momentum. As $\xi$ increases, the dipolar structure transforms into a more complex pattern with higher-order multipole features, including prominent quadrupole-like shapes. The emergence of negative regions, particularly for $\xi = 0.25$ and $\xi = 0.5$, is indicative of quantum interference effects and underscores the non-classical nature of Wigner distributions. These negative regions are sensitive to the relative phase structure of the light-front wave functions and are often connected to T-odd effects such as the Boer–Mulders or Sivers-type behavior.

In the mixed phase space (Figs.~\ref{rho_TU}(g)-(i)), which shows the distribution in $(b_y, k_y)$, the structure is strongly asymmetric along the $k_y$ direction and becomes increasingly localized in $b_y$ with larger skewness. The ridge-like structure observed at $\xi = 0$ sharpens as $\xi$ increases, while the transverse width narrows. Notably, the antisymmetric pattern about $b_y = 0$ persists throughout all values of $\xi$, highlighting a consistent spin–momentum correlation that is sensitive to the target’s transverse spin orientation. This antisymmetry is characteristic of the torque-like effect arising from the transverse polarization of the target, consistent with expectations from generalized parton correlators.

\subsection{Longitudinally Polarized quark in transversely polarized target}
%\section*{Analysis of the Wigner Distribution \texorpdfstring{$\rho_{TL}^x$}{rho\_TL\^x}}

\begin{figure}[!htp]
\begin{minipage}[c]{1\textwidth}
\small{(a)}\includegraphics[width=5cm,height=4cm,clip]{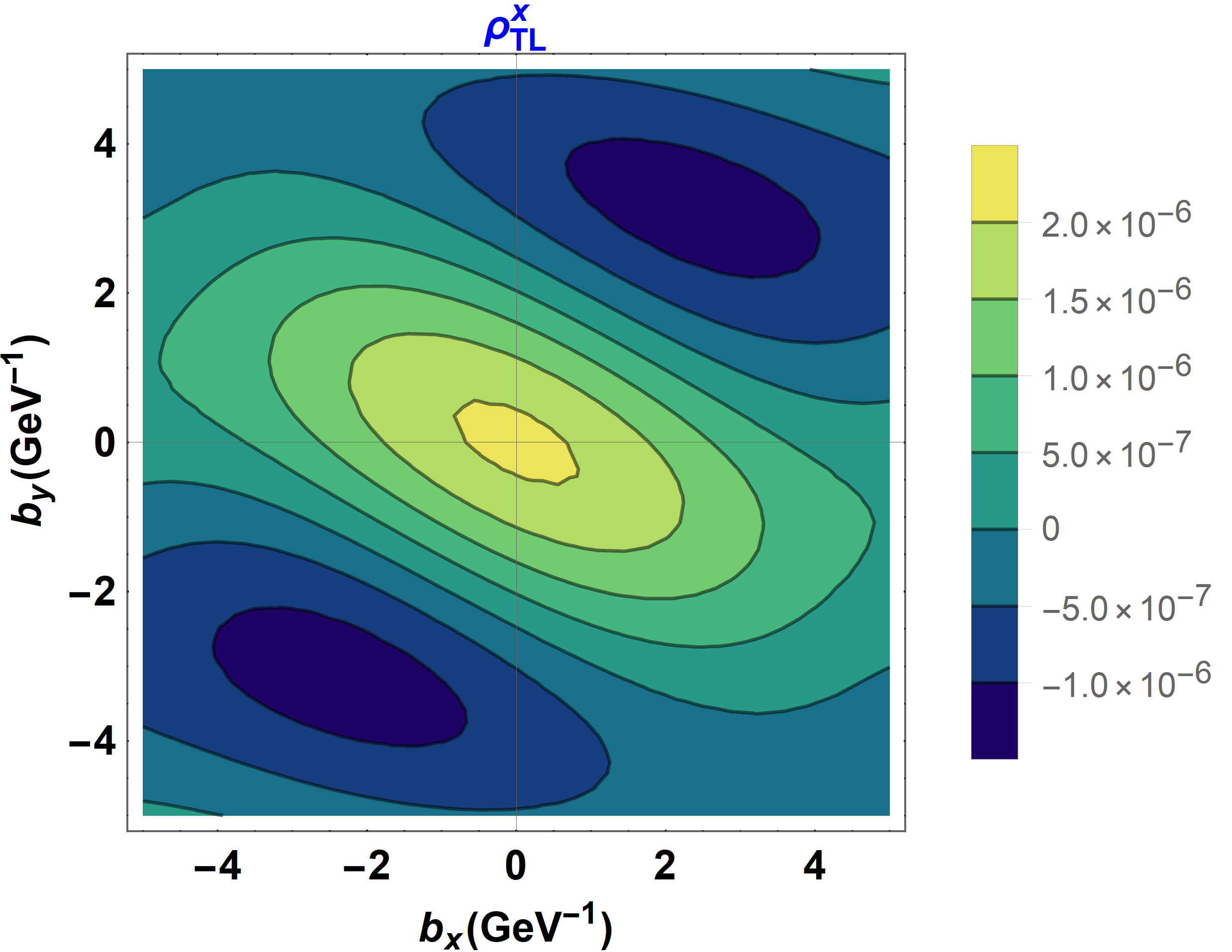}
\small{(b)}\includegraphics[width=5cm,height=4cm,clip]{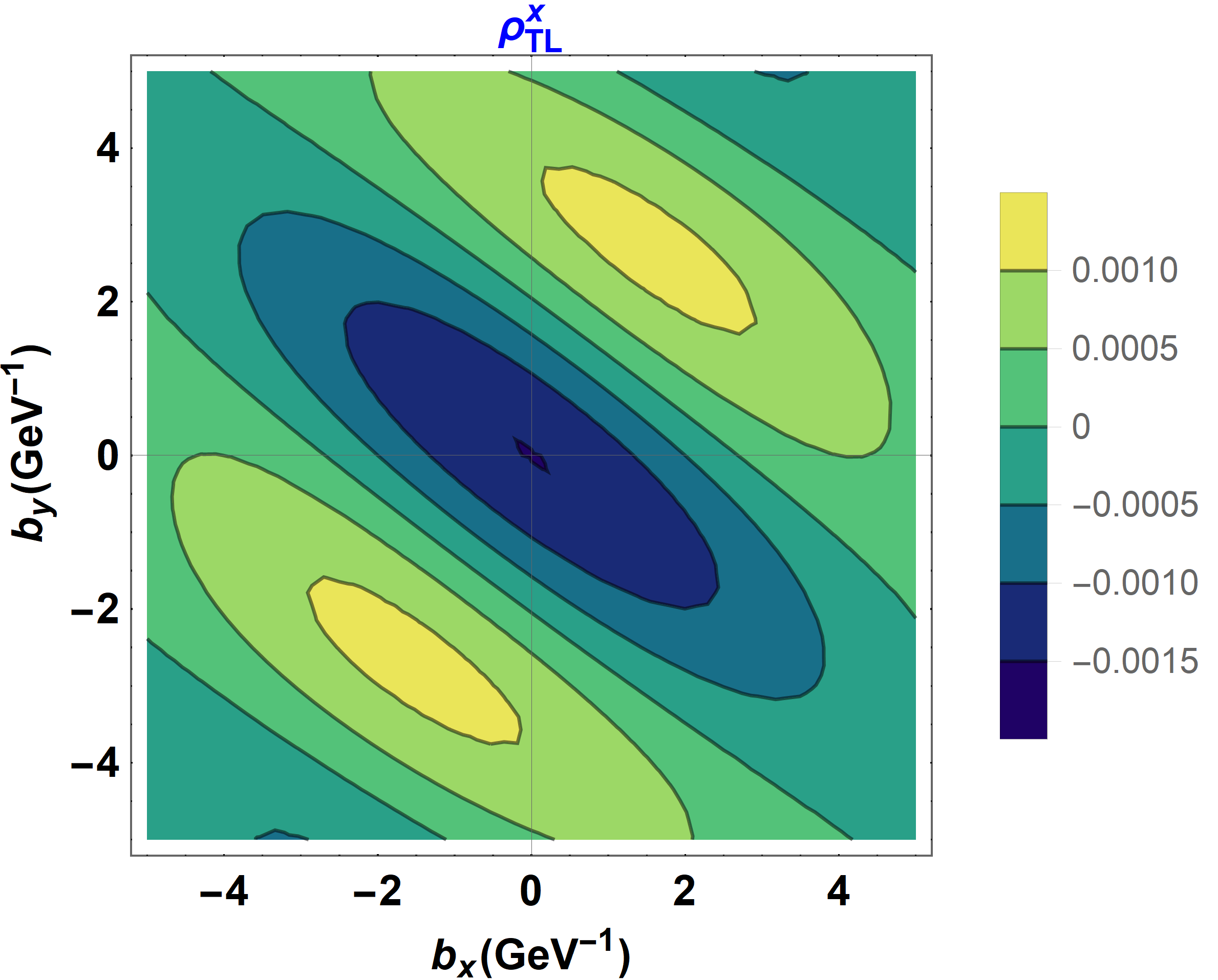} 
\small{(c)}\includegraphics[width=5cm,height=4cm,clip]{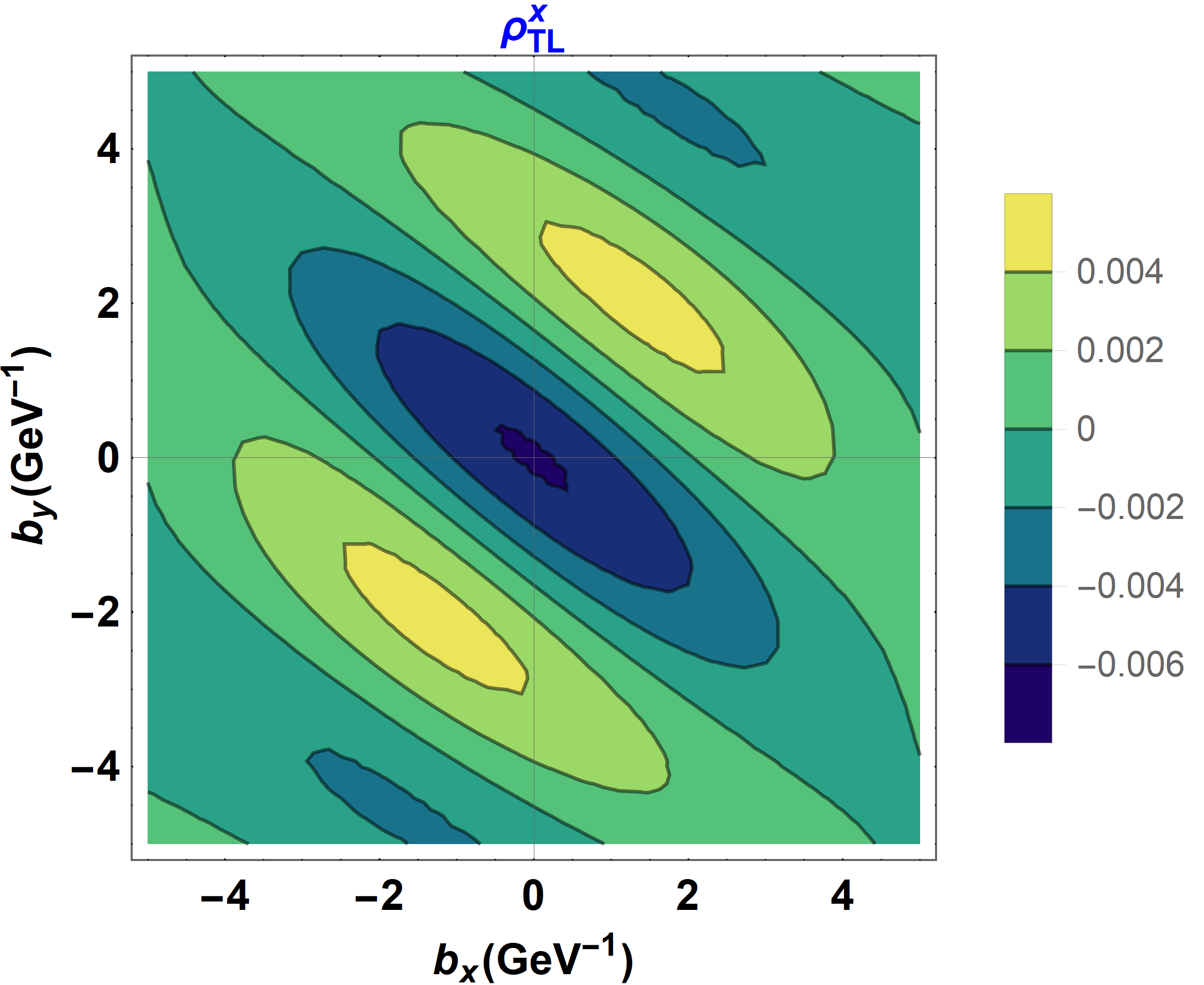}\\
\small{(d)}\includegraphics[width=5cm,height=4cm,clip]{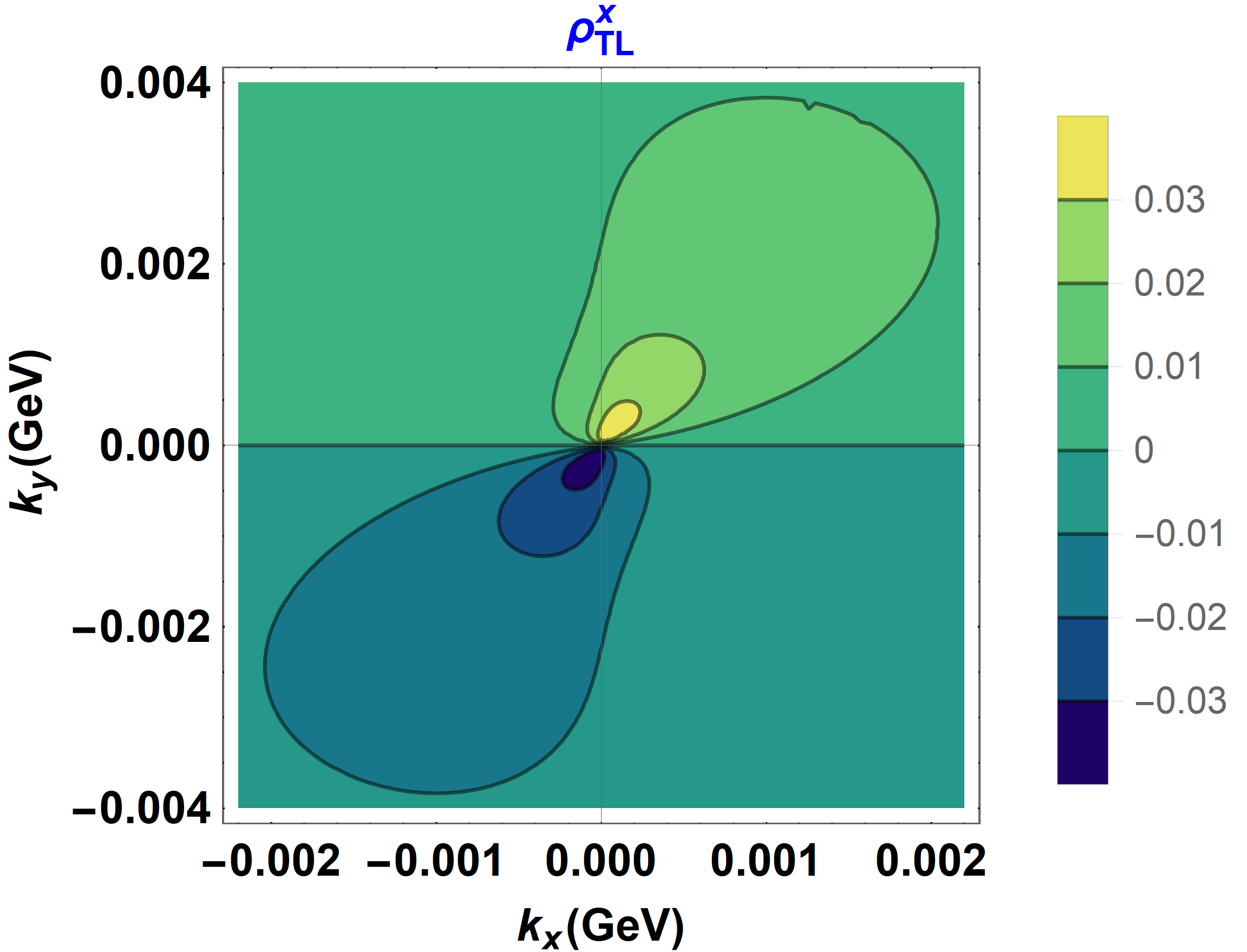} 
\small{(e)}\includegraphics[width=5cm,height=4cm,clip]{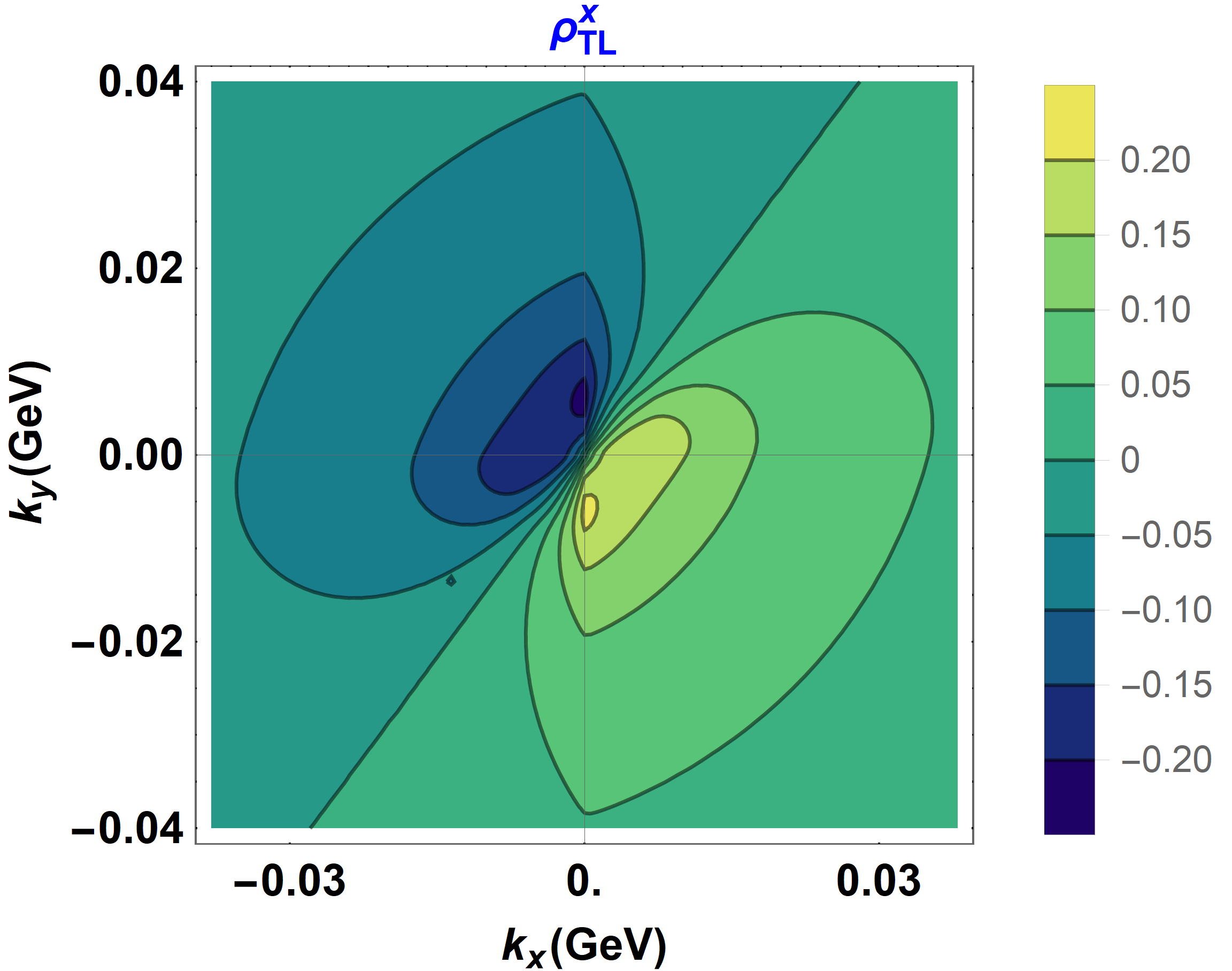}
\small{(f)}\includegraphics[width=5cm,height=4cm,clip]{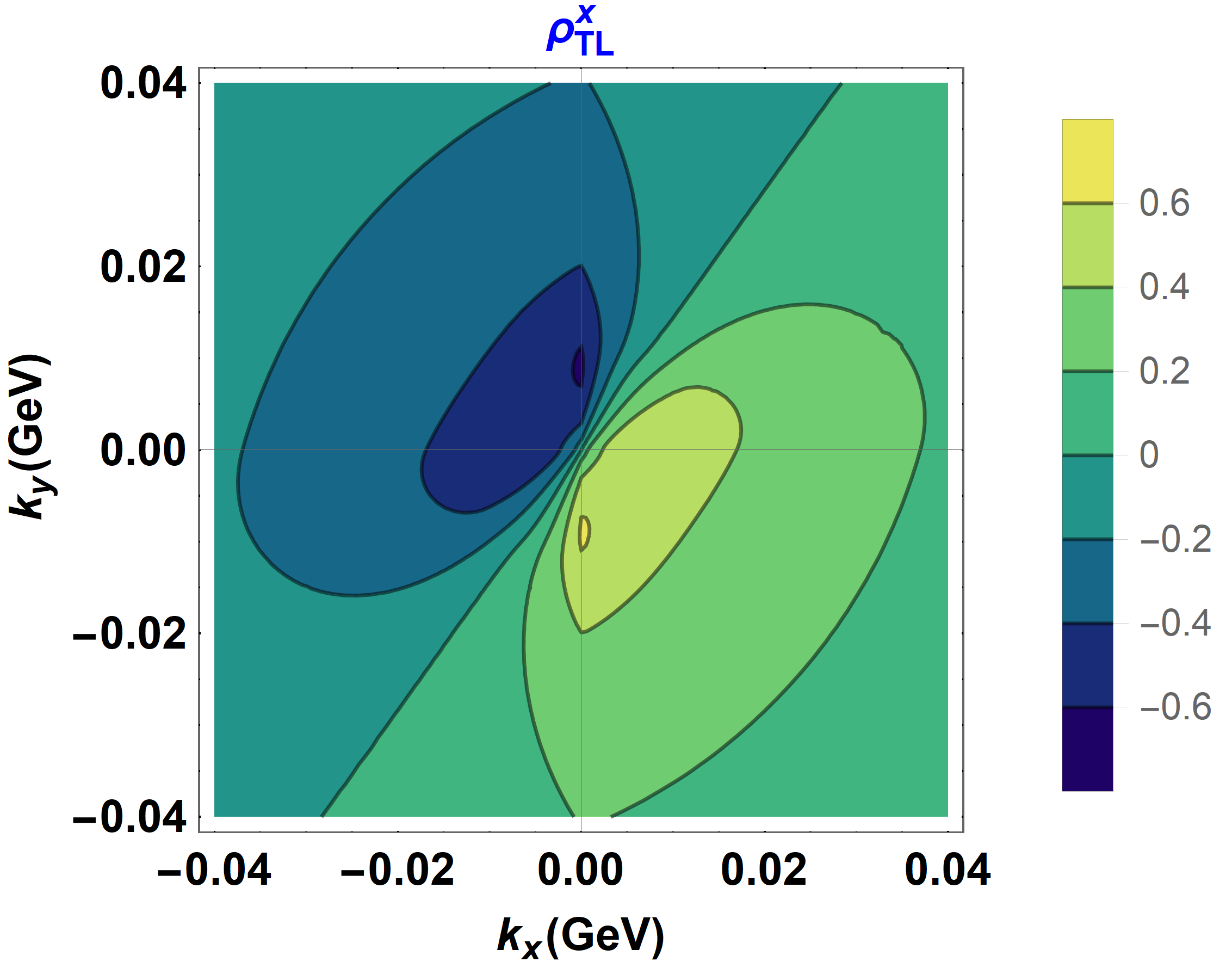} \\
\small{(g)}\includegraphics[width=5cm,height=4cm,clip]{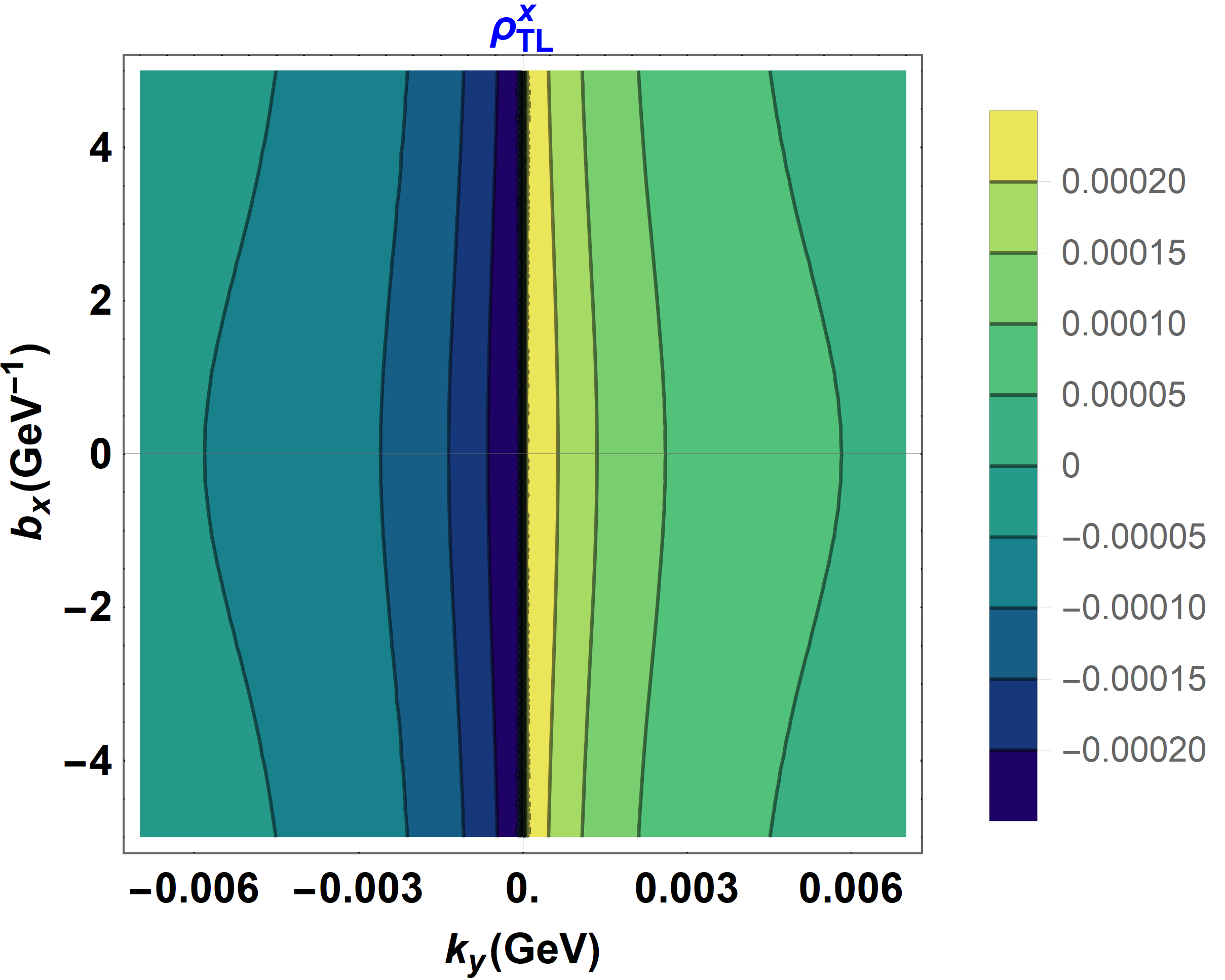} 
\small{(h)}\includegraphics[width=5cm,height=4cm,clip]{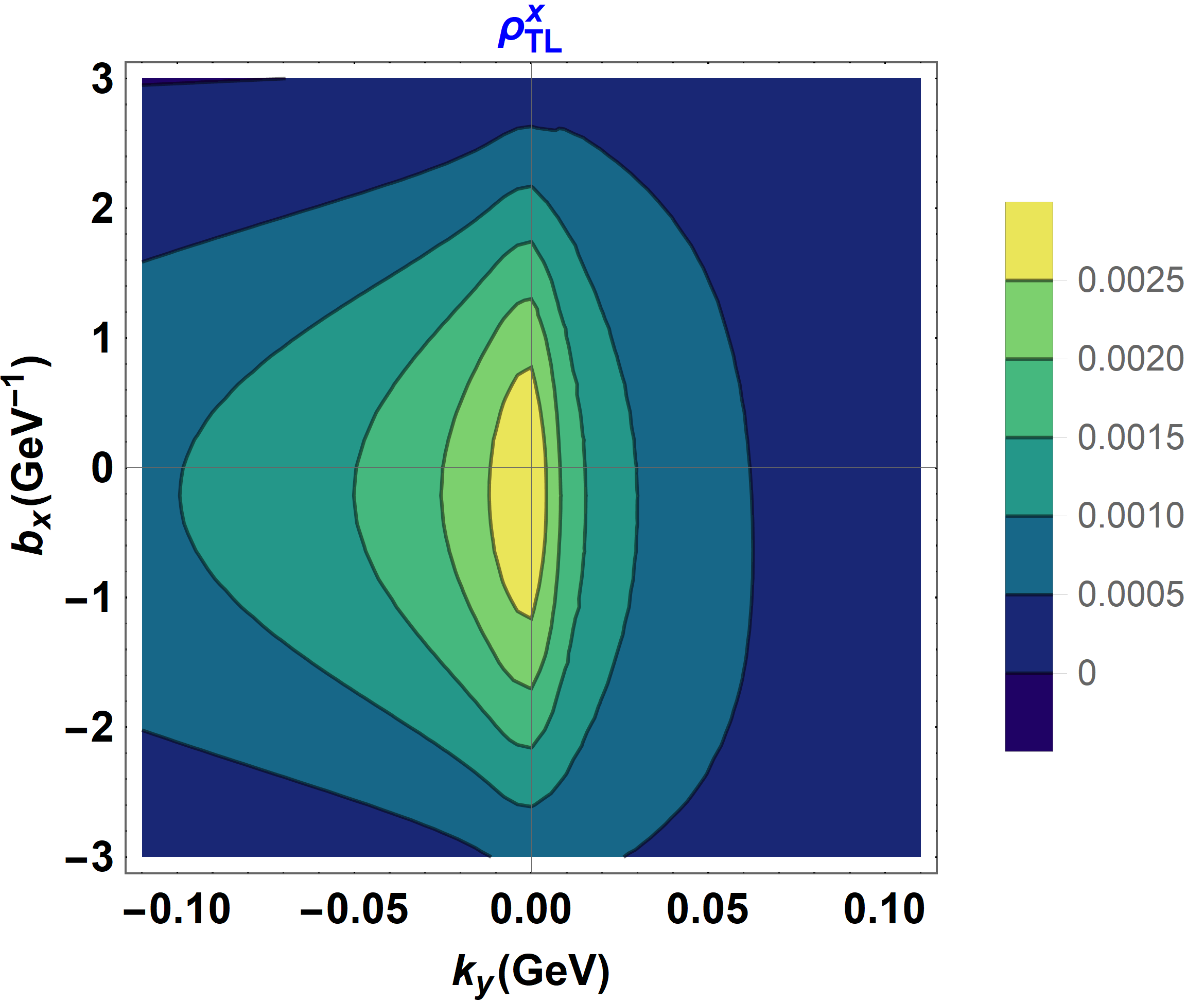}
\small{(i)}\includegraphics[width=5cm,height=4cm,clip]{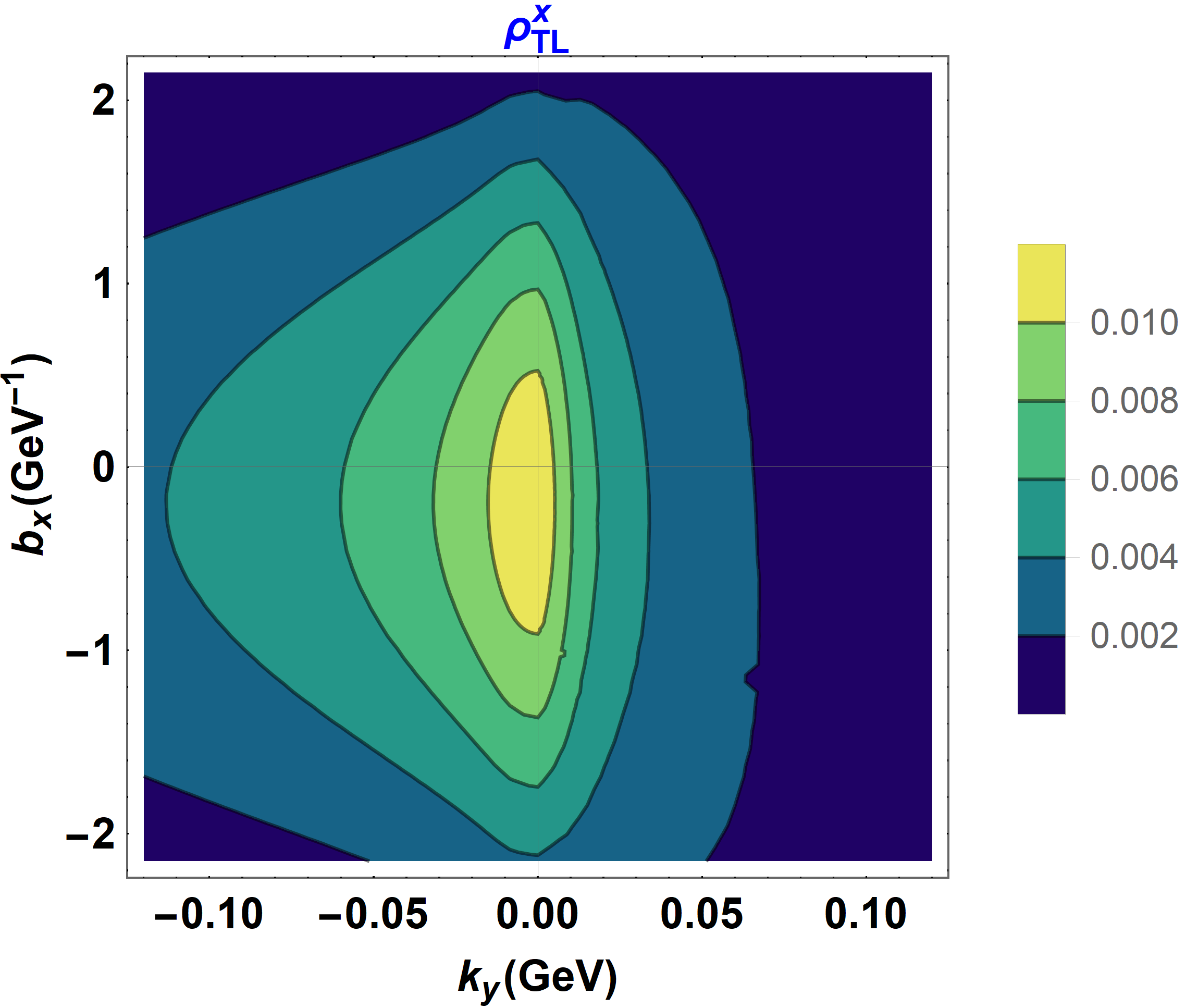} \\
\end{minipage}
\caption{\label{rho_TL}The quark Wigner distribution $\rho_{TL}^x$ in the transverse impact parameter plane, the transverse momentum plane, and the mixed plane. The left, middle, and right panels present the results for $\xi = 0$, $\xi = 0.25$, and $\xi = 0.5$, respectively. }
\end{figure}

Figure~\ref{rho_TL} displays the Wigner distribution $\rho_{TL}^x$, which characterizes the phase-space correlation between the longitudinally polarized quark and the transversely polarized target (polarized along the $\hat{x}$-direction). %The distributions are visualized in three different projections: the transverse impact parameter space $(b_x, b_y)$, the transverse momentum space $(k_x, k_y)$, and the mixed space $(b_y, k_y)$, for skewness values $\xi = 0$, $0.25$, and $0.5$.
In the transverse impact parameter space (Figs.~\ref{rho_TL}(a)–(c)), the distributions exhibit an overall dipole-like structure that becomes increasingly pronounced with nonzero skewness. At $\xi = 0$, the distribution is nearly symmetric, with a weak modulation in $b_x$. However, as $\xi$ increases, a stronger left–right asymmetry emerges, signaling the onset of spin–orbit interference. The alternating positive and negative lobes suggest the presence of interference between light-front wave functions differing by one unit of orbital angular momentum, a hallmark of transverse spin correlations. These structures become sharper and more spatially compressed as $\xi$ grows, indicating enhanced sensitivity to the off-forward nature of the Wigner function.

In the transverse momentum space (Figs.~\ref{rho_TL}(d)–(f)), the effect of the target’s transverse polarization on the quark’s longitudinal spin becomes evident. At $\xi = 0$, the distribution shows a weak dipolar distortion, but as $\xi$ increases, a clear asymmetry develops in the $k_x$ direction, leading to a prominent skewed dipole structure. Notably, for $\xi = 0.5$, the distribution acquires significant negative regions in the upper-left quadrant and positive lobes in the lower-right quadrant. This reflects nontrivial spin–momentum correlations and interference between wave function components with different orbital angular momentum projections. The increasing complexity of the structure with $\xi$ again underscores the role of longitudinal momentum transfer in modifying the internal dynamics of the nucleon.

The mixed space distributions (Figs.~\ref{rho_TL}(g)–(i)) reinforce these observations, revealing strongly asymmetric structures along the $b_y$ direction. For $\xi = 0$, the distribution is sharply peaked along $k_y = 0$, forming a narrow ridge consistent with coherent correlations between transverse position and momentum. As $\xi$ increases, this ridge broadens, and side lobes appear, leading to more intricate patterns with alternating signs. These features are indicative of a breakdown of coherence in the spin-dependent correlations, driven by skewness-induced phase shifts between the initial and final target states.

\section{Conclusion\label{sec:con}}
%\section{Summary and Conclusion}

In this work, we have presented a detailed investigation of quark Wigner distributions in a light-front dressed quark model, incorporating the effects of nonzero skewness. By using the analytical expression all leading-twist GTMDs, Wigner distributions were obtained for various polarization configurations: unpolarized, longitudinally polarized, and transversely polarized quark and the target. We have provided a multidimensional visualization of quark phase-space dynamics in transverse momentum, impact parameter, and mixed representations.

The inclusion of nonzero skewness distinguishes this analysis from earlier studies conducted in the forward limit. Skewness, defined as the longitudinal momentum asymmetry between the initial and final states, enables access to off-forward kinematic regimes and allows for a more complete exploration of quark-gluon correlations. This work thereby contributes to bridging the gap between two limiting frameworks: GPDs and TMDs, offering new perspectives on their unification through Wigner phase-space representations.

Our results reveal a range of novel and physically significant features. The unpolarized Wigner distribution $\rho_{UU}$ shows circular symmetry at $\xi = 0$, which becomes increasingly distorted as skewness grows, indicating the breaking of rotational and time-reversal symmetry in the transverse plane. The emergence of dipole and quadrupole structures in distributions involving polarized quarks or target reflects strong spin-orbit correlations and the presence of quantum interference between light-front wave function components with different orbital angular momenta. These multipole patterns become more pronounced with increasing $\xi$, accompanied by sign-changing regions and nodal structures, signaling reduced coherence in the spin-dependent phase-space correlations.

The transverse momentum space representations of several distributions exhibit asymmetric features consistent with T-odd effects, analogous to the Sivers and Boer-Mulders functions in the TMD framework. These effects are enhanced at larger skewness, highlighting the sensitivity of Wigner distributions to the dynamics of both longitudinal and transverse momentum transfer. Furthermore, the mixed-space representations reveal how transverse position and momentum are jointly modulated by polarization and skewness. The observed ridge-like and lobed structures undergo significant deformation with $\xi$, providing additional insight into the interplay between spatial and momentum distributions within the target state.

Overall, our findings extend previous results obtained in the forward limit and underscore the rich internal structure of the target revealed through Wigner distributions. The inclusion of nonzero skewness plays a pivotal role in unveiling intricate correlations involving spin, orbital angular momentum, and spatial-momentum entanglement. These results demonstrate the versatility of Wigner distributions as powerful tools for hadron tomography and offer valuable theoretical input for future experimental programs and lattice QCD studies aimed at mapping the multidimensional structure of hadrons.

Future extensions of this work may incorporate higher Fock components and gluonic degrees of freedom, facilitating the study of gluon Wigner distributions as the function of skewness and deepening our understanding of nonperturbative QCD dynamics. The framework employed here thus lays the foundation for more sophisticated explorations of the spin and momentum structure of hadrons in off-forward regimes.

\appendix
\section{Analytical Expressions of GTMDs for Nonzero Skewness}

 The analytic expressions for the leading-twist generalized transverse momentum-dependent distributions (GTMDs) for quarks, including the effects of nonzero skewness parameter $\xi$. These expressions are derived within the framework of a light-front dressed quark model at leading order. The full details of the computation are provided in \cite{ojha2023quark}.

\begin{itemize}
    \item \textbf{Unpolarized quark:}
    \begin{align}
        F_{1,1} &= \frac{\alpha(1-\xi^2)}{2(1-x)^3} \Big[ 4(1+x^2-(3+x^2)\xi^2+2\xi^4)\vec{k}_\perp^{\,2} + 4m^2(1-x)^4 \nonumber \\
        & \quad + 4(1-x)\xi(1+x^2 - 2\xi^2) \vec{k}_\perp \cdot \vec{\Delta}_\perp - (1-x)^2(1+x^2 - 2\xi^2)\vec{\Delta}_\perp^{\,2} \Big], \\
        F_{1,2} &= -\beta \Big[ ((1+x)\xi(k_2 \Delta_1 - k_1 \Delta_2) - 2m^2 x(1-x)) \vec{\Delta}_\perp^{\,2} + 4m^2 \xi(1+x) \vec{k}_\perp \cdot \vec{\Delta}_\perp \Big], \\
        F_{1,3} &= \frac{\beta}{4(1-x)^2} \Big[ (k_1 \Delta_2 - k_2 \Delta_1)\big((1+x^2-2\xi^2)(4(1-\xi^2)\vec{k}_\perp^{\,2} + (1-x)^2 \vec{\Delta}_\perp^{\,2}) \nonumber \\
        & \quad - 8\xi(1-\xi^2)(1-x)\vec{k}_\perp \cdot \vec{\Delta}_\perp - 4m^2(1-x)^4\big) \nonumber \\
        & \quad + 8m^2(1-x)^2 \big(2(1+x)\xi \vec{k}_\perp - (1-x)x \vec{\Delta}_\perp \big) \cdot \vec{k}_\perp \Big], \\
        F_{1,4} &= \frac{\alpha}{(1-x)} \Big[ 2m^2(1+x)(1-\xi^2) \Big].
    \end{align}

    \item \textbf{Longitudinally polarized quark:}
    \begin{align}
        G_{1,1} &= -\frac{\alpha}{(1-x)} \Big[ 2m^2(1+x)(1-\xi^2) \Big], \\
        G_{1,2} &= \beta \Big[ 2m^2 \big( (1-x)\xi \vec{\Delta}_\perp^{\,2} - 2(x+\xi^2) \vec{k}_\perp \cdot \vec{\Delta}_\perp \big) - (1+x)(k_2 \Delta_1 - k_1 \Delta_2)\vec{\Delta}_\perp^{\,2} \Big], \\
        G_{1,3} &= \frac{\beta}{4(1-x)^2} \Big[ (k_2 \Delta_1 - k_1 \Delta_2) \big( \xi(1+x^2 - 2\xi^2)(4(1-\xi^2)\vec{k}_\perp^{\,2} - (1-x)^2 \vec{\Delta}_\perp^{\,2}) \nonumber \\
        & \quad + 4(1-\xi^2)(1-x)(1-x^2 + 2\xi^2) \vec{k}_\perp \cdot \vec{\Delta}_\perp \big) \nonumber \\
        & \quad + 4m^2(1-x)^2 \big( 4(x+\xi^2)\vec{k}_\perp^{\,2} - \xi(1-x)\big(2 \vec{k}_\perp \cdot \vec{\Delta}_\perp + (1-x)(k_2 \Delta_1 - k_1 \Delta_2) \big) \big) \Big], \\
        G_{1,4} &= \frac{\alpha(1-\xi^2)}{4(1-x)^3} \Big[ (1+x^2 - 2\xi^2) \big( 4(1-\xi^2)\vec{k}_\perp^{\,2} + (1-x)\big( 4\xi \vec{k}_\perp - (1-x)\vec{\Delta}_\perp \big) \cdot \vec{\Delta}_\perp \big) \nonumber \\
        & \quad - 4m^2(1-x)^4 \Big].
    \end{align}

    \item \textbf{Transversely polarized quark:}
    \begin{align}
        H_{1,1} &= \beta \Big[ 2m^2(1-\xi^2)(4\xi \vec{k}_\perp \cdot \vec{\Delta}_\perp - (1-x) \vec{\Delta}_\perp^{\,2}) \Big], \\
        H_{1,2} &= -\beta \Big[ 2m^2(1-\xi^2)(4\xi \vec{k}_\perp^{\,2} - (1-x) \vec{k}_\perp \cdot \vec{\Delta}_\perp) \Big], \\
        H_{1,3} &= \frac{\alpha}{(1-x)^3 \vec{k}_\perp \cdot \vec{\Delta}_\perp} \Big[ \vec{k}_\perp \cdot \vec{\Delta}_\perp \big( 4(x - \xi^2)(1 - \xi^2)\vec{k}_\perp^{\,2} + 2\xi(1-x)^2(3 + \xi^2)(k_1 \Delta_2 - k_2 \Delta_1) \big) \nonumber \\
        & \quad + (1 - x)(x - \xi^2) \big( 4\xi(k_1^2 \Delta_1^2 + k_2^2 \Delta_2^2) - (1 - x) \vec{\Delta}_\perp^{\,2} (\vec{k}_\perp \cdot \vec{\Delta}_\perp) \big) \nonumber \\
        & \quad + \xi(1-x)^2 \big( 4(k_1^2 \Delta_2^2 + k_2^2 \Delta_1^2) - \xi(1-x)\vec{\Delta}_\perp^{\,2}(k_1 \Delta_2 - k_2 \Delta_1) \big) \nonumber \\
        & \quad - 8\xi(1 - x)(1 - 2x + \xi^2)k_1k_2\Delta_1\Delta_2 \Big], \\
        H_{1,4} &= \frac{\beta}{\vec{k}_\perp \cdot \vec{\Delta}_\perp} \Big[ m^2 \xi \vec{\Delta}_\perp^{\,2} \big( 2(1+\xi^2) \vec{k}_\perp \cdot \vec{\Delta}_\perp + 4(k_1 \Delta_2 - k_2 \Delta_1) - (1 - x)\xi \vec{\Delta}_\perp^{\,2} \big) \Big], \\
        H_{1,5} &= -\beta \Big[ m^2 \big( 2\xi(3 + \xi^2)\vec{k}_\perp \cdot \vec{\Delta}_\perp + 8\xi(k_1 \Delta_2 - k_2 \Delta_1) - (1 - x)(1 + \xi^2)\vec{\Delta}_\perp^{\,2} \big) \Big], \\
        H_{1,6} &= \frac{\beta}{\vec{k}_\perp \cdot \vec{\Delta}_\perp} \Big[ m^2 \big( 4\xi \vec{k}_\perp^{\,2}(\vec{k}_\perp \cdot \vec{\Delta}_\perp - (k_2 \Delta_1 - k_1 \Delta_2)) - (1 - x)(k_1^2(\Delta_1^2 + \xi^2 \Delta_2^2) \nonumber \\
        & \quad + k_2^2(\xi^2 \Delta_1^2 + \Delta_2^2)) - (1 - x)(1 - \xi^2)k_1k_2\Delta_1\Delta_2 \big) \Big], \\
        H_{1,7} &= -\beta \Big[ m^2(1 - \xi^2)\big(2(1 + \xi^2) \vec{k}_\perp \cdot \vec{\Delta}_\perp - (1 - x)\xi \vec{\Delta}_\perp^{\,2} \big) \Big], \\
        H_{1,8} &= \beta \Big[ m^2(1 - \xi^2)\big(2(1 + \xi^2) \vec{k}_\perp^{\,2} - (1 - x)\xi \vec{k}_\perp \cdot \vec{\Delta}_\perp \big) \Big].
    \end{align}
\end{itemize}

The and auxiliary functions $\alpha$, and $D$ are defined as:
\begin{align}
    D(\vec{k}_\perp, x) &= m^2 - \frac{m^2 + \vec{k}_\perp^{\,2}}{x} - \frac{\vec{k}_\perp^{\,2}}{1 - x}, \\
    \alpha(x, \xi, \vec{k}_\perp^{\,2}, \vec{\Delta}_\perp^{\,2}, \vec{k}_\perp \cdot \vec{\Delta}_\perp) &= \frac{N}{D(q_\perp, y) D^*(q_\perp', x')(x^2 - \xi^2)},
\end{align}
where $N = \frac{g^2 C_f}{2(2\pi)^3}$, with $g$ being the strong coupling constant and $C_f$ the color factor. For notational simplicity, we denote $\alpha(x, \xi, \vec{k}_\perp^{\,2}, \vec{\Delta}_\perp^{\,2}, \vec{k}_\perp \cdot \vec{\Delta}_\perp) \equiv \alpha$, and define a new function $\beta$ as:
\begin{align}
    \beta \equiv \frac{\alpha}{(1-x)(k_2 \Delta_1 - k_1 \Delta_2)}.
\end{align}

\bibliographystyle{unsrt}
\bibliography{Ref.bib}

\begin{thebibliography}{10}

\bibitem{Accardi:2012qut}
A.~Accardi et~al.
\newblock {Electron Ion Collider: The Next QCD Frontier}: {Understanding the glue that binds us all}.
\newblock {\em Eur. Phys. J. A}, 52(9):268, 2016.

\bibitem{Anderle:2021wcy}
Daniele~P. Anderle et~al.
\newblock {Electron-ion collider in China}.
\newblock {\em Front. Phys. (Beijing)}, 16(6):64701, 2021.

\bibitem{gluck1998dynamical}
M~Gl{\"u}ck, E~Reya, and A~Vogt.
\newblock Dynamical parton distributions revisited.
\newblock {\em The European Physical Journal C-Particles and Fields}, 5(3):461--470, 1998.

\bibitem{gluck1995dynamical}
Moshe Gl{\"u}ck, Ewald Reya, and Andreas Vogt.
\newblock Dynamical parton distributions of the proton and small-x physics.
\newblock {\em Zeitschrift f{\"u}r Physik C Particles and Fields}, 67(3):433--447, 1995.

\bibitem{martin1998parton}
Alan~D Martin, RG~Roberts, WJames Stirling, and RS~Thorne.
\newblock Parton distributions: A new global analysis.
\newblock {\em The European Physical Journal C-Particles and Fields}, 4(3):463--496, 1998.

\bibitem{Lorce:2025aqp}
C{\'e}dric Lorc{\'e}, A.~Metz, B.~Pasquini, and P.~Schweitzer.
\newblock {Parton Distribution Functions and their Generalizations}.
\newblock 7 2025.

\bibitem{Karr:2020wgh}
Jean-Philippe Karr, Dominique Marchand, and Eric Voutier.
\newblock {The proton size}.
\newblock {\em Nature Rev. Phys.}, 2(11):601--614, 2020.

\bibitem{Gao:2017yyd}
Jun Gao, Lucian Harland-Lang, and Juan Rojo.
\newblock {The Structure of the Proton in the LHC Precision Era}.
\newblock {\em Phys. Rept.}, 742:1--121, 2018.

\bibitem{Diehl:2023nmm}
Stefan Diehl.
\newblock {Experimental exploration of the 3D nucleon structure}.
\newblock {\em Prog. Part. Nucl. Phys.}, 133:104069, 2023.

\bibitem{Constantinou:2020pek}
Martha Constantinou.
\newblock {The x-dependence of hadronic parton distributions: A review on the progress of lattice QCD}.
\newblock {\em Eur. Phys. J. A}, 57(2):77, 2021.

\bibitem{belitsky2005unraveling}
Andrei~V Belitsky and AV~Radyushkin.
\newblock Unraveling hadron structure with generalized parton distributions.
\newblock {\em Physics reports}, 418(1-6):1--387, 2005.

\bibitem{diehl2003generalized}
Markus Diehl.
\newblock Generalized parton distributions.
\newblock {\em Physics Reports}, 388(2-4):41--277, 2003.

\bibitem{ji1997deeply}
Xiangdong Ji.
\newblock Deeply virtual compton scattering.
\newblock {\em Physical Review D}, 55(11):7114, 1997.

\bibitem{Mukherjee:2013yf}
Asmita Mukherjee, Sreeraj Nair, and Vikash Kumar~Ojha.
\newblock {Generalized Parton Distributions of the Photon with Helicity Flip}.
\newblock {\em Phys. Lett. B}, 721:284--289, 2013.

\bibitem{brodsky2002final}
Stanley~J Brodsky, Dae~Sung Hwang, and Ivan Schmidt.
\newblock Final-state interactions and single-spin asymmetries in semi-inclusive deep inelastic scattering.
\newblock {\em Physics Letters B}, 530(1-4):99--107, 2002.

\bibitem{bacchetta2007semi}
Alessandro Bacchetta, Markus Diehl, Klaus Goeke, Andreas Metz, Piet~J Mulders, and Marc Schlegel.
\newblock Semi-inclusive deep inelastic scattering at small transverse momentum.
\newblock {\em Journal of High Energy Physics}, 2007(02):093, 2007.

\bibitem{barone2002transverse}
Vincenzo Barone, Alessandro Drago, and Philip~G Ratcliffe.
\newblock Transverse polarisation of quarks in hadrons.
\newblock {\em Physics reports}, 359(1-2):1--168, 2002.

\bibitem{mulders1996complete}
PJ~Mulders and RD~Tangerman.
\newblock The complete tree-level result up to order 1/q for polarized deep-inelastic leptoproduction.
\newblock {\em Nuclear Physics B}, 461(1-2):197--237, 1996.

\bibitem{Boussarie:2023izj}
Renaud Boussarie et~al.
\newblock {TMD Handbook}.
\newblock 4 2023.

\bibitem{Echevarria:2016scs}
Miguel~G. Echevarria, Ignazio Scimemi, and Alexey Vladimirov.
\newblock {Unpolarized Transverse Momentum Dependent Parton Distribution and Fragmentation Functions at next-to-next-to-leading order}.
\newblock {\em JHEP}, 09:004, 2016.

\bibitem{wigner1932quantum}
Eugene Wigner.
\newblock On the quantum correction for thermodynamic equilibrium.
\newblock {\em Physical review}, 40(5):749, 1932.

\bibitem{ji2003viewing}
Xiangdong Ji.
\newblock Viewing the proton through “color” filters.
\newblock {\em Physical review letters}, 91(6):062001, 2003.

\bibitem{Belitsky:2003nz}
Andrei~V. Belitsky, Xiang-dong Ji, and Feng Yuan.
\newblock {Quark imaging in the proton via quantum phase space distributions}.
\newblock {\em Phys. Rev. D}, 69:074014, 2004.

\bibitem{radhakrishnan2022wigner}
Ramkumar Radhakrishnan and Vikash~Kumar Ojha.
\newblock Wigner distribution of sine-gordon and kink solitons.
\newblock {\em Modern Physics Letters A}, 37(37n38):2250236, 2022.

\bibitem{meissner2009generalized}
Stephan Mei{\ss}ner, Andreas Metz, and Marc Schlegel.
\newblock Generalized parton correlation functions for a spin-1/2 hadron.
\newblock {\em Journal of High Energy Physics}, 2009(08):056, 2009.

\bibitem{lorce2013structure}
C{\'e}dric Lorce and Barbara Pasquini.
\newblock Structure analysis of the generalized correlator of quark and gluon for a spin-1/2 target.
\newblock {\em Journal of High Energy Physics}, 2013(9):1--30, 2013.

\bibitem{Lorce:2011dv}
Cedric Lorce, Barbara Pasquini, and Marc Vanderhaeghen.
\newblock {Unified framework for generalized and transverse-momentum dependent parton distributions within a 3Q light-cone picture of the nucleon}.
\newblock {\em JHEP}, 05:041, 2011.

\bibitem{lorce2012quark}
C{\'e}dric Lorc{\'e}, Barbara Pasquini, Xiaonu Xiong, and Feng Yuan.
\newblock Quark orbital angular momentum from wigner distributions and light-cone wave functions.
\newblock {\em Physical Review D}, 85(11):114006, 2012.

\bibitem{lorce2011quark}
Cedric Lorce and Barbara Pasquini.
\newblock Quark wigner distributions and orbital angular momentum.
\newblock {\em Physical Review D}, 84(1):014015, 2011.

\bibitem{mukherjee2015wigner}
Asmita Mukherjee, Sreeraj Nair, and Vikash~Kumar Ojha.
\newblock Wigner distributions for gluons in a light-front dressed quark model.
\newblock {\em Physical Review D}, 91(5):054018, 2015.

\bibitem{mukherjee2014quark}
Asmita Mukherjee, Sreeraj Nair, and Vikash~Kumar Ojha.
\newblock Quark wigner distributions and orbital angular momentum in light-front dressed quark model.
\newblock {\em Physical Review D}, 90(1):014024, 2014.

\bibitem{kaur2018wigner}
Satvir Kaur and Harleen Dahiya.
\newblock Wigner distributions and gtmds in a proton using light-front quark--diquark model.
\newblock {\em Nuclear Physics B}, 937:272--302, 2018.

\bibitem{liu2015quark}
Tianbo Liu and Bo-Qiang Ma.
\newblock Quark wigner distributions in a light-cone spectator model.
\newblock {\em Physical Review D}, 91(3):034019, 2015.

\bibitem{chakrabarti2016wigner}
D~Chakrabarti, T~Maji, C~Mondal, and A~Mukherjee.
\newblock Wigner distributions and orbital angular momentum of a proton.
\newblock {\em The European Physical Journal C}, 76(7):1--16, 2016.

\bibitem{chakrabarti2017quark}
D~Chakrabarti, T~Maji, C~Mondal, and A~Mukherjee.
\newblock Quark wigner distributions and spin-spin correlations.
\newblock {\em Physical Review D}, 95(7):074028, 2017.

\bibitem{more2017quark}
Jai More, Asmita Mukherjee, and Sreeraj Nair.
\newblock Quark wigner distributions using light-front wave functions.
\newblock {\em Physical Review D}, 95(7):074039, 2017.

\bibitem{more2018wigner}
Jai More, Asmita Mukherjee, and Sreeraj Nair.
\newblock Wigner distributions for gluons.
\newblock {\em The European Physical Journal C}, 78:1--15, 2018.

\bibitem{Jana:2023btd}
Sujit Jana, Vikash~Kumar Ojha, and Tanmay Maji.
\newblock {Gluon generalized TMDs and Wigner distributions in boost invariant longitudinal space}.
\newblock {\em Nucl. Phys. A}, 1053:122958, 2025.

\bibitem{ojha2023quark}
Vikash~Kumar Ojha, Sujit Jana, and Tanmay Maji.
\newblock Quark generalized tmds at skewness and wigner distributions in boost invariant longitudinal space.
\newblock {\em Physical Review D}, 107(7):074040, 2023.

\bibitem{maji2022leading}
Tanmay Maji, Chandan Mondal, and Daekyoung Kang.
\newblock Leading twist gtmds at nonzero skewness and wigner distributions in boost-invariant longitudinal position space.
\newblock {\em Physical Review D}, 105(7):074024, 2022.

\bibitem{Guo:2023ahv}
Yuxun Guo, Xiangdong Ji, M.~Gabriel Santiago, Kyle Shiells, and Jinghong Yang.
\newblock {Generalized parton distributions through universal moment parameterization: non-zero skewness case}.
\newblock {\em JHEP}, 05:150, 2023.

\bibitem{Mamo:2022jhp}
Kiminad~A. Mamo and Ismail Zahed.
\newblock {Quark and gluon GPDs at finite skewness from strings in holographic QCD: Evolved and compared with experiment}.
\newblock {\em Phys. Rev. D}, 108(8):086026, 2023.

\bibitem{Rinaldi:2017roc}
Matteo Rinaldi.
\newblock {GPDs at non-zero skewness in ADS/QCD model}.
\newblock {\em Phys. Lett. B}, 771:563--567, 2017.

\bibitem{Chakrabarti:2024hwx}
Dipankar Chakrabarti, Poonam Choudhary, Bheemsehan Gurjar, Tanmay Maji, Chandan Mondal, and Asmita Mukherjee.
\newblock {Gluon generalized parton distributions of the proton at nonzero skewness}.
\newblock {\em Phys. Rev. D}, 109(11):114040, 2024.

\bibitem{Chen:2019lcm}
Jiunn-Wei Chen, Huey-Wen Lin, and Jian-Hui Zhang.
\newblock {Pion generalized parton distribution from lattice QCD}.
\newblock {\em Nucl. Phys. B}, 952:114940, 2020.

\bibitem{Riberdy:2023awf}
Michael~Joseph Riberdy, Herv{\'e} Dutrieux, C{\'e}dric Mezrag, and Pawe{\l} Sznajder.
\newblock {Combining lattice QCD and phenomenological inputs on generalised parton distributions at moderate skewness}.
\newblock {\em Eur. Phys. J. C}, 84(2):201, 2024.

\bibitem{Mamo:2024vjh}
Kiminad~A. Mamo and Ismail Zahed.
\newblock {String-based parametrization of nucleon GPDs at any skewness: A comparison to lattice QCD}.
\newblock {\em Phys. Rev. D}, 110(11):114016, 2024.

\bibitem{Bhattacharya:2018lgm}
Shohini Bhattacharya, Andreas Metz, Vikash~Kumar Ojha, Jeng-Yuan Tsai, and Jian Zhou.
\newblock {Exclusive double quarkonium production and generalized TMDs of gluons}.
\newblock {\em Phys. Lett. B}, 833:137383, 2022.

\bibitem{Bhattacharya:2023yvo}
Shohini Bhattacharya, Duxin Zheng, and Jian Zhou.
\newblock {Accessing the gluon GTMD F1,4 in exclusive {\ensuremath{\pi}}0 production in ep collisions}.
\newblock {\em Phys. Rev. D}, 109(9):096029, 2024.

\bibitem{harindranath1996introduction}
Avaroth Harindranath.
\newblock An introduction to light-front dynamics for pedestrians.
\newblock {\em arXiv preprint hep-ph/9612244}, 1996.

\bibitem{zhang1994light}
Wei-Min Zhang.
\newblock Light-front dynamics and light-front qcd.
\newblock {\em arXiv preprint hep-ph/9412244}, 1994.

\bibitem{brodsky2001light}
Stanley~J Brodsky, Markus Diehl, and Dae~Sung Hwang.
\newblock Light-cone wavefunction representation of deeply virtual compton scattering.
\newblock {\em Nuclear Physics B}, 596(1-2):99--124, 2001.

\bibitem{kumar2015single}
Narinder Kumar and Harleen Dahiya.
\newblock Single transverse spin asymmetries in semi-inclusive deep inelastic scattering in a spin-1 diquark model.
\newblock {\em The European Physical Journal A}, 51(4):51, 2015.

\bibitem{bacchetta2008transverse}
Alessandro Bacchetta, Francesco Conti, and Marco Radici.
\newblock Transverse-momentum distributions in a diquark spectator model.
\newblock {\em Physical Review D}, 78(7):074010, 2008.

\bibitem{bacchetta2016electron}
Alessandro Bacchetta, Luca Mantovani, and Barbara Pasquini.
\newblock Electron in three-dimensional momentum space.
\newblock {\em Physical Review D}, 93(1):013005, 2016.

\bibitem{harindranath1999nonperturbative}
A~Harindranath, Rajen Kundu, and Wei-Min Zhang.
\newblock Nonperturbative description of deep inelastic structure functions in light-front qcd.
\newblock {\em Physical Review D}, 59(9):094012, 1999.

\bibitem{harindranath1999orbital}
A~Harindranath and Rajen Kundu.
\newblock Orbital angular momentum in deep inelastic scattering.
\newblock {\em Physical Review D}, 59(11):116013, 1999.

\bibitem{zhang1993light}
Wei-Min Zhang and Avaroth Harindranath.
\newblock Light-front qcd. ii. two-component theory.
\newblock {\em Physical Review D}, 48(10):4881, 1993.

\bibitem{maji2016light}
Tanmay Maji and Dipankar Chakrabarti.
\newblock Light front quark-diquark model for the nucleons.
\newblock {\em Physical Review D}, 94(9):094020, 2016.

\bibitem{mukherjee2013generalized}
Asmita Mukherjee, Sreeraj Nair, and Vikash~Kumar Ojha.
\newblock Generalized parton distributions of the photon with helicity flip.
\newblock {\em Physics Letters B}, 721(4-5):284--289, 2013.

\end{thebibliography}
\end{document}